\documentclass[12pt]{article}
\usepackage{epsfig,amsfonts,amssymb,}
\usepackage{cite}
\usepackage{amsmath, leftidx}
\usepackage{cite}
\usepackage{comment}
\usepackage{IEEEtrantools}
\usepackage{graphicx}

\usepackage[dvipsnames]{xcolor}
\usepackage{tikz}
\usetikzlibrary{shapes.misc}

\def\ZZZ{{\hbox{ Z\kern-1.6mm Z}}}
\def\RRR{{\hbox{ R\kern-2.4mm R}}}
\def\CCC{{\hbox{ C\kern-2.0mm C}}}
\def\zzz{{\hbox{z\kern-1mm z}}}

\newcommand{\qeq}{{\hbox{=\kern-2.3mm ? \kern.5mm }}}
\renewcommand{\qeq}{=}

\newcommand{\hdelta}{\hat{\delta}}

\newcommand{\be}{\begin{equation}}
\newcommand{\ee}{\end{equation}}
\newcommand{\ben}{\begin{eqnarray}\displaystyle}
\newcommand{\een}{\end{eqnarray}}
\newcommand{\rsp}{&\mathrel{\phantom{=}}}

\def\one{{\hbox{ 1\kern-.8mm l}}}
\def\zero{{\hbox{ 0\kern-1.5mm 0}}}

\newcommand{\bea}[1]{\begin{eqnarray}\label{#1} }
\newcommand{\eea}{\end{eqnarray}}

\usepackage{bm}

\def\ol{\overline{l}}

\def\hcr{{\cal R}}

\def\tp{\tilde{p}}
\def\tipo{\tilde{p}_{1}}
\def\tipt{\tilde{p}_{2}}
\def\po{p_{1}}
\def\pt{p_{2}}
\def\amp{{\cal A}_{5}(\tilde{p}_{1},\, \tilde{p}_{2}\, \rightarrow\, p_{1},\, p_{2},\, k)}
\def\ramp{{\cal M}_{5}(\tilde{p}_{1},\, \tilde{p}_{2}\, \rightarrow\, p_{1},\, p_{2},\, k)}
\def\usafour{{\cal A}_{4}(\tilde{p}_{1},\, \tilde{p}_{2}\, \rightarrow\, p_{1},\, p_{2})}

\begin{document}

\baselineskip 24pt

\begin{center}

{\Large \bf Soft Radiation from Scattering Amplitudes Revisited}

\end{center}

\vskip .6cm
\medskip

\vspace*{4.0ex}

\baselineskip=18pt

\centerline{\large \rm Manu A$^{a}$, Debodirna Ghosh$^{b}$, Alok Laddha$^{c}$ and Athira P V $^{d}$}

\vspace*{4.0ex}

\centerline{\large \it Chennai Mathematical Institute, Siruseri, Chennai, India}



\vspace*{1.0ex}
\centerline{\small E-mail:  manu@cmi.ac.in, debodirna@cmi.ac.in, aladdha@cmi.ac.in, athira@cmi.ac.in}

\vspace*{5.0ex}

\centerline{\bf Abstract} \bigskip

We apply the recently developed formalism by Kosower, Maybee and O'Connell (KMOC) \cite{kosower} to analyse the  soft electromagnetic and soft gravitational radiation emitted by particles without spin in $D\, \geq\, 4$ dimensions.  We use this formalism in conjunction with quantum soft theorems to derive radiative electro-magnetic and gravitational fields in low frequency expansion and upto next to leading order in the coupling. We show that in \emph{all} dimensions, the classical limit of sub-leading soft (photon and graviton) theorems is consistent with the classical soft theorems proved by Sen et al in a series of papers. In particular in \cite{aab} Saha, Sahoo and Sen proved classical soft theorems for electro-magnetic and gravitational radiation in $D\, =\, 4$ dimensions. For the class of scattering processes that can be analyzed using KMOC formalism,  we show that the classical limit of quantum soft theorems is consistent with the $D\, =\, 4$ classical soft theorems, paving the way for their proof from scattering amplitudes. 



\vfill \eject

\baselineskip 18pt

\tableofcontents

\section{Introduction}
Soft theorems in quantum field theories are universal statements about factorisation of scattering amplitudes in gauge theories and gravity \cite{weinberg, bern1, ashoke1, cachazo1, ashoke2}.
Classical Soft theorems \cite{ashoke1801,ashoke1906} are exact statement about low frequency radiation emitted during  generic scattering processes.   As such they are a consequences of the under-lying gauge invariance of the theory and capture the universality of low frequency radiation \cite{ashoke1906}.   In $D\ >\ 4$ dimensions, these theorems were first derived as classical limits of quantum soft theorems \cite{ashoke1801}. It was shown in \cite{ashoke1801}  that for a class of scattering processes which could be classified by  either  large impact parameter or   (in the case of $2\, \rightarrow\, 2$ scattering) so-called probe scatterer approximation (an approximation in which ratio of scatterer mass to probe mass is large), quantum soft theorems could be used to compute low frequency classical radiation. More in detail, it was shown that   extremizing the probability distribution of emitted soft quanta in a given frequency bin is tantamount to taking the classical limit and results in classical radiation arranged in soft frequency expansion. The probability distribution was in turn obtained from the multi-soft graviton theorem \cite{ashoke17}. The final result is rather simple to state. The radiative field at long distances is  proportional to the  ``classical limit" of a single soft factor  where momentum and angular momentum operators in quantum theory are replaced by their classical counter-parts. Hence such low frequency radiative fields are called classical soft factors.

 In \cite{ashoke1804}, these ideas were used to propose a definition of classical soft factor in $D\ =\ 4$ dimensions. The essential departure from higher dimensions was the long range infra-red effect which causes scattering particles to radiate even asymptoticallly. The soft expansion then contained a new term which was proportional to $\ln\omega$ (where $\omega$ is the frequency of radiation).  It was explicitly checked in a number of examples that in the soft expansion of classical radiation in four dimensions, this term was indeed present. The soft factor is called classical log soft factor.  In a seminal paper, Sahoo and Sen \cite{sahoo} showed that soft theorems in QED and quantum gravity were loop corrected in $D\, =\, 4$ dimensions. Although the leading Weinberg soft factor remained un-effected, the tree-level soft expansion breaks down at sub-leading order in the soft expansion due to a new term which is proportional to $\ln\omega$. Just like Weinberg soft factor, this term was shown to be universal and one loop exact, resulting in a new factorisation theorem for loop corrected Scattering amplitudes in QED and quantum gravity. 

In \cite{aab}, Saha, Sahoo and Sen extended the proof of \cite{ashoke1906} to four  dimensions and proved the proposal in \cite{ashoke1804}. However unlike in $D\, >\,4$ dimensions, where the classical soft radiation can be derived from quantum soft theorems by a careful analysis of classical limit,  no such derivation exists in four dimensions. And the proof is likely to be more intricate then the corresponding proof in higher dimensions. In higher ($D\, >\, 4$) dimensions, the classical and quantum soft factors were related by simply replacing the linear and angular momentum operators in quantum theory with the classical counterparts. However  the quantum log soft factor derived in \cite{sahoo} was sum of two terms in which one term is precisely the classical log soft factor proposed in \cite{ashoke1804}.  The other term however is absent in the classical radiation. This term is not manifestly quantum (in the sense of being higher order in $\hbar$) and the precise reasons for it's disappearence in the classical limit remains unclear. But as the classical limit of soft theorem is subtle \cite{ashoke1801}, it is expected that this term would vanish under careful analysis of the classical limit.\footnote{In $D\, >\, 4$ dimensions, the classical soft factor was essentially obtained by taking quantum soft factor and replacing quantum operators by their classical counter parts, this substitution did not produce $D\, =\, 4$ classical log soft factor from the quantum counter part as the quantum log soft factor had certain additional terms.}
 
A novel formalism to obtain classical radiation and other classical observables such as momentum impulse  from scattering amplitude was developed in \cite{kosower} by Kosower, Maybee and O'Connell (KMOC).  The central idea of the KMOC formalism could be summarised in two steps. In a $2\, \rightarrow\, 2$ scattering, we start with a wave packet in the far past which is peaked around certain momenta of the two particles. We then evolve the state using the S-matrix and use the final state to compute expectation values of quantum observables.  
Classical limit of the expectation value is obtained by interpreting classical expansion as a large impact parameter expansion.\footnote{This  formalism assumes that we are in  the large  impact parameter regime  and to the best of our knowledge, it is not clear how to generalise it to other scenarios, e.g. the Probe scatterer approximation.}  

The formalism synthesized various recent developments of obtaining classical observables from quantum amplitudes in a coherent framework. Power of the formalism lies in the fact that the classical limit is taken already at the level of loop integrands contained in the perturbative expansion of the scattering amplitude. On one hand, this drastically simplifies the ``quantum" computation as only a subset of Feynman diagrams contribute in this limit and on the other hand, the powerful techniques available for analysing higher loop amplitudes could be used to perform the computations. Thanks to these advances and a beautiful relationship between adiabatic invariants in a bound binary system with observables for classical scattering processes ( for a rigorous derivation of this relationship in the classical theory itself, see \cite{portok,portok1} ) striking results in analysing various aspects of the conservative dynamics of the  spinning binary systems have been obtained in recent years.  For a sampling of some of these results we refer the reader to following papers and references therein \cite{cfq1,cfq2,cfq2.4,cfq2.5,cfq3,cfq3.5,cfq3.6,cfq4,cfq4.5,cfq5,cfq6,cfq7,cfq8}. 

In this paper, we analyse radiative (as opposed to conservative) sector in soft frequency expansion. The radiative sector has been relatively less studied using modern tools of scattering amplitudes. Notable exceptions are ( \cite{wi}, \cite{gb, gb1}). The question we ask  is if we can use the KMOC formalism to prove classical soft theorem from quantum soft theorem in $D\, \geq\, 4$ dimensions in QED and gravity.  As we work in the large impact parameter regime, where the contribution of spin angular momenta to the soft factor is sub-dominant compared to orbital angular momenta, we work with particles without spin.  We show that the classical limit of soft photon/graviton theorem produces the classical log soft factor upto next to leading order (NLO) in the coupling. We believe this provides important first steps towards giving a comprehensive proof of the classical soft theorem from quantum soft theorem.  We note that at leading order in the frequency, that is when we consider Weinberg soft photon theorem in the quantum amplitude,  this result was already established in a seminal paper by Bautista and Guevara \cite{gb}. We generalise this result to sub-leading order in the soft expansion.  

Our analysis also reveals a rather nice surprise when using KMOC formalism to analyse soft radiation. Namely that in $D\, =\, 4$ dimensions,  even tree-level scattering amplitudes produce soft radiation that has logarithmic dependence on radiation frequency. The log dependence on soft frequency arises due to integration over phase space of initial scattering states.  We also remark that apriori, there is a puzzling aspect to the KMOC formalism in that as the amplitude is contructed with Feynman propagators, it is unclear how the classical limit of quantum radiation will match with a classical computation based on retarded propagator. In fact, as was argued in \cite{sahoo}, it was precisely this difference  that was responsible for the discrepancy between classical and quantum log soft factors as shown in \cite{sahoo, aab}.  However the reason, poles of Feynman propagator do not directly contribute in the classical limit is precisely due to the fact that all the states are on-shell. In the classical limit, this constraint ensures that the corresponding poles have vanishing residues.\footnote{This will becomes empirically clear through the number of computations we do in the main sections of the paper.} 

We would like to emphasize that none of our results are new. They merely re-affirm (in the context of large impact parameter scattering) the results establlished in \cite{ashoke1801,ashoke1804, ashoke1906, sahoo, aab}. However we believe that the KMOC formalism sheds new light on the relationship between quantum and classical soft theorems and  provides a potentially powerful framework to analyse higher order terms in soft expansion directly from scattering amplitudes. Our work is a small step in this direction. 

The paper is organised as follows. In section \ref{cspfdg4}, we  derive the soft electro-magnetic radiation at $O(\omega^{0}$) in $D\, >\, 4$ dimensions  by starting with the set up of \cite{golrid, chen}.  In section \ref{csptdg4fq}, we show that one obtains the same result via KMOC formalism when we use sub-leading soft photon theorem to evaluate tree-level scattering amplitude. In section \ref{nloem4d}, we extend the computation of \ref{csptdg4fq} to  four  dimensions and show that one obtains soft radiation which scales as $\ln\omega$ with soft frequency. This result matches with the classical log soft factor obtained in \cite{aab} at leading order in the coupling. In section \ref{nloem4d}, we analyse the soft electromagnetic radiation using KMOC formalism at next to leading order (NLO), which requires computation of one loop soft amplitude. We use quantum soft theorems in four dimensions of Sahoo and Sen \cite{sahoo} and show that the resulting classical limit is in agreement with classical log soft factor at NLO. In sections (\ref{scsgfde4}, \ref{csgfde4nlo}) we repeat this analysis for gravity. We end with some discussion on open issues. Appendices contain proof of certain key identities used in the main text of the paper. 

\section*{Set up}  
Classical soft theorems are stated in terms of initial and final momenta.  Our analysis is based on the set up proposed in \cite{golrid} in which in the classical theory, one starts with initial momenta and use the equations of motion to determine the final momenta and computes the radiation in small deflection (large impact parameter) regime. It is this set up which is the basis of KMOC formalism. Due to this, there are several technical differences with the computations of \cite{ashoke1801, aab}. 

In particular, the soft theorems as stated are exact statements and seen from the perspective of the set up used in \cite{golrid, kosower}, they are obtained by resumming the perturbative expansion of final momenta in terms of initial momenta. Hence a complete derivation of the soft theorem from perturbative amplitudes appears to be formidable. We do not meet this challenge in this paper and only confine ourself to give a ``perturbative evidence" for the proof of classical soft theorem from quantum amplitudes.

\section{Brief Review of classical Soft theorems} 
In this section we review the classical soft theorems derived by Sen and his collaborators in a series of papers. Our primary focus is on the remarkable soft theorems proved by Saha, Sahoo and Sen in $D\, =\, 4$ dimensions. \cite{aab}

We first review the classical soft photon theorem in $D\, \geq\, 4$ dimensions. Given a scattering process, where incoming  classical particles\footnote{These particles can have infinitely many multipole moments and hence also describe composite objects like stars and black holes.} with momenta $\{ p_{1},\, \dots,\, p_{n}\}$  and charges $\{q_{1},\, \dots\, q_{n}\, \}$ scatter into outgoing states with momenta $\{p_{1}^{\prime},\, \dots,\, p_{m}^{\prime}\}$ and charges $\{q_{1}^{\prime}\, \dots,\, q_{m}^{\prime}\}$, the theorem states that the radiative gauge field at sub-leading order in frequency is given by,
\begin{flalign}\label{revissfcl}
J^{\mu}(\omega, \hat{k})\, \sim\, f_{D}(\omega)\, \big(\, \sum_{a=1}^{n}\, q_{a}\, S^{(1) \mu}(\{ p_{a}\}, \hat{k})\, +\, \sum_{a=1}^{m}\, q_{a}^{\prime}\, S^{(1) \mu}(\{ p^{\prime}_{a}\}, \hat{k})\, \big)
\end{flalign}
where we have suppressed the leading order term in the soft expansion given by the Weinberg's soft photon factor.  $f_{D}(\omega)\, =\, \omega^{0}$ for $D\, >\, 4$ and $=\, \ln\omega$ for $D\, =\, 4$.\footnote{Strictly speaking the frequency dependence in $D = 4$ dimensions is more subtle. It is $\ln(\omega \pm i\epsilon)$ for incoming/out-going particles respectively. This detail will be important in the main section of the paper, but we suppress it in eqn.(\ref{revissfcl}).} $S^{(1) \mu}(\{ p_{a}\}, \hat{k})$ is known as classical sub-leading soft photon factor and is defined as,
\begin{flalign}\label{ssfrevialld}
\begin{array}{lll}
S^{(1) \mu}(\{p_{a}\}, \hat{k})&\approx\, \frac{J_{a}^{\mu\nu} k_{\nu}}{p_{a} \cdot k}\ \textrm{if}\, D\, >\, 4\\[0.5em]
&=\, \frac{1}{4\pi}\, \sum_{b\vert \sigma(a,b)\, =\, 1}\, q_{a}q_{b}\, \frac{1}{ (\, (p_{a} \cdot p_{b})^{2}\, -\, m_{a}^{2} m_{b}^{2}\, )^{\frac{3}{2}}}\, \frac{k_{\rho}}{p_{a} \cdot k}\, ( p_{a} \wedge p_{b} )^{\mu\rho}\ \textrm{if}\, D\, =\, 4
\end{array}
\end{flalign}
where in the first line in eqn.(\ref{ssfrevialld}) $J_{a}^{\mu\nu}$ is the total angular momentum of the $a$-th particle. In the second line $\sigma(a,b)\, =\, 1$ depending on whether the pair of particles $(a, b)$ are both incoming or both outgoing.\\ 
The approximation sign in the first equation in eqn. (\ref{ssfrevialld}) is to emphasize that the soft factor is not universal \cite{elvang}. The non-universal terms depend on higher derivative contact interactions that may be present. However in the large impact parameter regime, these terms are sub dominant and upto sub-leading order in the frequency, the radiative gauge field is universal, depending only on the asymptotic linear momentum and angular momentum of scattering particles. 
In $D\, =\, 4$ dimensions the situation is significantly more subtle, although the result is even stronger than in  higher dimensions. The new soft factor at order $\ln\omega$ is due to Coulombic interactions which persist even when particles are far apart in the asymptotic region. The log soft factor is universal and does not change under addition of higher dimensional operators in the Lagrangian.\\
The classical soft graviton theorems at sub-leading order are statements regarding universality of low frequency gravitational field in the radiation regime. If we denote the radiative field as $J^{\mu\nu}(\omega, \hat{k})$ then,
\begin{flalign}
J^{\mu\nu}(\omega, \hat{k})\, \sim\, f_{D}(\omega)\, \big(\, \sum_{a=1}^{n}\, S^{(1) \mu\nu}(\{ p_{a}\}, \hat{k})\, +\, \sum_{a=1}^{m}\, S^{(1) \mu\nu}(\{ p^{\prime}_{a}\}, \hat{k})\, \big)
\end{flalign}
$f_{D}(\omega)\, =\, \kappa\, \omega^{0}$ for $D\, >\, 4$ and $=\, \frac{\kappa^{2}}{16\pi}\, \ln\omega$ for $D\, =\, 4$ with $\kappa\, =\, \sqrt{32 \pi G}$.\\ 
$S^{(1) \mu\nu}(\{ p_{a}\}, \hat{k})$ is known as classical sub-leading soft graviton factor and is defined as,
\begin{flalign}\label{ssfrevialldg}
\begin{array}{lll}
S^{(1) \mu\nu}(\{p_{a}\}, \hat{k})&=\, \frac{p_{a}^{(\mu} J_{a}^{\nu)\rho} \hat{k}_{\rho}}{p_{a} \cdot \hat{k}}\, \textrm{if}\, D\, >\, 4\\[0.5em]
&=\, \sum_{b\vert \sigma(a,b)\, =\, 1}\, \frac{p_{a} \cdot p_{b}}{ (\, (p_{a} \cdot p_{b})^{2}\, -\, m_{a}^{2} m_{b}^{2}\, )^{\frac{3}{2}}}\, \{\, \frac{3}{2} p_{a}^{2} p_{b}^{2}\, -\, ( p_{a} \cdot p_{b} )^{2}\, \}\, \frac{k_{\rho}p_{a}^{(\mu}}{p_{a} \cdot k}\, ( p_{a} \wedge p_{b} )^{\nu)\rho}\\[0.5em]
&\hspace*{3.9in} \textrm{if}\, D\, =\, 4
\end{array}
\end{flalign}
As we have emphasized before, soft theorems are exact statements describing electro-magnetic or gravitational radiation in soft frequency expansion. However in the more standard approach to classical radiation (see \cite{golrid} and references therein), one starts with an initial configuration of scattering particles with certain boundary conditions and then computes outgoing radiation in the far future using equations of motion. Seen from this perspective, the soft factors are really ``re-summed results" obtained from classical perturbation theory once we know the exact final momentum of a particle in terms of initial momenta. That is, consider a $2\, \rightarrow\, 2$ scattering process with large impact parameter. These processes can be studied within perturbation theory (with respect to $q$ or $\kappa$).  If $p_{a}^{f}$ is the final momentum of a particle with initial momenta $p_{a}$ then for gravitational scattering,  
\begin{flalign}
p_{a}^{f \mu}\, =\, p_{a}^{\mu}\, +\, \sum_{n=1}^{\infty}\, \kappa^{2n}\, \overset{(n)}{\triangle p_{a}^{\mu}}
\end{flalign}
where $\overset{(1)}{\triangle p_{a}^{\mu}}$ is the leading order (LO) impulse and $\kappa^{2n}$-th term is the $\textrm{N}^{n}$LO order impulse. Thus when we compute soft radiation perturbatively in the coupling, a necessary condition for consistency with the soft factor is that the radiation at any perturbative order is consistent with classical soft factor. 
\section{Brief Review of the KMOC  formalism}\label{kmorev}
In this section we give a cursory review of the KMOC formalism introduced in \cite{kosower}.  We can not do justice to the several nuances and technicalities in their work and hence limit ourselves to the bare essentials which are directly needed in the main sections of the paper. Interested reader is encouraged to consult the original reference as well as \cite{maybee}. 

It appears to be a rather convoluted idea to compute classical observables like  flux of radiation or Scattering angle by first quantizing the theory and then taking  classical limit of quantum observables. However for  past two decades it has been recognised that computing classical observables  using scattering amplitudes  offer enormous simplifications. The reason this is possible is because of the realisation that only a subset of Feynman diagrams contribute in the classical limit and hence the main idea is to isolate this set of diagrams before doing the integration over loop momenta ! The recent computations of gravitational potential at third and fourth Post Minkowskian orders are some of the most striking outcomes of this endeavour \cite{cfq4} \cite{cfq5}.

In \cite{kosower}, Kosower, Maybee and O'Connell synthesized these ideas in a formalism using which classical observables can be computed from scattering amplitudes.  Their  basic idea is to take wave packets for incoming (classical) particles, evolve them using quantum S-matrix operator and then compute expectation value of an observable in the final state. Classical limit was obtained by recognising that in large impact parameter regime, the small $ \vert {\bf q}\vert$ (${\bf q}$ being the momentum transfer) expansion is precisely the classical expansion. Their beautiful analysis has many caveats but in a nutshell, it turns the intuition of defining classical limit as a small $\vert {\bf q} \vert$ expansion  into concrete formulae. 

In four dimensions, any classical observable (e.g. linear momentum impulse suffered by one of the scattering states or flux of radiation emitted in a given frequency bin) is obtained from a quantum field theory computation through following  formula, 
\begin{flalign}\label{kos1}
O^{A}(p_{1},\, p_{2}\, \dots)\, =\, \lim_{\hbar \rightarrow\, 0}\, \hbar^{\beta_{O}}\, [\, \leftidx{_\textrm{in}}{\langle}\, \Psi\vert\, S^{\dagger}\, \hat{O}^{A}\, S\, \vert \Psi\rangle_{\textrm{in}}\, -\, \leftidx{_\textrm{in}}{\langle}\, \Psi\vert\, \hat{O}^{A}\, \vert \Psi\rangle_{\textrm{in}}\, ]
\end{flalign}
This formula expresses expectation value of any observable in a final state which is obtained by evolving an initial 2  particle coherent state in which the 2 particles are separated by an impact parameter $b$. 

The index $A$ on $O^{A}(p_{1},\, p_{2}\, \dots)$ is an abstract index as $O^{A}$ maybe a vector as in the case of momentum impulse or a tensor as in the case of angular momentum impulse. The dots on the right hand side indicate possible dependence of $O$ on other degrees of freedom such as spin. $\beta_{O}$ is the exponent that depends on the observable $O$ and $\vert \Psi\rangle_{\textrm{in}}$ is the incoming  two particle coherent state in which the particles are separated by impact parameter $b$ and their momenta are localised around $p_{1},\, p_{2}$. In the large impact parameter regime, the expectation value of the momenta of the two particles are also centered around $p_{1},\, p_{2}$ respectively. The spread in the initial coherent state is responsible for the momentum transfer between the two particles. (As the impact parameter is large, we expect the momentum transfer to be small as compared to the incoming momenta of the particles.) We now describe the initial state in slightly more detail.
 
If we choose origin of the co-ordinate system to coincide with the initial position of the second particle (that has momentum $p_{2}$ )
\begin{flalign}
\vert \Psi\rangle_{\textrm{in}}\, =\, \int d\mu(p^{\prime}_{1})\, d\mu(p^{\prime}_{2})\, \phi_{1}(p^{\prime}_{1})\, \phi_{2}(p^{\prime}_{2})\, e^{\frac{i b \cdot p^{\prime}_{1}}{\hbar}}\, \vert p^{\prime}_{1},\, p^{\prime}_{2} \rangle
\end{flalign}
	where the measure,
\begin{equation}
\begin{array}{lll}
d\mu(p_{i}^{\prime})\,=\,\frac{d^{4}p_{i}^{\prime}}{(\,2 \pi\,)^{4}}\,(\,2 \pi \,)\theta(p_{i}^{\prime 0})\,\delta(\,p_{i}^{\prime 2}-m_{i}^{2}\,)
\end{array}
\end{equation}
$\phi_{i}(p_{i}^{\prime})$ are relativistic generalisation of non-relativistic Gaussian coherent state, defined as,
\begin{flalign}
\phi_{1}(p_{1}^{\prime})\, =\, \frac{{\cal N}(\zeta)}{m_{1}}\, \exp^{-\, \frac{p_{1}^{\prime} \cdot  p_{1}}{m_{1}^{2}\, \zeta}}
\end{flalign}
This  exponent in the wave function is linear in $p_{1}^{\prime}$.  But it can be readily verified by going to rest frame of $p_{1}$ that in the non-relativistic limit, it reduces to the familiar Gaussian. $p_{1}$ is the 4 momenta ``around which the wave packet is peaked".\\
$\zeta$ is the classicality parameter used in the non-relativistic Gaussian coherent states, $\zeta\, :=\,( \frac{l_{c}}{l_{w}})^{2}$. where $l_{c}$ is the Compton wavelength associated to the particle and $l_{w}$ is the spread and ${\cal N}$ is a normalisation constant.\footnote{As $p_{1}^{2}\, =\, m_{1}^{2}$, it can be shown that the wave function is normalisable with respect to Lorentz invariant measure, \cite{kosower}.}\\
The master formula in eqn.(\ref{kos1}) looks rather abstract. The Right hand side of the equation involves perturbative expansion of the S-matrix . It would be incredibly complicated were it not for the happy facts that, (1) there have been remarkable advances in computing the scattering amplitude at high loop orders in gravity and gauge theories and (2) in the KMOC formalism, one only sums over those Feynman diagrams that dominate when the momentum exchange and loop momenta scale with $\hbar$ in the classical limit.\footnote{In the KMOC formalism, along with taking the small  exchange momentum  limit, one also takes the limit where loop momenta become small as $q^{\mu}\ =\, \hbar\, \overline{q}^{\mu}$. This is motivated by the fact, in the large impact parameter regime, if one considers inelastic scattering then the radiated massless quanta has small momenta. Unitarity constraints then motivate us to scale loop momenta with $\hbar$ as well.} 
  Two examples analysed in great detail in \cite{kosower} are momentum impulse in electro-magnetic scattering and the electro-magnetic radiation at leading order in the coupling. In \cite{maybee}, KMOC formalism was also used to compute the angular momentum impulse in scattering at leading order in the coupling.\\
In the case of linear momentum impulse,   let $\triangle p_{1}^{\mu}$ be the impulse associated to the first particle. Then, as was shown in \cite{kosower}, 
\begin{flalign}\label{kos2}
\begin{array}{lll}
\triangle p_{1}^{\mu}&=\, \lim_{\hbar\, \rightarrow\, 0}\, i\hbar^{2}\int_{l_{1},\, l_{2}}^{\textrm{on-shell}}\, e^{\frac{-i b \cdot l_{1}}{\hbar}}\, {\cal I}^{\mu}\\[0.6em]
{\cal I}^{\mu}&=\, \hbar^{2}\, l_{1}^{\mu}\, {\cal A}_{4}(\, p_{1} + l_{1},\, p_{2} + l_{2}\, \rightarrow\, p_{1},\, p_{2})\, +\, O({\cal A}\, \cdot {\cal A}^{\star})
\end{array}
\end{flalign}
Our notation for exchange momenta is $l^{\mu}$ rather then the standard $q^{\mu}$. This will help us in comparing our integrands in the classical limit with results in \cite{golrid, aab}. We will denote loop momenta as $q^{\mu}$ instead.\\
${\cal A}_{4}$ is the unstripped amplitude. $O({\cal A}\, \cdot {\cal A}^{\star})$ denotes terms which are quadratic in the amplitude. At leading order in the coupling only the first term contributes and is proportional to the tree-level amplitude. 
The $\int_{l_{1},\, l_{2}}^{\textrm{onshell}}$ measure is defined as,
\begin{flalign}
\int_{l_{1}, l_{2}}^{\textrm{onshell}}\, :=\, \int\, \prod_{i}\, \frac{d^{4} l_{i}}{(2\pi)^{4}}\, \hdelta( 2 p_{i} \cdot l_{i} + l_{i}^{2})\, \theta(p_{i}^{0} + l_{i}^{0})
\end{flalign}
It ensures that in the incoming coherent state, one is only summing over on-shell states.  While taking the $\hbar\, \rightarrow\, 0$ limit, one first scales the exchange momenta (and loop momenta) with $\hbar$ as $l^{\mu}\, =\, \hbar\, \overline{l}^{\mu}$, keeps only the leading order terms and integrates over $\overline{l}^{\mu},\, \overline{q}^{\mu}$. The final integration is over wave numbers $\overline{l}^{\mu},\, \overline{q}^{\mu}$ and produces the classical limit. 

Another important result in \cite{kosower} which will be of central importance to us is that of computing emitted radiation. 

For simplicity we review their formula in the case of electro-magnetic scattering, although in section \ref{scsgfde4} we will use the formalism to compute gravitational radiation. To compute radiation, an important intermediate quantity introduced by KMOC is the so-called  radiation kernel ${\cal R}^{\mu}(k, X)$. Radiation kernel is simply  the gauge field radiated at momentum $k^{\mu} =: (\omega,\, \omega\hat{k})$ and is a result of in-elastic scattering where the out-going states can include in addition to the two massive particles and a photon, additional states which are collectively denoted as $X$.  
 ${\cal R}^{\mu}(k, X)$ is associated to  the radiation emitted in a given bin ${\cal J}^{\mu}$ as,
\begin{flalign}
{\cal J}^{\mu}\ =\, \int\, d\mu(\overline{k})\, \overline{k}^{\mu}\, \sum_{X}  \vert\, \epsilon_{\mu}(\overline{k}) \cdot {\cal R}^{\mu}(\overline{k}, X)\, \vert^{2}
\end{flalign}
The reason ${\cal R}^{\mu}$ was introduced is because it's formula has the following compact expression.
\begin{flalign}\label{kos3}
{\cal R}^{\mu}(\overline{k}, X)\, =\, \lim_{\hbar \rightarrow\, 0}\, \hbar^{\frac{3}{2}}\, \int_{l_{1}, l_{2}}^{\textrm{on-shell}}\, \delta^{4}(l_{1} + l_{2} - k - r_{X})\, {\cal A}_{5+X}(\tp_{1},\, \tp_{2}\ \rightarrow\, p_{1},\, p_{2},\, k,\, X) 
\end{flalign}
where $\tp_{i}= p_{i}\,+\,l_{i}$. 
Kosower, Maybee and O' Connell also showed in their paper that classical limit of the  electro-magnetic radiation kernel at leading order in the coupling equals the clasical result computed directly from equations of motion \cite{golrid}. 

KMOC formalism in \cite{kosower} was developed to evaluate classical observables from quantum field theory in four dimensions. But one can readily generalise their formulae to arbitrary dimensions. For example, the formula for electro-magnetic radiation kernel in eqn.(\ref{kos3}) can be generalised as,
\begin{flalign}\label{kos4}
{\cal R}^{\mu}(\overline{k}, X)\, =\, \lim_{\hbar \rightarrow\, 0}\, \hbar^{\frac{3}{2} - (D - 4)}\, \int_{l_{1}, l_{2}}^{\textrm{on-shell}}\, \delta^{D}(l_{1} + l_{2} - k - r_{X})\, {\cal A}_{5+X}(\tp_{1},\, \tp_{2}\ \rightarrow\, p_{1},\, p_{2},\, k,\, X) 
\end{flalign}
where the measure is the on-shell momentum space measure in $D$ space-time dimensions.\\

\subsubsection*{A disclaimer about notation} 
Eqns. (\ref{kos2}), (\ref{kos3}) and (\ref{kos4}) will feature prominently in this paper. Although to take the classical limit, one needs to express all the massless momenta in terms of wave numbers (and hence the integration over momentum exchange and loop momenta is over the wave numbers $\overline{l}^{\mu}$, $\overline{q}^{\mu}$) we will not explicitly introduce the wave numbers in our formula. This is simply to avoid the notational clutter, but it will always be understood that in the integrands that we evaluate for computing classical observables the integration is over wave numbers.\\
Finally in the KMOC formalism a double bracket notation $\langle\langle O^{A}\rangle\rangle$ is used to denote classical limit of a quantum observables. This notation symbolizes integration over the initial momenta weighted by Gaussian wave packets. The result of this integration are the final formulae of the kind in eqn. (\ref{kos3}). As these are the formulae we will directly use in the paper, we refrain from explicitly displaying the double brackets to indicate classical limit. 
\section{Revisiting the classical sub-leading soft photon factor in $D\ >\ 4$}\label{cspfdg4}
In this section we review the derivation of  Electro-magnetic radiation in $D\ >\ 4$ dimensions upto sub-leading order \cite{ashoke1801}.   Our set up is the same as the one considered in \cite{golrid, kosower}. That is, we consider a scattering of 2 charges $q_{1},\, q_{2}$ with masses $m_{1},\, m_{2}$.  We assume that the particles do not have any spin. As in \cite{golrid}, we work in the large impact parameter (small deflection limit) defined via $b\ >>\ m_{i}^{-1}$. The trajectories of both the particles are parametrized as 
\begin{equation}\label{ridgebnd}
\begin{array}{lll}
x^{\mu}_{i}(\sigma) &=\, b_{i}^{\mu}\, +\, v_{i}^{\mu}\, \sigma\, +\, z_{i}^{\mu}(\sigma)\\ [0.4em]
z_{i}^{\mu}(-\infty)& =\, 0 
\end{array}
\end{equation}
where, $i=1,2 $.  $z_{i}^{\mu}(\sigma) $ is the correction to the free trajectory of the $i $th particle.

The boundary conditions ensure that the particles are free in the far past  with initial velocities given by $v^{\mu}_{i}$.
  The equations of motion of the two particles can be written as \cite{golrid}
\begin{equation}\label{zacc}
m_{i}\, \frac{d^{2}\, z_{i}^{\mu}}{d\, \sigma_{i}^{2}}\, =\, i\, q_{i}\, \sum_{j\neq i}\, q_{j}\, \int_{l}\ e^{-i\,  l \cdot x_{i}(\sigma_{i})}\, G_{r}(l)\, \hdelta(p_{j}\cdot l)\, [\, l\, \wedge\, p_{j}\, ]^{\mu\nu}\, p_{i \nu}
\end{equation}
where $\hdelta(x)\, =\, 2\pi\, \delta(x)$, $\int_{l}\,=\,\int \frac{d^{D}l}{(2 \pi)^{D}} $ and $G_{r}(l)$ is the retarded propagator. 
We now compute the radiative gauge field at sub-leading order in soft expansion and verify if it satisfies the classical soft photon theorem\cite{ashoke1801}.
We start with the equation for radiative gauge field as given in \cite{kosower}.\footnote{This formulae are written in 4 dimensions but the integral expressions hold in all dimensions as can be readily checked.} 
It is convenient to work in the center of mass frame with the origin of the co-ordinate system chosen such that $b_{2}^{\mu}\, =\, 0$. 
\begin{flalign}
\begin{array}{lll}
{\cal R}^{\mu}(k)\, =\, 4\, q_{1}^{2}\, q_{2}\,  \int \frac{d^{D} l_{1}}{(2\pi)^{D}}\, \frac{d^{D} l_{2}}{(2\pi)^{D}}\, \hdelta(2p_{1} \cdot l_{1})\, \hdelta(2p_{2} \cdot l_{2}) \,e^{i b\cdot l_{1}}\, \hdelta^{D}(\,l_{1} + l_{2} - k\,)\,G_{r}(l_{2})\\[0.5em]
\hspace*{1.7in}  [\, p_{2}^{\mu}\, -\, \frac{(\,p_{1}\cdot p_{2}\,)\, l_{2}^{\mu}}{p_{1}\cdot k}\, -\,  p_{1}^{\mu}\, \frac{p_{2}\cdot k}{p_{1}\cdot k}\, +\, \frac{(\,l_{2}\cdot k\,)\, (\,p_{1}\cdot p_{2}\,)\, p_{1}^{\mu}}{(\,p_{1}\cdot k\,)^{2}}\,]\, + \, 1\, \leftrightarrow 2
\end{array}
\end{flalign}
We can re-write this expression as, 
\begin{flalign}\label{ssl5ptD}
\begin{array}{lll}
\hspace*{-0.3in}
{\cal R}^{\mu}(k)\, =\,
4\, \left\{  q_{1}^{2}\, q_{2}\int \, \frac{d^{D} l}{(2\pi)^{D}}\, \hdelta(2p_{1} \cdot (k-l))\, \hdelta(2p_{2} \cdot l) e^{i b\cdot (k-l)} G_{r}(l)\right.\\[0.4em]
\hspace*{2in} [\, p_{2}^{\mu}\, -\, \frac{(\,p_{1}\cdot p_{2}\,) l^{\mu}}{p_{1}\cdot k}\, -\, p_{1}^{\mu}\, \frac{p_{2}\cdot k}{p_{1}\cdot k}\, +\, \frac{(\,l\cdot k\,) \,(p_{1}\cdot p_{2}) p_{1}^{\mu}}{(p_{1}\cdot k)^{2}}\, ]\\ [0.4em]
\hspace*{0.7in}+\,q_{2}^{2}\, q_{1}\int\, \frac{d^{D} l}{(2\pi)^{D}}\, \hdelta(2p_{1} \cdot l)\, \hdelta(2p_{2} \cdot (k-l))\, e^{i b\cdot l}\ G_{r}(l)\\[0.4em]
\hspace*{2in}\left.  [\, p_{1}^{\mu}\, -\, \frac{(\,p_{1}\cdot p_{2}\,) \,l^{\mu}}{p_{2}\cdot k}\, -\, p_{2}^{\mu}\, \frac{p_{1}\cdot k}{p_{2}\cdot k}\, +\, \frac{(\,l\cdot k\,) \,(\,p_{1}\cdot p_{2}\,) \,p_{2}^{\mu}}{(\,p_{2}\cdot k\,)^{2}}\, ]\, \right\}
\end{array}
\end{flalign}
We can write the delta function to sub-leading order in  momentum $ k^{\mu}$ as,
\begin{flalign}
\begin{array}{lll}
\hdelta(p_{1}\cdot (\,k-l\,))\,=\,\hdelta(p_{1}\cdot k)\,-\,p_{1}\cdot k\,\hdelta^{\prime }(p_{1}\cdot l)
\end{array}
\end{flalign}
Hence the  soft radiation is given by,
\begin{flalign}\label{rcistep}
\begin{array}{lll}
&{\cal R}^{\mu}(k)\,=\\[0.4em]&\hspace{.5in}
4\, \int\, \frac{d^{D} l}{(2\pi)^{D}}\,  G_{r}(l)\,\Big(
\{ e^{-ib \cdot l}\, (\, q_{1}^{2}\, q_{2}\, \hdelta(2p_{1} \cdot l)\, \hdelta(2p_{2} \cdot l)\, [\,p_{2}^{\mu}\,  -\, p_{1}^{\mu}\, \frac{p_{2}\cdot k}{p_{1}\cdot k}\, ]\,)\,
\\[0.4em]& \hspace{4in}
+\, e^{ib \cdot l}\, (\, 1\leftrightarrow 2 \,)\,\} \\[0.4em]
&\hspace{.5in}-\,\{ e^{-ib \cdot l}\, (\, q_{1}^{2}\, q_{2}\,\hdelta^{\prime}(2p_{1} \cdot l)\, \hdelta(2p_{2} \cdot l)\, [\,- (\,p_{1}\cdot p_{2}\,)\,l^{\mu}\, +\, \frac{(\,l\cdot k\,)\,(\,p_{1}\cdot p_{2}\,)\, p_{1}^{\mu}}{(\,p_{1}\cdot k\,)}\, ]\, )\\[0.4em]& \hspace{4in} +\, e^{i b \cdot l}\,(\, 1\leftrightarrow 2\,)\,\}\\[.4em]&\hspace{.1in}
 \hspace{.5in}+\, e^{-ib \cdot l} \, q_{1}^{2} q_{2}\, (\, i b\, \cdot\, k\, )\, \, \hdelta(2p_{1} \cdot l)\  \, \hdelta(2p_{2} \cdot l)\, [\,- \frac{(p_{1}\cdot p_{2})\, l^{\mu}}{p_{1}\cdot k}\, +\, \frac{(l\cdot k)\, (p_{1}\cdot p_{2}) p_{1}^{\mu}}{(p_{1}\cdot k)^{2}}\ ]\, \Big)
\end{array}
\end{flalign}
\begin{itemize}
\item In $D\, =\, 4$ dimensions, the boundary conditions in the far past make the analysis more subtle. This is because unlike in higher dimensions, the Coulombic interactions cause particles to accelerate even in the far past and far future. Thus to ensure the boundary conditions in eqn.(\ref{ridgebnd}), we need to use $i \epsilon$ prescription \cite{chen}.  In appendix \ref{d=4ssrqed}, we compute the  sub-leading soft radiation kernel  in four dimensions, essentially reviewing the computation of soft electromagnetic radiation in \cite{aab}, but adjusted to the set up in which outgoing momenta are not independent of initial momenta and are determined from the initial momenta using equations of motion.
\end{itemize}
Let us compare the integral expression given in eqn.(\ref{rcistep}) with the one we obtain by a direct computation of classical soft factor defined in \cite{ashoke1801}. We will denote this soft factor as $S^{(1) \mu}$ where the super-script indicates that it is the sub-leading expansion in photon frequency. 
\begin{flalign}
S^{(1) \mu}\,=\, \sum_{i}\, q_{i}\, [\, \frac{1}{p_{ + i}\, \cdot k}\, J_{+ i}^{\mu\nu}\, k_{\nu} -\, \frac{1}{p_{- i}\ \cdot k}\, J_{ - i}^{\mu\nu}\, k_{\nu}\, ]
\end{flalign}
where $p_{\pm i} $ are the initial and final momenta of the $i$th particle,  $J_{ \pm i}^{\mu\nu}$ are the  initial and final (classical) angular momenta defined  with respect to a choice of the origin.\footnote{It was shown in \cite{ashoke1801} that the choice of origin is gauge choice due to total momentum conversation.} 
We will focus on the contribution of the first particle to $S^{(1) \mu}$. 
Due to the parametrization of the trajectory, the initial orbital angular momentum of the first particle is given by,
\begin{flalign}
J_{-1}^{\mu\nu}\, =\, b^{\mu}\, p_{1}^{\nu}\, -\, b^{\nu}\, p_{1}^{\mu}
\end{flalign}
On the other hand, with respect to the choice of origin used in defining $J_{-1}$, the final angular momentum is given by,
\begin{equation}
\begin{array}{lll}
J_{+ 1}^{\mu\nu}\, =\, (b^{\mu}\, +\, z_{1}^{\mu}(0)\, )\, p_{+ 1}^{\nu}\, -\, \mu\leftrightarrow\, \nu
\end{array}
\end{equation}
Hence at leading order in the coupling, the ``angular momentum impulse" is given by,
\begin{equation}
J_{+ 1}^{\mu\nu}\, -\, J_{- 1}^{\mu\nu}\, =\, b^{\mu}\, \triangle p_{1}^{\nu}\, -\, b^{\nu} \, \triangle p_{1}^{\mu}\, +\, (\, z_{1}^{\mu}(0)\, p_{1}^{\nu}\, -\, z_{1}^{\nu}(0)\, p_{1}^{\mu}\, )
\end{equation}
Here $\triangle p_{1}^{\mu}$ is the linear impulse suffered by the first particle \cite{kosower}. 
Thus the classical sub-leading soft factor is given by,
\begin{flalign}\label{nov16min1}
\begin{array}{lll}
S^{(1) \mu}&=\, q_{1}\, [\, \frac{1}{p_{+ 1}\, \cdot k}\, J_{+ 1}^{\mu\nu}\, k_{\nu}\, -\, \frac{1}{p_{- 1} \cdot k}\, J_{-1}^{\mu\nu}\, k_{\nu}\, ]\\[0.4em]
&=\, q_{1}\, [\, \frac{1}{p _{+1}\cdot k}\, (\, J_{+1 }^{\mu\nu}\, -\, J_{-1}^{\mu\nu} \,)\, k_{\nu} \, +\, (\, \frac{1}{p_{+1}\cdot k}\, -\, \frac{1}{p_{-1}\cdot k} \,)\, J_{-1}^{\mu\nu} \, k_{\nu}\,]\\[0.4em]
&=\, q_{1}\, [\, \frac{1}{(\,p_{1}\, +\, \triangle p_{1}\,)\cdot k}\, (\, J_{+1 }^{\mu\nu}\, -\, J_{-1}^{\mu\nu} \,) \, k_{\nu}\, + \,(\, \frac{1}{(\,p_{1} \,+\, \triangle p_{1}\,)\cdot k}\, -\, \frac{1}{p_{1}\cdot k}\, )\, J_{-1}^{\mu\nu}\, k_{\nu}\,] \,\\[0.4em]
&=\, q_{1}\, [\,  \frac{1}{p_{1} \cdot k}\, (\, J_{+1 }^{\mu\nu}\, -\, J_{-1}^{\mu\nu} \,)\,k_{\nu} \, -\, \frac{\triangle p_{1}\cdot k}{(\,p_{1}\cdot k\,)^2}\, J_{-1}^{\mu\nu}\, k_{\nu} \,  ]\\[0.4em]
&=\, q_{1}\, [\,  \frac{1}{p_{1} \cdot k}\, \{\, (\, b\, \wedge\, p_{1}\, )^{\mu\nu}\, k_{v}\, +\,  (\, z_{1}(0)\, \wedge\, p_{1}\, )^{\mu\nu}\, k_{\nu}\, \}\, -\, \frac{\triangle p_{1}\cdot k}{(\,p_{1}\cdot k\,)^2} \,(\,b^{\mu}\, p_{-1}^{\nu}\, -\, b^{\nu}\, p_{-1}^{\mu} \,)\, k_{\nu} \,  ]
\end{array}
\end{flalign}
where in the second last line, we have expanded the second term to leading order in the coupling. 
 The linear impulse was computed in \cite{kosower} and is given by,
\begin{equation}\label{nov16-0}
\begin{array}{lll} 
\triangle p_{1}^{\mu}\, =\, i\, q_{1}\, q_{2}\, \int \frac{d^{D} l}{(2\pi)^{D}}\, \hdelta (2p_{1}\cdot l)\, \hdelta(2p_{2}\cdot l)\,e^{-i b\cdot l}\, G_{r}(l)\, 4 (\,p_{1}\cdot p_{2}\,)\, l^{\mu}
\end{array}
\end{equation}
In the same way, the deflected trajectory at $\sigma\, =\, 0$ is given by,
\begin{flalign}\label{nov16-1}
z_{1}^{\mu}(0)\, =\,  q_{1}\, q_{2}\, \int \frac{d^{D} l}{(2\pi)^{D}}\, \frac{1}{(p_{1}\, \cdot\, l)_{+}^{2}}\hdelta(2p_{2}\cdot l)\,e^{-i b\cdot l}\, G_{r}(l)\, (\, l\, \wedge\, p_{2}\, )^{\mu\nu}\, p_{1 \nu}\\
=\, 2\, q_{1}\, q_{2}\, p_{1 \nu}\, (p_{2}\, \wedge\, \frac{\partial}{\partial p_{1}}\, )^{\mu\nu}\, \int \frac{d^{D} l}{(2\pi)^{D}}\, \frac{1}{(p_{1}\, \cdot\, l)_{+}}\hdelta(2p_{2}\cdot l)\,e^{-i b\cdot l}\, G_{r}(l)\\
=\, 2 q_{1} q_{2}\, p_{1 \nu}\, (p_{2}\, \wedge\, \frac{\partial}{\partial p_{1}}\, )^{\mu\nu}\, \int \frac{d^{D} l}{(2\pi)^{D}}\, \hdelta(2p_{1} \cdot l) \hdelta(2p_{2}\cdot l)\,e^{-i b\cdot l}\, G_{r}(l)\\
=\, 4\, q_{1} q_{2}\, p_{1 \nu}\, (p_{2}\, \wedge\, \frac{\partial}{\partial p_{1}}\, )^{\mu\nu}\, \int \frac{d^{D} l}{(2\pi)^{D}}\, P(\frac{1}{p_{1} \cdot l})\, \hdelta(2p_{2}\cdot l)\,e^{-i b\cdot l}\, G_{r}(l)
\end{flalign}
In going from second to the third line, we have used 
\begin{equation}
\frac{1}{(p_{1} \cdot l)_{+}}\, =\, P(\frac{1}{p_{1} \cdot l})\, -\, i\, \pi\, \delta(p_{1} \cdot l)
\end{equation}
and, $\pi\, \delta(p_{1} \cdot l)\, =\, \hdelta(2 p_{1} \cdot l)$ respectively.\\
The second term in RHS of eqn.(\ref{nov16-1}) vanishes. This can be most easily seen by working in the rest frame of $p_{1}$ so that $P(\frac{1}{p_{1} \cdot l})\, =\, \frac{1}{m_{1}}\, P(\frac{1}{l_{0}})$. The integral is then an odd function of $l_{0}$ (as $b$ is spatial vector) and hence vanishes. 
Thus,
\begin{flalign}\label{nov16-2}
z_{1}^{\mu}(0) &=\, 4 q_{1} q_{2}\, p_{1 \nu}\, (p_{2}\, \wedge\, \frac{\partial}{\partial p_{1}}\, )^{\mu\nu}\, \int \frac{d^{D} l}{(2\pi)^{D}}\, \hdelta(2p_{1} \cdot l) \hdelta(2p_{2}\cdot l)\,e^{-i b\cdot l}\, G_{r}(l)
\end{flalign}
Contribution of $z_{1}^{\mu}(0)$ to the final angular momentum $J^{\mu\nu}_{+ 1}$ can be evaluated as,
\begin{flalign}\label{nov17-1}
(\, p_{1}\, \wedge\, z_{1}\, )^{\mu\nu}\, &=\nonumber\\[0.4em]
&4\, q_{1}\, q_{2}\, [\, (\, p_{1}\, \wedge\, p_{2}\, )^{\mu\nu}\, p_{1}\, \cdot\, \frac{\partial}{\partial p_{1}}\, -\, (\, p_{1} \cdot p_{2}\, )\, (\, p_{1}\, \wedge\, \frac{\partial}{\partial p_{1}}\, )^{\mu\nu}\, ]\nonumber\\
&\hspace*{1.6in} \int \frac{d^{D} l}{(2\pi)^{D}}\, \hdelta(2p_{1} \cdot l) \hdelta(2p_{2}\cdot l)\,e^{-i b\cdot l}\, G_{r}(l)
\end{flalign}
We can now use the identity 
\begin{flalign}
p_{1}\ \cdot\, \frac{\partial}{\partial p_{1}}\, \int\, \frac{d^{D}l}{(2\pi)^{D}}\, \hdelta(2p_{1} \cdot l)\, \hdelta(2p_{2} \cdot l)\, e^{-i b\cdot l}\, G_{r}(l)\, =\, - \int\, \frac{d^{D}l}{(2\pi)^{D}}\, \hdelta(2p_{1} \cdot l)\, \hdelta(2p_{2} \cdot l)\, e^{-i b\cdot l}\, G_{r}(l)
\end{flalign}
and substitute eqn.(\ref{nov17-1}) in the RHS of eqn.(\ref{nov16min1}) to readily verify that it agrees with eqn.(\ref{rcistep}). 
 We thus see that $S^{(1) \mu}$ equals  the integral expression for ${\cal R}^{\mu}(k)$ obtained in eqn.(\ref{rcistep}). In the next section we will confirm these classical results  for the radiation kernel ${\cal R}^{\mu}(k)$ by using sub-leading soft photon theorem in KMOC framework.

\section{From quantum to classical sub-leading soft photon theorem}\label{csptdg4fq}
In this section we compute the classical radiation kernel from soft expansion of tree-level amplitudes using KMOC formalism.  That is, we consider scattering of two incoming states with masses $m_{1},\, m_{2}$ which scatter into two outgoing states and one photon. In the usual statement of classical soft theorem, given the initial and the final states of the particles, one can compute soft radiation without using equations of motion. However in the KMOC  formalism,  we only know the scattering states in the far past.  Hence the computation of soft radiation in KMOC formalism depends on the details of the scattering amplitude without the photon. We consider tree-level amplitudes in scalar QED and hence our scattering particles have zero spin. But the analysis can be generalised to higher spin  cases as well. \cite{gb, cfq2.5,  cfq4.5, cfq4, maybee, deb-prep}. 

Our idea is to take soft limit before the classical limit (as in \cite{ashoke1801}) and hence we first write the  tree level five point amplitude via quantum soft theorem and then take the classical limit.  As we show, this reproduces the classical soft theorem upto sub-leading  order.  We note that, as the KMOC  set up  is such that the impact parameter $b$ is larger then the Schwarzchild radius of the particles, we expect the results upto sub-leading order to match with the so-called universal soft factors.\footnote{Sub-leading soft photon theorem is not universal \cite{elvang}. However the non-universal terms arise via higher derivative interaction terms, all of which are sub dominant in large impact parameter regime.}
\\
As we reviewed in section \ref{kmorev}, the primary quantity of interest is the radiation kernel ${\cal R}^{\mu}(k)$ whose classical limit is the radiative gauge field. In order to obtain the leading order (in the coupling) classical radiation, we start with the quantum radiation kernel generated by tree-level amplitude 
\begin{equation}\label{rcdm4}
\begin{array}{lll}
{\cal R}^{\mu}(k)\, =\\[0.4em]
\hbar^{\frac{3}{2}-D+4}\, \int\, \prod_{i}\, \frac{d^{D}l_{i}}{(2\pi)^{D}}\, \hdelta(2 p_{i}\cdot l_{i}\, +\, l_{i}^{2})\, \theta(p_{i}^{0}\, +\, l_{i}^{0})\, e^{\frac{ib\cdot l_{1} }{\hbar}}\, 
&\\[0.4em] &\hspace*{-1in}
{\cal A}^{\mu}_{5}(p_{1}\, +\, l_{1},\, p_{2} + l_{2} \, \rightarrow\, p_{1}, p_{2}, k)
\end{array}
\end{equation}
We start by quickly reviewing the tree-level soft  photon theorems in  scalar QED.  It is convenient to write the five point amplitude in terms of the stripped amplitude ${\cal M}_{5}^{\mu}$ as,
\begin{flalign}
{\cal A}_{5}^{\mu}(\,p_{1}+l_{1},\, p_{2} + l_{2}\, \rightarrow\,& p_{1},\, p_{2},\, k\,)\, =\\[0.4em]&\, \delta^{D}(\,l_{1} + l_{2} - k\,)\, {\cal M}^{\mu}_{5}(\,p_{1}+l_{1},\, p_{2} + l_{2}\, \rightarrow\, p_{1},\, p_{2},\, k\,)
\end{flalign}
The sub-leading soft photon theorem for tree-level amplitudes is stated as follows.
\begin{equation}\label{sspt2t}
\begin{array}{lll}
\hdelta^{D}(\,l_{1} + l_{2} - k\,)\, {\cal {M}}_{5}^{\mu}(\,p_{1}+l_{1},\, p_{2} + l_{2}\, \rightarrow\, p_{1},\, p_{2},\, k\,)\, =\\[0.4em]
\hdelta^{D}(\,l_{1} + l_{2}\,)\, {S}^{(1) \mu}\,{\cal M}_{4}(\,p_{1}+l_{1},\, p_{2} + l_{2}\, \rightarrow\, p_{1},\, p_{2}\,)\\ [0.4em] \hspace*{1.5in} -\, S^{(0) \mu}\, k \cdot \partial\, (\, \hdelta^{D}(\,l_{1} + l_{2})\,)\, {\cal M}_{4}(\,p_{1}+l_{1},\, p_{2} + l_{2}\, \rightarrow\, p_{1},\, p_{2}\,)
\end{array}
\end{equation}
where $S^{(0) \mu}$ and ${S}^{(1) \mu}$ are the leading and sub-leading soft  photon factors. 
\begin{flalign}
S^{(0) \mu}\, =\, \sum_{i} q_{i}\, [\, \frac{p_{i}^{\mu}}{p_{i}\cdot k}\, -\, \frac{(\,p_{i} + l_{i}\,)^{\mu}}{(\,p_{i} + l_{i}\,) \cdot k}\, ]&&,\, 
{S}^{(1) \mu}\, =\, i\, \sum_{i}\, q_{i}\,[\, \frac{\hat{J}_{+ i}^{\mu\nu}\, k_{\nu}}{p_{i}\cdot k}\, +\, \frac{\hat{J}_{- i}^{\mu\nu}\, k_{\nu}}{(\,p_{i} + l_{i}\,)\cdot k}\, ]&&
\end{flalign}
To leading order in the momentum mis-match $l^{\mu}$,
\begin{flalign}\label{sfsall}
\begin{array}{lll}
S^{(0) \mu}\, =\, \sum_{i}\, q_{i}\, (\, -\, \frac{l_{i}^{\mu}}{p_{i} \cdot k}\, +\, \frac{ l_{i} \cdot k}{(\,p_{i} \cdot k\,)^{2}}\, p_{i}^{\mu}\, )&&
\end{array}
\end{flalign}
The sub-leading soft photon factor is linear in the angular momentum operator. \footnote{We note that the sub-leading soft factor consists of terms with relative positive sign between the in and the out states. This is simply because the action of these operators on in-coming and out-going states differ by a sign.}
\begin{flalign}
\begin{array}{lll}
\hat{J}_{+ i}^{\mu\nu}\, =\,   -\,i\, (\, p_{i}^{\mu}\, \frac{\partial}{\partial p_{i}^{\nu}}\, -\,  p_{i}^{\nu}\, \frac{\partial}{\partial p_{i}^{\mu}} \,)&&\\[0.4em]
\hat{J}_{- i}^{\mu\nu}\, =\,  -\,i\,(\, (\,p_{i} + l_{i}\,)^{\mu}\, \frac{\partial}{\partial (\,p_{i} + l_{i}\,)^{\nu}}\, -\,  (\,p_{i} + l_{i}\,)^{\nu}\, \frac{\partial}{\partial (\,p_{i} + l_{i})^{\mu}}\, )
\end{array}
\end{flalign}
And finally the soft factor acts on the four point amplitude, 
\begin{equation}
{\cal A}_{4}\, =\,  \, q_{1}\,q_{2}\, \hdelta^{D}(\,l_{1} + l_{2}\,)\,G_{F}(l_{2})\, (\,2p_{1} + l_{1}\,)\, \cdot (\,2p_{2} +  l_{2}\,)
\end{equation}
where $G_{F}(l_{2})\,=\,\frac{1}{l_{2}^{2}\,+\,i \epsilon} $ is the Feynman propagator.
We can now use the sub-leading soft photon theorem to evaluate the quantum radiation kernel. As the soft theorem is sum of two terms (proportional to ${S}^{(1)}$ and $S^{(0)}$), we decompose the radiation kernel as,
\begin{equation}\label{qrkqedalld}
\begin{array}{lll}
{\cal R}^{\mu}(k)\, =\, {\cal R}^{\mu}_{1}(k)\, +\, {\cal R}^{\mu}_{2}(k)
\end{array}
\end{equation}
where $\hcr^{\mu}_{1}(k),\, \hcr^{\mu}_{2}(k)$ are defined as,
\begin{flalign}
\begin{array}{lll}
{\cal R}^{\mu}_{1}(k)\, :=\\[0.4em]
\hbar^{\frac{3}{2}-D+4}\, \int\,\prod_{i}\, \frac{d^{D}l_{i}}{(2\pi)^{D}}\, \hdelta(\,2 p_{i}\cdot l_{i}\, +\, l_{i}^{2}\,)\, \theta(\,p_{i}^{0}\, +\, l_{i}^{0}\,)\, e^{\frac{ib\cdot l_{1} }{\hbar}}\, \hdelta^{D}(\,l_{1} \, +\, l_{2}\,)\\[0.4em]
\hspace*{3in}\, {S}^{(1) \mu}\, {\cal M}_{4}(\,p_{1} + l_{1},\, p_{2} + l_{2} \, \rightarrow\, p_{1},\, p_{2}\,)\\[0.4em]
{\cal R}_{\textrm{2}}^{\mu}(k)\, :=\\[0.4em]
-\, \hbar^{\frac{3}{2}-D+4}\, \int\, \prod_{i}\, \frac{d^{D}l_{i}}{(2\pi)^{D}}\, \hdelta(\,2 p_{i}\cdot l_{i}\, +\, l_{i}^{2}\,)\, \theta(\,p_{i}^{0}\, +\, l_{i}^{0}\,)\, e^{\frac{ib\cdot l_{1} }{\hbar}}\\[0.4em]
\hspace*{1.9in}S^{(0) \mu} \, k \cdot \partial\, (\, \hdelta^{D}(\,l_{1} + l_{2})\,)\, {\cal M}_{4}(\,p_{1}+l_{1},\, p_{2} + l_{2}\, \rightarrow\, p_{1},\, p_{2}\,)
\end{array}
\end{flalign}
We can now compute $\hcr^{\mu}_{\textrm{1}}(k)$ to leading order in $l^{\mu}$ using the following approximate identity.
\begin{flalign}
{S}^{(1) \mu}\, {\cal M}_{4}(\,p_{1} + l_{1},\, p_{2} + l_{2} \, \rightarrow\, p_{1},\, p_{2})\, \approx\, i\, \sum_{i=1}^{2}\, q_{i}\,\frac{1}{l_{i+1}^{2}}\, \frac{\hat{J}_{i}^{\mu\nu}\, k_{\nu}}{p_{i} \cdot k}\, \{\, 4 \,q_{1}\, q_{2}\, ( \,p_{1} \cdot p_{2} \,)\, \}
\end{flalign}
where the propagator is indexed modulo 2 and the approximation sign indicates that the identity holds only to leading order in $l^{\mu}$. 

This identity is based on the following observation. Action of $S^{(1) \mu}$  on the stripped amplitude is sum of the two terms acting on particles 1 and 2. When the soft factor associated to particle 1 acts on the amplitude, we can express propagator in terms of $l_{2}^{\mu}$ vice versa.\footnote{We note that the total action of $S^{(1) \mu}$ on unstripped amplitude which also involves action on the momentum conserving delta function is unaffected by such re-labellings of the propagator.}  Thus the action of $S^{(1) \mu}$ is simply on the numerator of the four point amplitude. It is now simple to verify the approximate identity  and use it to compute $\hcr^{\mu}_{1}(k)$.
\begin{flalign}\label{qrkqeddg4}
\hcr^{\mu}_{\textrm{1}}(k)\, =\, i\, \hbar^{(\,-D+4\,)}\, \sum_{m}\, q_{m}\, q_{1}\,q_{2}\, \frac{k_{\nu}\, \hat{J}_{m}^{\mu\nu}}{p_{m}\cdot k}\, (\,p_{1} \cdot p_{2}\,)\,
\int\, \frac{d^{D}l}{(2\pi)^{D}}\, \prod_{i}\,\hdelta(p_{i}\cdot l)\, e^{\frac{ib\cdot l }{\hbar}}\, \frac{1}{l^{2} + i\epsilon}
\end{flalign}
The integral in eqn.(\ref{qrkqeddg4}) can be evaluated directly. The pole of the Feynman propagator has trivial residue due to the on-shell delta function constraints and we can write $\frac{1}{l^{2} + i\epsilon}\, =\, \frac{1}{l^{2}}$.\\ 
Let $l^{\mu}\, =\, \hbar\, \ol^{\mu}$. Then 
\begin{flalign}\label{mifipp}
\begin{array}{lll}
\, \int\, \frac{d^{D} \ol}{(2\pi)^{D}}\, \hdelta(p_{1} \cdot \ol)\, \hdelta(p_{2} \cdot \ol)\, e^{i b \cdot \ol}\, \frac{1}{\ol^{2}}\, =\,\, \alpha_{D}\, \frac{1}{\sqrt{(\,p_{1}\cdot p_{2}\,)^{2}\, -\, m_{1}^{2} \,m_{2}^{2}}}\, \frac{1}{(\,\vec{b}\cdot \vec{b}\,)^{\frac{D-4}{2}}}
\end{array}
\end{flalign}
where $\alpha_{D}\, :=\, -\, \frac{1}{4\pi^{\frac{D-2}{2}}}\, \Gamma[\frac{D-2}{2}-1]$.
We have put an arrow sign on the impact parameter to emphasise that it is a spatial vector in a plane transversal to the one spanned by $p_{1},\, p_{2}$. 
Using eqn.(\ref{mifipp}) in eqn.(\ref{qrkqeddg4}) we can evaluate ${\cal R}^{\mu}_{1}(k)$. For simplicity, we choose to focus on the radiation kernel emitted by the first particle.
\begin{equation}\label{rmunuslf}
\begin{array}{lll}
{\cal R}^{\mu}_{1,1}(k)\, =\, \alpha_{D}\, q_{1}^{2}\,q_{2}\, \frac{p_{1}^{\mu}\,(\,p_{2}\cdot k\,)\, -\, p_{2}^{\mu}\,(\,p_{1}\cdot k\,)}{p_{1}\cdot k}\, \frac{1}{\sqrt{(\,p_{1} \cdot p_{2}\,)^{2}\, -\, m_{1}^{2} \,m_{2}^{2}}}\, \frac{1}{(\,\vec{b} \cdot \vec{b}\,)^{\frac{D-4}{2}}}\, 
\end{array}
\end{equation}
where the additional subscript indicates that we are only considering radiation emitted by  particle with final momentum $p_{1}^{\mu}$.

We now evaluate ${\cal R}^{\mu}_{2}(k)$ in the classical limit.  In the interest of pedagogy, we skip a few intermediate steps  by dropping higher order terms in $l^{\mu}$.\footnote{Although while taking the classical limit, the exchange momenta scales as $l^{\mu}\, \rightarrow\, \hbar\overline{l}^{\mu}$, we will drop the bar and always indicate the exchange moementum at $l^{\mu}$. We believe that from the context it becomes clear if we are working with the quantum radiation kernel or it's classical limit.}\footnote{As the step function $\theta(p_{i}^{0} + l_{i}^{0})$ become identity in classical limit and hence we just drop them to avoid the clutter}
\begin{flalign}
\begin{array}{lll}
{\cal R}_{2}^{\mu}(k)\, :=\\[0.4em]
\, 4 \,q_{1} \,q_{2}\, \int\, \prod_{i}\, \frac{d^{D}l_{i}}{(2\pi)^{D}}\, \hdelta(2 p_{i}\cdot l_{i})\, \big[\, e^{\frac{ib\cdot l_{1} }{\hbar}}\,
S^{(0) \mu}\, \{\delta^{D}(\,l_{1} + l_{2} - k\,)\, -\, \delta^{D}(\,l_{1} + l_{2}\,)\, \}\, \frac{p_{1} \cdot p_{2}}{l_{2}^{2}}\, +\, O(l_{2}^{\mu})\, \big]
\end{array}
\end{flalign}
The minus sign in front of the equation is because we have expressed $k \cdot \partial\, \delta^{D}(l_{1} + l_{2})$ as $-\, \{\, \delta^{D}(l_{1} + l_{2} - k)\, -\, \delta^{D}(l_{1} + l_{2})\, \}$.\\
As $S^{(0)\mu}$ is sum over the two particles, we can analyse contribution of both the particles separately. With out loss of generality, we focus on the first particle and denote the corresponding contribution to radiation kernel as ${\cal R}_{2, 1}^{\mu}(k)$. Denoting $l_{2}^{\mu}$ as $l^{\mu}$ and solving for $l_{1}$ in terms of $k, l_{2}$ we get, 
\begin{flalign}
\begin{array}{lll}
{\cal R}_{2, 1}^{\mu}(k)\,=\\[0.4em]
4\, q_{1}\, q_{2}\, \int\, \frac{d^{D}l}{(2\pi)^{D}}\, \hdelta(\,2 p_{2} \cdot l\,)\\[0.4em]
\hspace*{1.0in}\big[\, S_{1}^{(0) \mu}\, \{\, \hdelta(\,2 p_{1} \cdot (\,k - l\,) \,)\, e^{\frac{ i b \cdot ( k - l)}{\hbar}} \,-\, \hdelta(\,2 p_{1} \cdot l\,)\, e^{-i \frac{b \cdot l}{\hbar}}\, \}\, \frac{p_{1} \cdot p_{2}}{l^{2}}\, +\, O(l^{\mu})\, \big]\\[0.4em]
=\,  q_{1}\, q_{2}\, \int\ \frac{d^{D}l}{(2\pi)^{D}}\, \hdelta(p_{2} \cdot l)\\[0.4em]
\hspace*{1.0in}\big[\, S_{1}^{(0) \mu}\, \{\, -(\,p_{1} \cdot k\,)\, \delta^{\prime}(\,p_{1} \cdot l\,)\, +\,  \frac{i}{\hbar}\, b \cdot k\, \hdelta(\, p_{1} \cdot l\,)\, \}\,e^{\frac{ - i b \cdot l}{\hbar}}\,\frac{p_{1} \cdot p_{2}}{l^{2}}\, +\, O(l^{\mu})\, \big]
\end{array}
\end{flalign}
We can now use the following identities to simplify the above result.
\begin{flalign}
\begin{array}{lll}
l^{\mu}\, \delta^{\prime}(\,p_{i} \cdot l\,)\, =\, \frac{\partial}{\partial p_{i}^{\mu}}\, \hdelta(\,p_{1} \cdot l\,)\\[0.4em]
l^{\mu}\, e^{- \frac{ i b \cdot l}{\hbar}}\, =\, i\, \hbar\, \frac{\partial}{\partial b^{\mu}}\, e^{- \frac{ i b \cdot l}{\hbar}}
\end{array}
\end{flalign}
And since $l_{1}^{\mu}\, =\, -\, l^{\mu}$
\begin{flalign}\label{sfsal}
S_{1}^{(0) \mu}\, =\, q_{1}\, (\, \frac{l_{1}^{\mu}}{p_{1} \cdot k}\, -\, \frac{ l_{1} \cdot k}{(\,p_{1} \cdot k\,)^{2}}\, p_{1}^{\mu}\, )\\[0.4em]
=\, q_{1}\, (\,-\, \frac{l^{\mu}}{p_{1} \cdot k}\, +\, \frac{ l \cdot k}{(\,p_{1} \cdot k\,)^{2}}\, p_{1}^{\mu}\, )
\end{flalign}
we can write ${\cal R}^{\mu}_{2, 1}(k)$ as, 
\begin{flalign}\label{rmu2do}
{\cal R}^{\mu}_{2,1}(k)\, =\, q_{1}^{2}\, q_{2}\, (\,p_{1} \cdot p_{2}\,)\, \{\, -\, \hat{O}_{1}^{\mu}\, +\, \frac{(\,b\cdot k\,)}{p_{1}\cdot k}\, \hat{O}_{2}^{\mu} \}\, \int\, \frac{d^{D}l}{(2\pi)^{D}}\, \hdelta(\,p_{1} \cdot l\,)\, \hdelta(\,p_{2} \cdot l\,)\, e^{-i \frac{b \cdot l}{\hbar}}\, \frac{1}{l^{2}}
\end{flalign}
where $\hat{O}_{1}^{\mu},\, \hat{O}_{2}^{\mu}$ are differential operators defined as,
\begin{equation}\label{rdioto}
\begin{array}{lll}
\hat{O}_{1}^{\mu}\, =\, [\, \frac{p_{1}^{\mu}}{p_{1}\cdot k}\, k \cdot \frac{\partial}{\partial p_{1}}\, -\, \frac{\partial}{\partial p_{1}^{\mu}}\, ]\\[0.4em]
\hat{O}_{2}^{\mu}\, =\, [\, \frac{p_{1}^{\mu}}{p_{1}\cdot k}\, k \cdot \frac{\partial}{\partial b}\, -\, \frac{\partial}{\partial b^{\mu}}\, ]
\end{array}
\end{equation}
Hence the classical soft radiation (at $O(\omega^{0})$) emitted by particle-1 is given by adding eqns. (\ref{qrkqeddg4}) and (\ref{rdioto}).
\begin{equation}\label{nov17-2}
\begin{array}{lll}
{\cal R}^{\mu}_{1,1}(k)\, +\, {\cal R}^{\mu}_{2,1}(k)\, =\\
q_{1}^{2}q_{2}\, \frac{1}{p_{1} \cdot k}\, (\, p_{1} \wedge\, p_{2})^{\mu\nu}\, k_{\nu}
\int\, \frac{d^{D}l}{(2\pi)^{D}}\, \prod_{i}\,\hdelta(p_{i}\cdot l)\, e^{b\cdot l}\, \frac{1}{l^{2}}\\
+\, q_{1}^{2}\, q_{2}\, (\,p_{1} \cdot p_{2}\,)\, \{\, -\, \hat{O}_{1}^{\mu}\, +\, \frac{(\,b\cdot k\,)}{p_{1}\cdot k}\, \hat{O}_{2}^{\mu} \}\, \int\, \frac{d^{D}l}{(2\pi)^{D}}\, \hdelta(\,p_{1} \cdot l\,)\, \hdelta(\,p_{2} \cdot l\,)\, e^{-i \frac{b \cdot l}{\hbar}}\, \frac{1}{l^{2}}
\end{array}
\end{equation}

The result matches with the classical radiation kernel in eqn.(\ref{rcistep}). This proves the sub-leading classical soft theorem in $D\, >\, 4$ dimensions. The radiation kernel can in fact be explicitly computed.\\
We can use eqn.(\ref{mifipp}) in conjunction with eqn.(\ref{rdioto}) to get 
\begin{equation}\label{rmunulf}
\begin{array}{lll}
{\cal R}^{\mu}_{2,1}(k)\, =\, 
-\, q_{1}^{2}\, q_{2}\, \alpha_{D}\, (p_{1} \cdot p_{2})\, 
\bigg[\, [\, \frac{(\,p_{1}\cdot p_{2}\,)}{(\,(\,p_{1}\cdot p_{2}\,)^{2}\, -\, m_{1}^{2}\, m_{2}^{2}\,)^{\frac{3}{2}}}\, (\, p_{2}^{\mu}\, -\, \frac{p_{2} \cdot k}{p_{1} \cdot k}\, p_{1}^{\mu}\, )\, ] \\[0.4em]
\hspace*{1.3in}-\, (D-4)\, \frac{b\cdot k}{p_{1}\cdot k}\, \frac{1}{(\, (\,p_{1}\cdot p_{2}\,)^{2}\, -\, m_{1}^{2}\, m_{2}^{2}\, )^{\frac{1}{2}}}\, [\, \frac{p_{1}^{\mu}}{p_{1}\cdot k}\, k \cdot b\, - b^{\mu}\, ]\, \frac{1}{(\,\vec{b}\cdot \vec{b}\,)} \,\bigg]\, \frac{1}{(\,\vec{b}\cdot \vec{b}\,)^{\frac{D-4}{2}}}
\end{array}
\end{equation}
We can now add right hand side of eqns. (\ref{rmunuslf}) and (\ref{rmunulf}) to get the classical radiation kernel at sub-leading order in $\omega$. 
\begin{equation}
\begin{array}{lll}
{\cal R}^{\mu}_{\textrm{particle 1}}(k)\, =\\[.4em]
-\, \alpha_{D}\, q_{1}^{2}\,q_{2}\,
\bigg[\, [\, \frac{m_{1}^{2}\, m_{2}^{2}}{(\,(\,p_{1}\cdot p_{2}\,)^{2}\, -\, m_{1}^{2} \,m_{2}^{2}\,)^{\frac{3}{2}}}\, (\, \frac{p_{2} \cdot k}{p_{1} \cdot k}\, p_{1}^{\mu}\, -\, p_{2}^{\mu}\, )\, ] \\[0.4em]
\hspace*{1.6in}-\, (D-4)\, \frac{\hat{b}\cdot k}{p_{1}\cdot k}\, [\, \frac{p_{1}\cdot p_{2}}{(\, (p_{1}\cdot p_{2}\,)^{2}\, -\, \frac{1}{2}\, m_{1}^{2}\, m_{2}^{2}\, )^{\frac{1}{2}}}\,(\,  \frac{p_{1}^{\mu}}{p_{1}\cdot k}\, k \cdot \hat{b}\, - \hat{b}^{\mu}\,) \,]\, \bigg]\, \frac{1}{(\,\vec{b}\cdot \vec{b}\,)^{\frac{D-4}{2}}}
\end{array}
\end{equation}
where $\hat{b}\, =\, \frac{\vec{b}}{\vert b\vert}$.\\
After some algebra, we can write the final expression in a more compact form as,
\begin{equation}
\begin{array}{lll}
{\cal R}_{\textrm{particle 1}}^{\mu}(k)\, =\\[.4em]
-\, \alpha_{D}\, q_{1}^{2}\,q_{2}\, \frac{1}{p_{1}\cdot k}\, \frac{1}{{\cal D}}\, [\, m_{1}^{2}\, m_{2}^{2}\, (\,p_{1} \wedge p_{2}\,)^{\mu\nu}\, k_{\nu}\, \frac{1}{{\cal D}^{2}}\,\\[0.4em]
\hspace*{2.1in} -\, (D-4)\, \frac{\hat{b}\cdot k}{p_{1}\cdot k}\, (\,p_{1}\cdot p_{2}\,)\, (\,p_{1}\ \wedge \hat{b}\,)^{\mu\nu}\, k_{\nu}\, ]\, \frac{1}{(\,\vec{b}\cdot \vec{b}\,)^{\frac{D-4}{2}}}
\end{array}
\end{equation}
where ${\cal D}\, =\, \{\, (p_{1}\cdot p_{2})^{2}\, -\, m_{1}^{2} m_{2}^{2}\, \}^{\frac{1}{2}}$.
${\cal R}^{\mu}_{2}(k)$ can be computed similarly by using $b_{2}^{\mu}\, =\, 0$.\\
We conclude this section with a few remarks. 
\begin{itemize}
	\item Our results are consistent with the interpretation of classical soft theorem given in \cite{ashoke1801}. That is, in the large impact parameter regime the soft expansion is really an expansion in $\omega b$.
	\item It may seem rather surprising that a quantum amplitude with Feynman propagator produces the same result as the one we obtain in classical theory via retarded propagator. But this is simply because all the external states are on-shell and hence the pole corresponding to Feynman propagator does not contribute in the classical limit.  The easiest way to see this is to work in center of mass frame with $p_{1}$ and $p_{2}$ along z axis. 
	\begin{equation}
	\begin{array}{lll}
	\int_{l} \,\hdelta(\,p_{1}\cdot l\,)\, \hdelta(\,p_{2}\cdot l\,)\, {\cal F}(l)\, &=\, \frac{1}{\sqrt{(\,p_{1}\cdot p_{2}\,)^{2}\, -\, m_{1}^{2}\, m_{2}^{2}}}\, \int_{l}\, \hdelta(l^{0})\, \hdelta(l^{3})\, {\cal F}(l)\\[0.4em]&
	=\, \frac{1}{\sqrt{(\,p_{1}\cdot p_{2}\,)^{2}\, -\, m_{1}^{2}\, m_{2}^{2}\,}}\, \int_{l}\, \hdelta(l^{0})\, \hdelta(l^{3})\, {\cal F}(l_{\perp})\\
	\end{array}
	\end{equation}
	where $l_{\perp}\ =\ (0,l_{x}, l_{y}, 0)$. 
	Thus the pole of the photon propagator does not contribute in the classical limit and hence RHS of eqn.(\ref{qrkqeddg4}) equals RHS of eqn.(\ref{rmunuslf}).  The vanishing residue from pole of the $G_{F}(l)$ is understood even at higher loop orders in \cite{cfq1}.
\end{itemize}
\begin{itemize}
	\item Although the master integral in eqn.(\ref{mifipp}) can be analytically evaluated, focussing on different integration regions sheds light on the origin of the classical soft theorem \cite{sahoo, aab}.  We first note that in the soft expansion the $l^{\mu}$ integration region is naturally restricted to $ \vert l\vert\, \geq\, \omega$.  One way to understand this is to notice that the soft expansion of the un-stripped amplitude is obtained by taylor expansion of the momentum conserving delta function $\delta^{D}(l_{1} + l_{2} - k)$ which implicitly assumes that $\omega\, <<\, \vert l\vert$.  
	Now if we evaluate the contribution from the lower limit of the integration then 
	\begin{equation}\label{idfaaet1}
	\begin{array}{lll}
	\int_{l}\, \frac{d^{D}l}{(\,2\pi\,5)^{D}}\, G_{F}(l)\, \hat{\delta}(\,p_{1}\cdot l\,)\, \hat{\delta}(\,p_{2}\cdot l\,)&=
	\, \frac{1}{\sqrt{(\,p_{1}\cdot p_{2}\,)^{2}\, -\, m_{1}^{2} \,m_{2}^{2}}}\, \int\, \frac{d^{D-2} l_{\perp}}{(\,2\pi\,)^{D-2}}\, \frac{1}{l_{\perp}^{2}}\\[0.4em]
	&\approx\, \omega^{D-4} 
	\end{array}
	\end{equation}
	In $D\, >\, 4$ dimensions this contribution is sub-subleading and hence does not contribute at the sub-leading order in $\omega$. The sub-leading contribution comes from the integration region $\vert l\vert\, \sim\, b^{-1}$. This is consistent with the known understanding of classical soft theorem in higher dimensions in \cite{ashoke1906}, where it was shown that  during scattering, the sub-leading contribution to the radiation comes from the ``outer" space-time region with size $\geq\, b$. 
\item In $D\, =\, 4$ space-time dimensions the contribution from the region of integration $\omega\, <<\, \vert l\vert\, << b^{-1}$ is of the order $\ln\omega$.  As we will see in section \ref{D=4case}, it is precisely this term that generates the classical log soft factor in four dimensions. We thus see that there is a ``reversal of order" as far as soft emission is concerned in $D = 4$, or $D\, >\, 4$ spacetime dimensions. It is the same integral that in  $D\, >\, 4$ produces $\omega^{0}$ term from ``UV region" characterised by $\vert l \vert\, \sim\, b^{-1}$ and higher order ($\omega^{D-4}$) terms from the ``IR" region $\vert l\vert\, \geq\, \omega$ whereas in $D\, =\, 4$ this integral produces $\ln\omega$ term from the IR region and  $\omega^{0}$ terms from the UV region.\footnote{We use UV and IR in the sense of their usage in effective field theory literature for binary systems \cite{porto}.}
 \item Although our analysis is for electro-magnetic radiation, it can be easily generalised  to the case soft gravitational radiation in $D\, >\, 4$ dimensions.
 \end{itemize}
\subsection{Soft radiation from soft photon theorem in $D\, =\, 4$}\label{D=4case}
The analysis in the previous section was based on tree-level sub-leading soft  photon theorem which resulted in eqn.(\ref{qrkqedalld}) defining the Radiation Kernel. Let us now analyse this formula in four dimensions.  The integration region in the soft limit is  $k\, <<\,\ \vert\, l\, \vert\, <\ b^{-1}$, where the upper limit is automatically imposed by the phase term in the integrand \cite{aab}.  As we show below, this region produces the classical log soft factor defined and analysed previously in \cite{ashoke1804, sahoo, aab}. 
This in turn implies that sub-leading soft photon theorem generates leading order  soft radiation in all dimensions.
The integrand in eqn.(\ref{qrkqedalld}) consists of two terms which we will referred to as ${\cal R}^{\mu}_{1}(k),\ {\cal R}^{\mu}_{2}(k)$ respectively. 
In four dimensions, instead of using the results of the full integral, we focus on specific integration region which has been shown to contribute to soft radiation at $\ln\omega$ order.  
The integrand in eqn.(\ref{qrkqedalld}) can  then be simplified by noting that 
\begin{itemize}
	\item $e^{\frac{i\,(k-l)\cdot b}{\hbar}}\, =\, 1$\vspace*{-0.1in}
	\item $\hdelta(\,p_{1}\cdot (k-l) \,)\, =\, \hdelta(p_{1}\cdot l)\, -\, (\, p_1 \cdot k \, ) \, \hat{\delta}'(\,p_{1}\cdot l\,)$
\end{itemize}
In appendix \ref{appA}, we show that,
\begin{equation}\label{eq1}
\begin{array}{lll}
{\cal R}^{\mu}(k)\, =\\[0.4em] 
q_{1}^{2}\, q_{2}\, \frac{k_{\nu}}{p_{1}\cdot k}\,  \left[ \, \big( \, p_{1}^{\mu} \, \frac{\partial}{\partial p_{1\nu}} \, - \,   p_{1}^{\nu} \, \frac{\partial}{\partial p_{1\mu}} \, \big) \, \right] \, 
\Big( \, (p_{1}\cdot p_{2})\, \int_{\omega\, <<\, \vert l\vert\, <<\, b^{-1}}\, G_{F}(l)\, \hat{\delta}(\,p_{1}\cdot l\,)\, \hat{\delta}(\,p_{2}\cdot l\,) \, \Big) \\[0.4em]\hspace*{5in}+\, \,(1 \leftrightarrow 2\, ) 
\end{array}
\end{equation} 
So, finally we are left with the following integral
\begin{equation} \label{int}
I \,=\, \int_{\omega\, <<\, \vert l\vert\, <<\, b^{-1}}\, G_{F}(l)\, \hat{\delta}(\,p_{1}\cdot l\,)\, \hat{\delta}(\,p_{2}\cdot l\,).
\end{equation}
This integral can be readily evaluated based on the analysis of \cite{aab}.  We work in a centre of mass frame with 
\begin{equation}
\begin{array}{lll}
p_{1}\, =\, (E,\,0,\,0,\,\vert p\vert)\\
p_{2}\, =\, (E,\,0,\,0,\,-\vert p\vert)
\end{array}
\end{equation}
We can do the integral  by changing the variables from  $  (\, l^{0},\,l^{1},\, l^{2},\, l^{3}\, ) $ to $( \, p_1 \cdot l,\, l^{1},\, l^{2},\, p_2 \cdot l \,) $. And, the Jacobian related to the change of variable can be given  as,
\begin{equation}
2 \, E \, |p| \, = \, \sqrt{\,(\,p_{1}\,\cdot \, p_{2})^{2} \, - \, m_{1}^{2} \,m_{2}^{2}}.
\end{equation}
With this change of variable we can rewrite the integral as,
\begin{equation} \label{int1}
I\, =\, \int_{\omega\, <<\, \vert l\vert\, <<\, b^{-1}}\, \frac{ \, d(\,p_1 \cdot  l\,) \, dl^{1} \, dl^{2} \, d(\,p_2 \cdot  l\,)\, }{2 \,  E \,  \vert p \vert} \, G_{F}(l)\, \hat{\delta}(\,p_{1}\cdot l\,)\, \hat{\delta}(\, p_{2}\cdot l\,).
\end{equation}
After doing the $( \, p_1 \cdot  l\, )$ and $(\,p_2 \cdot  l\, )$ integral we are left with a 2-dimensional integral 
\begin{equation} \label{eq2}
I = \frac{1}{\sqrt{(\,p_{1}\cdot p_{2}\,)^{2} - m_{1}^{2}\,m_{2}^{2}}} \,\int_{\omega\, <<\, \vert l_{\perp}\vert\, <<\, b^{-1}} \frac{d^{2}l_{\perp}}{(\,2\pi\,)^{2}} \frac{1}{(\,-l^{2}_{\perp} + i\epsilon\,)}
\end{equation} 
This integral can be easily done by going to polar coordinates and doing the radial integral in the $b^{-1} >> |l_{\perp}| >> \omega$ region. We get
\begin{equation}
I \,= \, \frac{ \ln (\omega b)}{2\pi} \frac{1}{\sqrt{(\,p_{1}\cdot p_{2}\,)^{2} - m_{1}^{2}\,m_{2}^{2}}}
\end{equation}
Plugging this into the \eqref{eq1} and evaluating the derivatives, we have 
\begin{flalign}
\begin{array}{lll}
\mathcal{R}^{\mu}_{1}(k)\, +\, \mathcal{R}^{\mu}_{2}(k)\, = \\[.4em] \,  -\, \frac{ 1}{2 \, \pi}\, \ln \,( \, \omega  \, b \, ) \, \frac{q_{1}^{2} \, q_{2}\, p_{1}^{2} \, p_{2}^{2}}{\{(\, p_{1}\cdot p_{2}\,)^{2}\, -\,  m_{1}^{2} \, m_{2}^{2}\}^{3/2}} \left[\,  \frac{k_{\nu}}{p_{1}\cdot k}\,  (\, p_{1}^{\mu}\, p_{2}^{\nu} \, - \, p_{1}^{\nu}\, p_{2}^{\mu} \,)\, \right] \,  +\,  (\, 1 \leftrightarrow 2 \, )
\end{array}
\end{flalign}
Hence the classical radiation kernel at the sub-leading order in frequency is given by,
\begin{equation}\label{crkemlo}
{\cal R}^{\mu}(k)\, =\,  -\, \frac{q_{1}^{2} q_{2}}{2 \, \pi}\, \ln \omega \, \frac{p_{1}^{2} \, p_{2}^{2}}{\{(\, p_{1} \cdot p_{2}\,)^{2}\, -\,  m_{1}^{2} \, m_{2}^{2}\}^{3/2}} \left[\,  \frac{k_{\nu}}{p_{1}\cdot k}\  (\, p_{1}^{\mu}\, p_{2}^{\nu} \, - \, p_{1}^{\nu}\, p_{2}^{\mu} \,)\, \right] \,  +\,  (\, 1 \leftrightarrow 2 \, )
\end{equation}
We now argue that in the large impact parameter regime, the result obtained here matches with the classical log soft factor obtained in \cite{aab}.  
\begin{itemize}
	\item For a scattering processes involving $n$ incoming particles with momenta $p_{1}\, \dots,\, p_{n}$ and $m$ out-going particles with momenta $p_{1}^{\prime},\, \dots\, p_{m}^{\prime}$ the classical log soft factor is defined in \cite{aab} as,
	\begin{equation}\label{aabsem}
	\begin{array}{lll}
	{\cal J}^{\mu}(k)\, =\, -\, \frac{1}{4 \pi}\, \ln \,( \, \omega+ i\epsilon)\, \sum_{a,b=1}^{n} q_{a}^{2}\, q_{b}  \frac{p_{a}^{2} \, p_{b}^{2}}{\{(\, p_{a}.p_{b}\,)^{2}\, -\,  m_{a}^{2} \, m_{b}^{2}\}^{\frac{3}{2}}\}} \left[\,  \frac{k_{\nu}}{p_{a}\cdot k}\  (\, p_{a}^{\mu}\, p_{b}^{\nu} \, - \, p_{a}^{\nu}\, p_{b}^{\mu} \,)\, \right]\\[0.4em]
	\hspace*{0.7in}-\, \frac{ 1}{4 \, \pi}\, \ln \,( \, \omega - i\epsilon)\, \sum_{a,b=1}^{m} q_{a}^{2} q_{b}\, \frac{p^{\prime 2}_{a} p^{\prime 2}_{b}}{\{\, (\, p^{\prime}_{a}.p^{\prime}_{b}\, )^{2}\, -\,  m^{\prime 2}_{a} m^{\prime 2}_{b}\}^{\frac{3}{2}}} \left[\,  \frac{k_{\nu}}{p^{\prime}_{a}\cdot k}\  (\, p^{\prime \mu}_{a}\, p^{\prime \nu}_{b}\, - \, p^{\prime \nu}_{a}\, p^{\prime \mu}_{b} \,)\, \right]
	\end{array}
	\end{equation}
	The overall minus sign is due to the fact that in \cite{aab}, all the incoming particles were thought of as out-going particles with sign of charges and momenta reversed. 
	\item We now see that in the case of $2\, \rightarrow\, 2$ scattering and in the limit of large impact parameter (i.e. when $p_{i}^{\prime}\, =\, p_{i}$ ), the result in eqn.(\ref{aabsem}) matches with the one obtained via KMOC formulation as
	\begin{equation}
	\ln(\omega + i\epsilon)\, +\, \ln(\omega - i\epsilon)\, =\, 2\, \ln\omega
	\end{equation}
	
\end{itemize}
\subsection{A caveat regarding counting the orders in coupling}
In the classical soft photon theorem proved in \cite{aab}, the leading order soft radiation is linear in the electro-magnetic coupling $e$. In $D\, >\, 4$ dimensions, even the sub-leading order soft radiation is at $O(e)$  \cite{ashoke1801}. However in four  dimensions, the sub-leading (that is, at order $\ln\omega$) soft radiation is at order $O(e^{3})$. Our results obtained from tree-level scattering amplitudes produces radiative gauge fields which are at the third order in the coupling, for any order in the soft expansion. The reason that these results are consistent with soft theorems is simply because statement of  soft theorem requires initial and final states are considered to be independent.  In the KMOC approach, the final states are determined from the initial states by equations of motion. This immediately implies that the most dominant contrbution to the soft field (proportional to $\frac{1}{\omega}$) vanishes at linear order in the coupling. This is because the final momenta differ from the initial momenta by  momentum impulse which is itself quadratic in the coupling. Hence the soft radiation that we obtain via KMOC approach is cubic in the charges of the external particles. This argument remains valid even at higher order in soft expansion in $D\, >\, 4$ dimensions. 

In $D\, =\, 4$ dimensions, both the classical log soft factor in \cite{aab} and the sub-leading radiation kernel obtained from tree-level amplitudes are at the same (cubic) order in the coupling. This is because for the classical log soft factor, there is a non-trivial contribution even as deflection  tends to zero, and hence is independent of the impulse.  This implies that when we expand the classical log soft factor in the coupling, the next-to-leading order (NLO) term which is linear in momentum impulse, occurs at fifth order in the coupling. We expect this term to be obtained by computing the NLO radiative field in the KMOC approach. In the next section we show that this is indeed the case. 


\subsection{From Quantum to classical sub-leading soft photon theorem at NLO}\label{nloem4d}
We now turn to the computation of  soft radiation kernel at next to leading order (NLO) in the coupling.  As we recall from section \ref{kmorev}, the classical radiation kernel at NLO is obtained from the quantum kernel by taking the classical limit, $\lim_{\hbar\, \rightarrow\, 0}\, {\cal R}^{\mu}_{\textrm{NLO}}(k)$. The NLO  (quantum) radiation kernel is defined as
\begin{flalign}\label{kos1lmf}
{\cal R}^{\mu}_{\textrm{NLO}}(k)\, =\, \hbar^{\frac{3}{2}}\, \int_{l_{1},\, l_{2}}^{\textrm{on-shell}}\, e^{i\frac{b\cdot l_{1}}{\hbar}}\, \delta^{4}(l_{1} + l_{2} - k)\, {\cal M}_{5}^{\textrm{1-loop}}(\tilde{p}_{1},\, \tilde{p}_{2},\, p_{1},\, p_{2},\, k)
\end{flalign}
There are two possible approaches to compute the NLO radiation kernel at sub-leading order in the soft expansion.
 Following the main premise of this paper, we can take the soft limit before taking the classical limit. This implies that we need to use the soft expansion of 1-loop amplitude upto sub-leading order in the soft expansion. The other possibility is to take the classical limit of the integrand in the first term in eqn.(\ref{kos1lmf}) and then take the soft limit. While the second possibility is expected to reproduce the classical soft theorem derived in \cite{aab}, our interest is in analysing the first possibiity.  As we show below,  the soft expansion of amplitude followed by the classical limit produces the radiative gauge field which satisfies the classical log soft theorem in four dimensions. 

Thus our starting point is  the loop corrected soft photon theorem for scattering amplitudes.  In $D\, =\, 4$ dimensions this theorem was derived by Sahoo and Sen in \cite{sahoo}. 
To state the  theorem  we first need to define the ``infra-red" finite part of the unstripped scattering amplitude given in \cite{gy, weinberg}.
\begin{flalign}
{\cal A}_{n}(p_{1},\, \dots,\, p_{n})\, =\, e^{K}\, ({\cal A}_{n}^{\textrm{tree}}(p_{1},\, \dots,\, p_{n})\, +\, {\cal A}_{n}^{\textrm{IR-fin}}(p_{1},\, \dots,\, p_{n})\, )
\end{flalign}
where $K$ is the infra-red divergent contribution due to virtual soft photons.\footnote{In \cite{sahoo} the infra-red finite part of the amplitude ${\cal A}_{n}^{\textrm{IR-finite}}$ was called ${\cal A}_{n}^{G}$ as it was obtained from the usual amplitude by replacing the Feynman propagator for the loop momentum with the so-called G-photon propagator.}   The detailed form of $K$ is not relevant for us. An important property of $K$ which is relevant (and was proved in \cite{sahoo}) is that $e^{K}$ is the same for an $n$-point amplitude without a photon and an $n+1$ point amplitude ${\cal A}_{n+1}(p_{1},\, \dots,\, p_{n},\, k)$ containing one additional photon.  This property of the QED amplitude leads to the loop-corrected soft photon theorem for the IR-finite part of the scattering amplitudes as,
\begin{flalign}
\begin{array}{lll}
{\cal A}_{n+1}^{\textrm{tree}}(p_{1},\, \dots,\, p_{n},\, k)\, +\, {\cal A}_{n + 1}^{\textrm{IR-fin}}(p_{1},\, \dots,\, p_{n}, k)\, =\\[0.4em]
\hspace*{0.8in} \{\, \frac{1}{\omega}\, S^{(0)}(\{p_{i}\})\, +\, \ln\omega\, {\cal S}_{(\ln)}\, \}\, (\, {\cal A}_{n}^{\textrm{tree}}(p_{1},\, \dots,\, p_{n})\, +\, {\cal A}_{n}^{\textrm{IR-fin}}(p_{1},\, \dots,\, p_{n})\, )
\end{array}
\end{flalign}
 It was shown in \cite{sahoo} that ${\cal A}_{n}^{\textrm{IR-fin}}(p_{1},\, \dots,\, p_{n})$  in fact vanishes. And the soft theorem can be written as,\footnote{Strictly speaking the proof in \cite{sahoo} was for a ``triangle loop". That is  when one replaces one of the photon propagators in the square loop in minimal scalar QED with a scalar quartic vertex. However the result is valid even in the minimal scalar QED, as can be verified.} 
 \begin{flalign}
 \begin{array}{lll}
{\cal A}_{n+1}^{\textrm{tree}}(p_{1},\, \dots,\, p_{n},\, k)\ +\, {\cal A}_{n + 1}^{\textrm{IR-fin}}(p_{1},\, \dots,\, p_{n}, k)\, =\\[0.4em]
\hspace*{0.8in} \{\, \frac{1}{\omega}\, S^{(0)}(\{p_{i}\})\, +\, \ln\omega\, {\cal S}_{(\ln)}\, \}\, {\cal A}_{n}^{\textrm{tree}}(p_{1},\, \dots,\, p_{n})
\end{array}
\end{flalign}
For our process of interest, the sub-leading soft photon theorem in four dimensions can be written as,
\begin{flalign}\label{5ptat1l}
\amp^{\textrm{IR-fin}}\, =\,  \ln\omega\, {\cal S}_{(\ln)}\, \usafour^{\textrm{tree}}
\end{flalign}
Our idea to compute radiation kernel at NLO is to use the infra-red finite five point amplitude in the integrand of the radiation kernel. Conceptually this differs from the set up of KMOC formalism where the scattering amplitude used to compute any classical quantity is always the standard (infra-red divergent) scattering amplitude. Naively one may think that if the scattering amplitude is infra-red divergent, the classical quantities computed from it may be ill-defined. However as was shown rather beautifully in \cite{kosower}, this is not true. For example, in the computation of NLO impulse in  \cite{kosower},  the loop-amplitude used in the impulse formula was the ``bare" infra-red divergent amplitude. However the procedure of taking classical limit prior to integration ensured that infra-red divergences present in individual Feynman diagrams cancelled  upon summing over all the relevant diagrams. It is certainly expected that if we compute NLO radiation where classical limit was taken prior to the soft limit, then infra-red divergences cancel in the end. However as we take the soft limit prior to taking classical limit, we need to work with infra-red finite amplitude for which soft limit is well defined.  

Thus it may appear that we are deviating from the KMOC formalism. But as the final  result in the classical theory is infra-red finite, one would expect that using bare amplitude or carefully defined infra-red finite amplitude should lead to the same final answer, and we choose to work with latter.\\
Although a detailed derivation of such a replacement (where we replace ``bare five point amplitude" with the infra-red finite amplitude) is outside the scope of this paper, it can be motivated in the following ways. 
\begin{itemize}
\item The radiation kernel (i.e. the radiative gauge field) is not an observable and is an intermediate quantity used to compute the emitted radiation. The formula for radiation in the KMOC formalism is in fact closely associated to the derivation of inclusive cross sections. We expect that if we compute the radiation as opposed to the radiation kernel, the ``virtual infra-red" divergence contained in $K$ will cancel with the real  soft photon emission contribution. This will perhaps be the most rigorous way to derive the classical log soft theorem in the KMOC approach. Although this approach may  obscure the relationship of the classical soft theorems with quantum soft theorems. 
\item In the the derivation of the formula for radiation kernel, the incoming coherent state is composed of free particle states. It is plausible to use the dressed states \cite{kf} to define the incoming  state which would lead to infra-red finite amplitude inside the integrand. 
\item  The dressed states alluded to above have so far remained rather ``formal objects" used to prove infra-red finiteness of S matrix but very rarely used in any concrete computations. A more robust way to compute infra-red finite S matrix  is the remarkable recent construction by Hannesdottir and Schwartz in  \cite{sh}. We believe that this formulation may be best suited to do higher loop computations in KMOC formalism. 
\end{itemize} 
We now proceed with the computation of  the radiation kernel using sub-leading soft photon theorem in eqn. (\ref{5ptat1l}).
We will denote the radiation kernel as ${\cal R}^{\mu}_{\ln}(k)$ (instead of ${\cal R}^{\mu}_{\textrm{NLO}}(k)$) to indicate that it is determined from  quantum log soft theorem. The log soft factor derived in \cite{sahoo} is a sum of two terms.\\
\begin{flalign}
{\cal S}_{\ln}\, =\, {\cal S}_{\ln}^{q}\, +\, {\cal S}_{\ln}^{\textrm{cl}}
\end{flalign}
These two factors are respectively given by,
\begin{flalign}\label{qsffdem}
S_{\ln}^{q}\, =\, \sum_{a,b=1}^{4}\, q_{a}^{2}\, q_{b}\, s^{q}(\tilde{p}_{a},\, \tilde{p}_{b})
\end{flalign}
where $\tilde{p}_{3}\, =\, -\, p_{1}$, $\tilde{p}_{4}\, =\, -\, p_{2}$ and $q_{3}\, =\, - q_{1}$, $q_{4}\, =\, -\, q_{2}$. 
$s^{q}(\tp_{a},\, \tp_{b})$ is defined as,
\begin{IEEEeqnarray}{cRl}\label{qsfocp}
s^{q}(\tp_{a},\, \tp_{b})&=&\nonumber\\
&&\frac{i}{4\pi^{2}}\, \frac{1}{\tp \cdot k}\, \frac{1}{(\tp_{a}\cdot \tp_{b})^{2}\, -\, m_{a}^{2}\, m_{b}^{2}}\, \{\, -\, \tp_{b}^{\mu}\, \tp_{a}\cdot k\, +\, \tp_{a}^{\mu}\, \tp_{b}\cdot k\, \}\nonumber\\
&&\hspace*{0.4in}\big[\, \frac{m_{a}^{2}m_{b}^{2}}{2} \ln[\, \frac{\tp_{a}\cdot \tp_{b}\, +\, \sqrt{(\tp_{a}\cdot \tp_{b})^{2}\, -\, m_{a}^{2}m_{b}^{2}}}{\tp_{a}\cdot \tp_{b}\, -\, \sqrt{(\tp_{a}\cdot \tp_{b})^{2}\, -\, m_{a}^{2}m_{b}^{2}}}\, ]\,  \frac{1}{(\, (\tp_{a}\cdot \tp_{b})^{2}\, -\, m_{a}^{2}\, m_{b}^{2})^{\frac{1}{2}}}\, -\, \tp_{a}\cdot \tp_{b}\, \big]\nonumber\\
&&\hspace*{-0.5in}=\, \frac{i}{4\pi^{2}}\, \frac{1}{(\tp_{a}\cdot \tp_{b})^{2}\, -\, m_{a}^{2}\, m_{b}^{2}}\nonumber\\
&&\bigg[\, \frac{m_{a}^{2}\, m_{b}^{2}}{2} \ln[\, \frac{\tp_{a}\cdot \tp_{b}\, +\, \sqrt{(\tp_{a}\cdot \tp_{b})^{2}\, -\, m_{a}^{2}m_{b}^{2}}}{\tp_{a}\cdot \tp_{b}\, -\, \sqrt{(\tp_{a}\cdot \tp_{b})^{2}\, -\, m_{a}^{2}m_{b}^{2}}}\, ]\,  \frac{1}{(\, (\tp_{a}\cdot \tp_{b})^{2}\, -\, m_{a}^{2}\, m_{b}^{2})^{\frac{1}{2}}}\, -\, \tp_{a}\cdot \tp_{b}\, \bigg]\, \tilde{s}_{q}(a, b)\nonumber\\[0.5em]
&&\hspace*{-0.5in}\textrm{where}\ \tilde{s}_{q}(a, b)\, :=\,  \{\, -\, \tp_{b}^{\mu}\, +\, \tp_{a}^{\mu}\, \frac{\tp_{b}\cdot k}{\tp_{a} \cdot k}\, \}\nonumber
\end{IEEEeqnarray}
\begin{IEEEeqnarray}{cRl}\label{csf4daab1}
{\cal S}_{\ln}^{\textrm{cl}} &=& \sum^{4}_{a,b, =\, 1\vert\, \sigma(a,b)\, =\, 0}\, q_{a}^{2}\, q_{b}\, s^{\textrm{cl}}(\tilde{p}_{a},\, \tp_{b})\nonumber\\
s^{\textrm{cl}}(\tilde{p}_{a},\tp_{b})\, &&=\, \frac{1}{4\pi}\, \frac{\tp_{a}^{2}\, \tp_{b}^{2}}{(\, (\tp_{a}\cdot \tp_{b})^{2}\, -\, \tp_{a}^{2} \tp_{b}^{2})^{\frac{3}{2}}}\, \frac{1}{\tp_{a}\cdot k}\, (\, \tp_{a}^{\mu}\, (\tp_{b}\, \cdot\, k)\, -\, \tp_{b}^{\mu}\, (\tp_{a}\, \cdot\, k)\, )\nonumber\\
&&=:\,  \frac{1}{4\pi}\, \frac{\tp_{a}^{2}\, \tp_{b}^{2}}{(\, (\tp_{a}\cdot \tp_{b})^{2}\, -\, \tp_{a}^{2} \tp_{b}^{2})^{\frac{3}{2}}}\, \tilde{s}^{\textrm{cl}}(a,b)
\end{IEEEeqnarray}
where
\begin{flalign}
\tilde{s}^{\textrm{cl}}(a,b)\, =\, \frac{1}{\tp_{a}\cdot k}\, (\, \tp_{a}^{\mu}\, (\tp_{b}\, \cdot\, k)\, -\, \tp_{b}^{\mu}\, (\tp_{a}\, \cdot\, k)\, )\end{flalign}
and, $\sigma(a,b)\, =\, 0$ indicates both particles are either incoming or outgoing . 
\begin{itemize}
\item ${\cal S}^{q}_{\ln}$ and ${\cal S}_{\ln}^{\textrm{cl}}$ differ from the corresponding expressions in \cite{sahoo} by an overall factor of $-i$. This is due to (1), our definition of four pt. amplitude ${\cal M}_{4}$ is $-i$ times the four point amplitude in \cite{sahoo}, (2) we use the opposite signature for space-time metric,  and (3) we define soft factor in terms of $\ln\omega$ as opposed to $\ln\omega^{-1}$ and (3) The polarisation vectors ${\cal M}_{5}$ is not contracted with the polarisation vectors which will absord the factor of  $i$ such that the ratio of ${\cal M}_{5}$ to ${\cal M}_{4}$ remains the same. 
\end{itemize}
A minor re-writing of ${\cal S}_{\ln}^{\textrm{cl}}$ turns out be useful for computation. 
\begin{flalign}
\begin{array}{lll}
{\cal S}_{\ln}^{\textrm{cl}}\, =\, \sum_{a,b=1}^{2}\, q_{a}^{2}\, q_{b}\, \big(\, 2\, s^{\textrm{cl}}(p_{a},\, p_{b})\, +\, \textrm{rest}(a,\, b)\, )
\end{array}
\end{flalign}
where $\textrm{rest}$ indicates all the terms which depend on the momentum mis-match $l_{i}^{\mu}$. 
\begin{flalign}
\textrm{rest}(a,\ b)\, =\, s^{\textrm{cl}}(\tilde{p}_{a},\, \tilde{p}_{b})\, -\, s^{\textrm{cl}}(p_{a},\, p_{b})
\end{flalign}
Let us now compute the contribution of this soft factor to the classical radiation kernel at fifth order in the coupling.  Let us recall the formula for ${\cal R}_{\ln}^{\mu}(k)$ once again. 
\begin{flalign}\label{rkfqed4dmf}
\begin{array}{lll}
{\cal R}_{ln}^{\mu}(k)\, =\, {\cal R}_{\ln \textrm{cl}}^{\mu}(k)\, +\, {\cal R}^{\mu}_{\ln \textrm{q}}(k)\\[0.4em]
{\cal R}_{\ln \textrm{cl}}^{\mu}(k)\, =\, \hbar^{\frac{3}{2}}\, \ln\omega\, \int_{l_{1},\, l_{2}}^{\textrm{on-shell}}\, e^{i\frac{b\cdot l_{1}}{\hbar}}\, \delta^{4}(l_{1} + l_{2})\\[0.4em] 
\hspace*{1.5in}\sum_{a,b=1}^{2}\, q_{a}^{2}\, q_{b}\, \big(\, 2\, s^{\textrm{cl}}(p_{a},\, p_{b})\, +\, \textrm{rest}(a,\, b)\, \big)\, {\cal M}^{\textrm{tree}}_{4}(\tilde{p}_{1},\, \tilde{p}_{2},\, p_{1},\, p_{2})\\[0.4em]
 {\cal R}^{\mu}_{\ln \textrm{q}}(k)\, =\, \hbar^{\frac{3}{2}}\, \ln\omega\, \int_{l_{1},\, l_{2}}^{\textrm{on-shell}}\, e^{i\frac{b\cdot l_{1}}{\hbar}}\, \delta^{4}(l_{1} + l_{2})\\[0.4em]
 \hspace*{1.5in}\sum_{a,b=1}^{2}\, q_{a}^{2}\, q_{b}\, \tilde{s}_{q}(\tilde{p}_{a},\, \tilde{p}_{b})\,  {\cal M}^{\textrm{tree}}_{4}(\tilde{p}_{1},\, \tilde{p}_{2},\, p_{1},\, p_{2})\\[0.4em]
\end{array}
\end{flalign}
We now compute ${\cal R}_{\ln q}^{\mu}(k)$ and ${\cal R}_{\ln \textrm{cl}}^{\mu}(k)$. But we first identify which terms can contribute in the classical limit via simple dimensional analysis. As $q_{i}\, \sim\, \frac{1}{\sqrt{\hbar}}$ and $l_{i}^{\mu}\, =\, \hbar\, \ol_{i}^{\mu}$,
\begin{flalign}
q^{5}\, \int_{l_{1}, l_{2}}^{\textrm{on-shell}}\, \delta^{4}(l_{1} + l_{2})\, \sim\, \frac{1}{\hbar}
\end{flalign}
and hence the integrand in the KMOC formula should scale as $O(\hbar)$. If integrand is more dominant as $\hbar\, \rightarrow\, 0$ than we will not have a well defined classical limit, and if the integrand is sub-dominant then it will generate no classical contribution. 
\subsection{Contribution of ${\cal R}_{\ln q}^{\mu}(k)$}
As all the external states are on-shell in the KMOC formula, we have the following identity.
\begin{flalign}\label{idfod}
\delta^{4}(l_{1} + l_{2})\, \prod_{i}\, \hdelta(2 p_{i} \cdot l_{i} + l_{i}^{2})\, (\tp_{a}\ \cdot \tp_{b})\, =\, (-1)^{\eta_{a}\, \cdot \eta_{b}}\, p_{a}\ \cdot p_{b} + O(l^{2})
\end{flalign}
where $\eta_{a}\, =\, 1$ for $a\, \in\, (1, 2)$ and $-1$ otherwise. 
Using eqn. (\ref{idfod}), ${\cal S}_{\ln}^{q}$ can be written in a more compact form to leading order in $l^{\mu}$ as,
\begin{flalign}\label{qsf4dsimp}
\begin{array}{lll}
\hspace*{-0.1in}{\cal S}^{q}_{\ln}&=&\\
&& \sum_{a,b=1 \vert a\, \neq\, b}^{2}\, q_{a}^{2}\, q_{b}\, \frac{i}{4\pi^{2}}\, \big[\, \frac{m_{a}^{2}m_{b}^{2}}{2} \ln[\, \frac{p_{a}\cdot p_{b}\, +\, \sqrt{(p_{a}\cdot p_{b})^{2}\, -\, m_{a}^{2}m_{b}^{2}}}{p_{a}\cdot p_{b}\, -\, \sqrt{(p_{a}\cdot p_{b})^{2}\, -\, m_{a}^{2}m_{b}^{2}}}\, ]\, \frac{1}{(\, (p_{a}\cdot p_{b})^{2}\, -\, m_{a}^{2}\, m_{b}^{2})^{\frac{1}{2}}}\, -\, p_{a}\cdot p_{b}\, \big]\\[0.4em]
&&\hspace*{1.4in} \frac{1}{{\cal D}^{2}}\, \{\, \tilde{s}_{q}(a, b)\, -\, \tilde{s}_{q}(a, b + 2)\, +\, \tilde{s}_{q}(a + 2, b)\, - \tilde{s}_{q}(a+2, b+2) \}\\[0.4em]
 &&-\, \frac{i}{(4\pi)^{2}}\, \sum_{a\, =\, 1}^{2}\, q_{a}^{3}\, \frac{p_{a}\cdot \tilde{p}_{a}}{\sqrt{(p_{a} \cdot \tp_{a})^{2}\, -\, p_{a}^{2} \tp_{a}^{2}}}\,  \{\, \tilde{s}_{q}(a, a + 2)\, +\, \tilde{s}_{q}(a + 2, a)\, \}\\[0.4em]
\end{array}
\end{flalign}
In the above equation, ${\cal D}\, =\, \sqrt{(p_{a} \cdot p_{b})^{2}\, -\, m_{a}^{2} m_{b}^{2}}$. In the first line the sum in fact also includes terms involving pairs $p_{a}, \tp_{a}$ (for $a\, =\, 1,\, 2$), but those vanish at leading order in $l^{\mu}$. 
It can now be verified that to leading order in $l^{\mu}$ 
\begin{flalign}
\sum_{a=1}^{3}\, \sum_{b=2}^{4}\, q_{a}^{2}\, q_{b}\, \{ \tilde{s}_{q}(\tilde{p}_{a},\, \tilde{p}_{b})\, -\, \tilde{s}_{q}(\tilde{p}_{a}, p_{b})\, \}\, =\, q_{1}^{2}\, q_{2}\, (\, \tilde{s}_{a}(\tp_{1},\, l_{2})\, -\, \tilde{s}_{q}(p_{1},\, l_{2})\, )=\, O(l^{2})
\end{flalign}
In the above equation we have displayed explicit dependence of $\tilde{s}_{q}$ on the momenta rather then labels.\\
Similar identity holds when $a$ and $b$ range over other values. Hence the first line of eqn.(\ref{qsf4dsimp}) vanishes. The second line vanishes because to sub-leading order in $l^{\mu}$, 
\begin{flalign}\label{rckql1}
\tilde{s}_{q}(a, a+2)\, =\, -\tilde{s}_{q}(a+2, a)
\end{flalign}
We have thus have shown that  ${\cal S}^{q}_{\ln}$ does not contribute to the classical radiation at next to leading order in the coupling and at sub-leading order in the soft expansion.  We end this section with a couple of remarks.
\begin{itemize}
\item At leading order, absence of  ${\cal S}^{q}_{\ln}$ in the classical radiation kernel was a consequence of the fact that the pole of the Feynman propagator in the momentum mismatch $l^{\mu}$ does not contribute in the classical limit. However, at leading order in the coupling (i.e. at zeroth order in $l^{\mu}$) even ${\cal S}^{q}_{\ln}$ manifestly vanishes, and hence the LO result obtained via KMOC formalism is rather expected. 
\item It may seem surprising that even at NLO ${\cal S}_{\ln}^{q}$ does not contribute in the classical limit. However a closer look at the soft factor itself (eqn. \ref{qsfocp}) shows that this is not surprising. If we expand ${\cal S}_{\ln}^{q}$ at next to leading order by expanding final momenta in terms of initial momenta and impulse then as at \emph{leading order} $\triangle p_{1} + \triangle p_{2}\, =\, 0$ and as $ p_{i} \cdot \triangle p_{i}\, =\, p_{i} \cdot b\, =\, 0$, ${\cal S}_{\ln}^{q}$ vanishes at NLO. The classical limit obtained via KMOC formalism is consistent with this result . 
\end{itemize}
\subsubsection{Contribution of ${\cal R}^{\mu}_{\ln \textrm{cl}}(k)$}\label{cofrlncl}
We split this contribution into two pieces $r^{\mu}_{1}(k),\, r^{\mu}_{2}(k)$ arising from $s^{\textrm{cl}}(p_{a},\, p_{b})$ and $\textrm{rest}(a,\, b)$ respectively. 
We first consider the contribution of $s^{\textrm{cl}}(p_{a},\, p_{b})$ to the radiation kernel.   The final result is obtained by taking classical $\hbar\, \rightarrow\, 0$ limit of $r^{\mu}_{1}(k)\, +\, r^{\mu}_{2}(k)$.
\begin{flalign}
\begin{array}{lll}
r^{\mu}_{1}(k)\, &=&\, \hbar^{\frac{3}{2}}\, \ln\omega\, \int_{l_{1},\, l_{2}}^{\textrm{on-shell}}\, e^{-i\frac{b\cdot l_{1}}{\hbar}}\, \delta^{4}(l_{1} + l_{2})\\[0.4em] 
&&\hspace*{1.7in}\sum_{a,b=1}^{2}\, 2\, q_{a}^{2} q_{b}\, s^{\textrm{cl}}(p_{a},\, p_{b})\, {\cal M}^{\textrm{tree}}_{4}(\tilde{p}_{1},\, \tilde{p}_{2},\, p_{1},\, p_{2})\\[0.4em]
&=&\, \sum_{a,b=1}^{2}\, 2\, q_{a}^{2}\, q_{b}\, s^{\textrm{cl}}(p_{a},\, p_{b})\, 4 q_{1}\, q_{2}\,  (p_{1}\cdot p_{2})\\[0.4em]
&&\hspace*{1.1in}\{\, \hbar^{\frac{3}{2}}\, \int\ \frac{d^{4}l_{1}}{(2\pi)^{4}}\, \frac{d^{4} l_{2}}{(2\pi)^{4}}\, \prod_{i=1}^{2}\, \theta(p_{i}^{0} + l_{i}^{0})\\[0.4em]
&&\hspace*{1.6in} \hdelta(2p_{1}\cdot l_{1} + l_{1}^{2})\, \hdelta(2 p_{2}\cdot l_{2} + l_{2}^{2})\, e^{-i\frac{b\cdot l_{1}}{\hbar}}\, \delta^{4}(l_{1} + l_{2})\, \frac{1}{l_{2}^{2}}\, \}
\end{array}
\end{flalign}
A simple power counting argument reveals that this term is super classical if we replace $l^{\mu}$ with $\hbar\ol^{\mu}$ and take the classical limit. Such a term would render the classical limit ill defined.\\
In order to eliminate the super-classical term, we use the on-shell delta function $\hdelta(2 p_{2}\cdot l_{2}\, +\, l_{2}^{2})$ to write $\frac{1}{l_{2}^{2}}\, =\, \frac{1}{-2 p_{2}\cdot l}$ before substituting $l^{\mu}$ in terms of the wave number $\ol^{\mu}$ (that is, before taking classical limit where $\delta (p_{i} \cdot \overline{l}\, +\, \overline{l}^{2})\, \approx\, \delta(p_{i}\cdot \ol)$).\footnote{In a more rigorous analysis where one essentially computes  inclusive cross section by summing over the additional $X$ states, we believe that such super-classical terms will cancel after summing over all the diagrams. In the absence of such a computation, we use on-shell delta functions to manipulate the denominator terms and check if modulo such ``on-shell substitutions" we can ensure that the most dominant term in any computation is $O(\hbar^{0})$.} It can now be checked that the resulting expression scales as $\hbar^{0}$ and the resulting classical limit is,
\begin{flalign}
\begin{array}{lll}
r^{\mu}_{1}(k)\, =\, \hbar^{\frac{3}{2}}\, \ln\omega\, \sum_{a,b=1}^{2}\, 2\,q_{a}^{2}\, q_{b}\, s^{\textrm{cl}}(p_{a},\, p_{b})\, 4 q_{1}\, q_{2}\,  (p_{1}\cdot p_{2})\\[0.4em]
\hspace*{1.8in} \int\ \frac{d^{4}l}{(2\pi)^{4}}\,  \hdelta( 2p_{1}\cdot l )\, \hdelta(2 p_{2}\cdot l)\, e^{-i\frac{b\cdot l}{\hbar}}\, \frac{1}{-2 p_{2} \cdot l}
\end{array}
\end{flalign}
This integral does not contribute in the classical limit. In order to prove this, we work  in the center of mass frame with $p_{2}\, =\, (\vert \sqrt{\vert p\vert^{2} + m_{2}^{2}},\, 0\, 0\, -\vert p\vert)$  the integral can be written as,
\begin{flalign}
\begin{array}{lll}
\textrm{Integral}\, =\, -\, \frac{1}{\sqrt{(p_{1} \cdot p_{2})^{2}\, -\, m_{1}^{2}m_{2}^{2}}}\, \int\, d(p_{2}\, \cdot\, l)\, \frac{1}{p_{2}\cdot l - i\epsilon}\hat{\delta}(p_{2} \cdot l)\, d^{2}\vec{l}_{\perp}\, e^{-i \frac{\vec{b} \cdot \vec{l}_{\perp}}{\hbar}}\\
[0.4em]
=\, -\, \delta^{2}(\vec{b}_{\perp})\,  \frac{1}{\sqrt{(p_{1} \cdot p_{2})^{2}\, -\, m_{1}^{2}m_{2}^{2}}}\, \int\, d(p_{2}\, \cdot\, l)\, \frac{1}{p_{2} \cdot l - i\epsilon}\, \hat{\delta}(p_{2} \cdot l)\, 
\end{array}
\end{flalign}
The above integral is a contact term which only contributes if the impact parameter vanishes. Hence $r^{\mu}_{1}(k)\, =\, 0$.\\
 We now compute $r^{\mu}_{2}(k)$. 
\begin{flalign}
\begin{array}{lll}
r^{\mu}_{2}(k)\, =\, \hbar^{\frac{3}{2}}\, \ln\omega\, \int_{l_{1},\, l_{2}}^{\textrm{on-shell}}\, e^{i\frac{b\cdot l_{1}}{\hbar}}\, \delta^{4}(l_{1} + l_{2})\\[0.4em] 
\hspace*{0.9in}\sum_{a,b=1}^{2}\, q_{a}^{2}\, q_{b}\, (\, s^{\textrm{cl}}(\tilde{p}_{a},\, \tilde{p}_{b})\, -\, s^{\textrm{cl}}(p_{a}, p_{b})\, )\, {\cal M}^{\textrm{tree}}_{4}(\tilde{p}_{1},\, \tilde{p}_{2},\, p_{1},\, p_{2})
\end{array}
\end{flalign}
Explicit expression for $s^{\textrm{cl}}(a,b)$ is given in eqn.(\ref{csf4daab1}).  We can use it along with the following equations which holds when all the external states are on-shell to compute $r^{\mu}_{2}(k)$. 
\begin{flalign}
\frac{1}{(\, (\tilde{p}_{a}\cdot \tilde{p}_{b})^{2}\, -\, m_{a}^{2}m_{b}^{2})^{\frac{3}{2}}}\ =\, \frac{1}{(\, (p_{a}\cdot p_{b})^{2}\, -\, m_{a}^{2}m_{b}^{2})^{\frac{3}{2}}}\\[0.4em]
\tilde{p}_{a}^{2}\, =\, m_{a}^{2} 
\end{flalign}
\begin{IEEEeqnarray}{rCl}
r^{\mu}_{2}(k)\, &=&  \hbar^{\frac{3}{2}}\, \ln\omega\, \{\, 4 q_{1}\, q_{2}\, (p_{1} \cdot p_{2} )\, \}\, \frac{1}{(\, (p_{a}\cdot p_{b})^{2}\, -\, m_{a}^{2}m_{b}^{2})^{\frac{3}{2}}}\nonumber\\
&&\int_{l_{1}, l_{2}}^{\textrm{on-shell}}\, \delta^{4}(l_{1} + l_{2})\, e^{-i\frac{b\cdot l_{1}}{\hbar}}\, \sum_{a\neq b}\, q_{a}^{2}q_{b}\, (\, s_{\textrm{cl}}(\tilde{p}_{a},\, \tilde{p}_{b})\, -\, s_{\textrm{cl}}(p_{a},\, p_{b})\, )\, \frac{1}{l_{2}^{2} + i\epsilon}
\end{IEEEeqnarray}
We can now use the fact that to leading order in the coupling, 
\begin{flalign}\label{impforkos}
\triangle p_{a}^{\mu}\, =\, \, \{\, 4 i\, q_{1}\, q_{2}\, (p_{1} \cdot p_{2} )\, \}\, \int_{l_{i}}^{\textrm{onshell}}\, e^{-i\frac{b_{i}\, \cdot l_{i}}{\hbar}}\, l_{i}^{\mu}\, \frac{1}{l_{i}^{2} + i\epsilon}
\end{flalign}
Using eqn.(\ref{impforkos}), we see that the corresponding contribution in the (classical) radiation kernel is 
\begin{flalign}\label{rckcl1}
\lim_{\hbar\, \rightarrow\, 0}\, {\cal R}^{\mu}_{\ln\, \textrm{cl}}(k)\, =\, -\, \frac{i}{4\pi}\, \ln\omega\, \sum_{a,b\vert a\neq b}\, \frac{1}{(\, (p_{a}\cdot p_{b})^{2}\, -\, m_{a}^{2}m_{b}^{2})^{\frac{3}{2}}}\, \triangle\, \{\, \frac{1}{p_{a}\cdot k}\, (\, p_{a}^{\mu}\, (p_{b} \cdot k)\, -\, p_{b}^{\mu} (p_{a}\ \cdot k)\, ) \}\end{flalign}
where $\triangle f(p_{a}, p_{b})\, :=\, \ \triangle p_{a}^{\mu}\, (\, \frac{\partial}{\partial p_{a}^{\mu}} f\, -\, \frac{\partial}{\partial p_{b}^{\mu}} f\, )$. 
Let us summarise the key results of this section.
\begin{itemize}
\item Combining eqns. (\ref{crkemlo}, \ref{rckcl1}), we see that the $\hbar^{0}$ term in the radiation kernel matches with the result of the radiative gauge field defined by classical log soft theorem, upto next to leading order in the coupling. 
\item  The contribution to the soft factor  resulting from $S^{(1) \mu\nu}$ action on $K_{q}$ in quantum soft theorem has trivial contribution to the classical radiative field. On the other hand ${\cal S}^{\textrm{cl}}_{\ln}$ also has a non-trivial sub-leading ($\omega\ln\omega$) contribution at NLO. Such contributions are expected to be non-universal (\cite{aab}) and we do not investigate them further in this paper.  
\end{itemize}
We can now compare ${\cal R}^{\mu}_{\ln}(k)$ with the radiative gauge field (denoted as $j^{\mu}(k)$) in \cite{aab}, when the final momenta are expanded in terms of initial momenta and impulse. A simple algebra reveals that
\begin{flalign}
j^{\mu}(k)\, =\, 
{\cal R}^{\mu}_{\ln}(k)\, +\, \text{terms proportional to}\, p_{a}\ \cdot \triangle p_{b}
\end{flalign}
However as the result is already at next to leading order, we can substitute $\triangle p_{b}\, =\, -\triangle p_{a}$. And as the impulse is orthogonal to the final momenta, the extra-term vanishes. Hence the radiation kernel computed using KMOC approach at NLO matches with the result in \cite{aab}. 
We thus see that  the classical limit of soft photon theorem in four dimensions match with the classical log soft theorem derived by Saha, Sahoo and Sen. The leading contribution to the classical log soft factor arises from tree-level subleading soft photon theorem and the NLO contribution arises due to loop-corrected quantum soft theorem. For the class of scattering processes amenable to the KMOC formalism, we believe that this derivation provides first step towards a perturbative proof of the classical log soft theorem from scattering amplitudes. 
\section{Soft Gravitational Radiation from sub-leading soft graviton theorems}\label{scsgfde4}
In this section, we consider  scattering of two scalar particles of masses $m_{1},\, m_{2}$ which emit a soft graviton with momentum $k^{\mu}$.  As before, our approach is to take soft limit of the scattering amplitude before taking the classical limit. We focus on the more intricate case of 4 dimensions but the generalisation of the analysis of sections (\ref{cspfdg4}, \ref{csptdg4fq}) to gravitational radiation in $D\, >\, 4$ dimensions is rather straightforward. 

In the soft expansion, the dominant term proportional to $\frac{1}{\omega}$  was derived in \cite{gb} and it was shown that it matches with the classical soft factor. At sub-leading  order in soft expansion (i.e. at $O(\ln\omega)$ in four dimensions)  and at leading order in the coupling,  derivation of classical radiation kernel using KMOC formalism  is fairly similar to the derivation in QED. However as we show below in section (\ref{ptd4v}), there is an interesting aside.
It was proved in \cite{aab}, the classical log soft graviton factor  has an additional ``phase" contribution which is absent in electro-magnetic radiation. This term arises due to the Coulombic drag on outgoing gravitational radiation. When we expand the soft factor in the coupling, the phase term vanishes at leading order.\\
We show that in the KMOC approach, such a term is indeed present in the soft expansion of the amplitude, but at sub-subleading order ! And it vanishes just as the classical phase term vanishes.  In section (\ref{ptd4v}), we will evaluate these terms separately by using  the double copy relations. \\
We also note that in general there is a contribution to the soft gravitational radiation at order $\ln\omega$ from the gravitational stress tensor. However in the KMOC approach to the radiation kernel,  the tree-level scattering amplitude does not involve hard graviton scattering and hence our results do not take into account the contribution of the gravitational stress tensor (or the three graviton coupling) to the radiation kernel. We believe that it is possible to extend our analysis such that the outgoing states contain not only massive scalars but also finite energy gravitons, but in this paper we have restricted ourselves to the simplest set up.\\
We denote the radiation kernel that contributes to soft radiation at the desired order  as
\begin{equation}\label{rkqfslssl}
{\cal R}^{\mu\nu}(k)\, =\, {\cal R}_{(0)}^{\mu\nu}(k)\, +\, {\cal R}_{(1)}^{\mu\nu}(k)
\end{equation}
where ${\cal R}_{(0)}$ is the radiation kernel at sub-leading order and ${\cal R}_{(1)}$ is the potential contribution to the radiation kernel from sub-leading terms in the amplitude, which eventually vanishes. 

But first we consider the contribution of sub-leading soft graviton theorem to the radiation kernel. Let ${\cal A}_{5}(\tilde{p}_{1},\, \tilde{p}_{2},\, p_{1},\, p_{2},\, k)$ be the (un-stripped)  tree-level 5 pt. amplitude. Soft expansion of ${\cal M}_{5}$ is given by,
\begin{equation}
\begin{array}{lll}
{\cal A}_{5}(\tilde{p}_{1},\, \tilde{p}_{2},\, p_{1},\, p_{2},\, k)\, =\, S^{(0)}\, {\cal A}_{4}(\tilde{p}_{1},\, \tilde{p}_{2},\, p_{1},\, p_{2})\, +\, {S}^{(1)}\, {\cal A}_{4}(\tilde{p}_{1},\, \tilde{p}_{2},\, p_{1},\, p_{2})\, +\, O(\omega)
\end{array}
\end{equation}
The soft factors are given by,
\begin{equation}
\begin{array}{lll}
{S}^{(1) \mu\nu}(\tilde{p}_{1}, \tilde{p}_{2}, p_{1}, p_{2})\, =\, i\, \frac{\kappa}{2}\, \sum_{i=1}^{2}(\, \tilde{S}_{(i)}^{(1) \mu\nu}\, +\, S_{(i)}^{(1) \mu\nu}\, )\\[0.4em]
\hspace*{1.4in}=\, i\, \frac{\kappa}{2}\,  \sum_{i=1}^{2}\, [\, \frac{\tilde{p}_{i}^{(\mu}\, \hat{\tilde{J}}_{i}^{\nu)\lambda} k_{\lambda}}{\tilde{p}_{i} \cdot k}\, +\, \frac{p_{i}^{(\mu}\, \hat{J}_{i}^{\nu)\lambda} k_{\lambda}}{p_{i} \cdot k}\, ]\\[0.4em]
S^{(0) \mu\nu}(\tilde{p}_{1}, \tilde{p}_{2}, p_{1}, p_{2})\, =\, \sum_{i=1}^{2}\, \frac{\tilde{p}_{i}^{(\mu}\, \tilde{p}_{i}^{\nu )}}{\tp_{i} \cdot k} + \frac{p_{i}^{(\mu}\, p_{i}^{\nu )}}{\tp_{i} \cdot k}\end{array}
\end{equation}
where as before, the angular momentum operators are defined with a relative minus sign between incoming and out-going states.

We now note the following. 
\begin{equation}\label{triond}
\begin{array}{lll}
(\, \tilde{p}_{i}^{\mu}\, \frac{\partial}{\partial \tilde{p}_{i}^{\nu}}\, -\,  \tilde{p}_{i}^{\nu}\, \frac{\partial}{\partial \tilde{p}_{i}^{\mu}}\, )\, \delta(\tilde{p}_{i}^{2} - m_{i}^{2})\, =\, 0
\end{array}
\end{equation}
The contribution of the sub-leading soft theorem to the radiation kernel can be then evaluated as,
\begin{equation}\label{qrktlg}
\begin{array}{lll}
{\cal R}_{(0)}^{\mu\nu}(k)\, &=\, \int\, \frac{d^{4} l_{1}}{(2\pi)^{4}}\, \frac{d^{4} l_{2}}{(2\pi)^{4}}\, \prod_{i}\, \theta(l_{i}^{0})\, \hdelta(\tilde{p}_{i}^{2} - m_{i}^{2})\, e^{\frac{i}{\hbar} b \cdot l_{1}}\, {S}^{(1)}\, {\cal A}_{4}(\tilde{p}_{1},\, \tilde{p}_{2},\, p_{1},\, p_{2})
\end{array}
\end{equation}
The computation of classical  radiation kernel is made easier by observing following (approximate) identity. 
\begin{flalign}\label{appiddiff}
\begin{array}{lll}
\hspace*{-0.9in}\int\, \frac{d^{4} l_{1}}{(2\pi)^{4}}\, \frac{d^{4} l_{2}}{(2\pi)^{4}}\, \prod_{i}\, \theta(l_{i}^{0})\, \hdelta(\tilde{p}_{i}^{2} - m_{i}^{2})\, e^{\frac{i}{\hbar} b \cdot l_{1}}\, {S}^{(1) \mu\nu}\, {\cal A}_{4}(\tilde{p}_{1},\, \tilde{p}_{2},\, p_{1},\, p_{2})\, \approx\\[0.4em]
\frac{i\kappa}{2}\, \sum_{i=1}^{2}\, \frac{p_{i}^{(\mu} \hat{J}_{i}^{\nu)\lambda} k_{\lambda}}{p_{i} \cdot k}\, \int\,  \frac{d^{4} l}{(2\pi)^{4}}\ \prod_{i}\,  \hdelta(2 p_{i} \cdot l)\, e^{\frac{i}{\hbar} b \cdot l}\,  {\cal M}^{\textrm{cl}}_{4}(p_{1},\, p_{2},\, l_{2})\,
\end{array}
\end{flalign}
In eqn. (\ref{appiddiff}), the approximation sign indicates that the integrands match upto leading order in $l^{\mu}$ and given order in frequency $\omega$. The right hand side of the identity has differential operators which only act on final external states and the $l^{\mu}$ is simply an integration variable. ${\cal M}_{4}^{\textrm{cl}}(p_{1}, p_{2}, l_{2})$ is the classical limit of the four point amplitude. Intuitively this identity simply states that soft and classical limit commute at this order in frequency.  We verify eqn. (\ref{appiddiff}) in appendix \ref{appiddiffp}.

The  reduced four point amplitude and it's classical limit are respectively  given by,
\begin{flalign}
\begin{array}{lll}
&{\cal M}_{4}( \tilde{p}_{1},\, \tilde{p}_{2},\, p_{1},\, p_{2})\, =\, {\cal M}_{4}^{\textrm{cl}}(p_{1}, p_{2}, l_{2}) +\, O(\vert l\vert)\\[0.4em]
\textrm{where} &{\cal M}_{4}^{\textrm{cl}}( p_{1},\, \tilde{p}_{2},\, l_{2})\, :=\, -\, \kappa^{2}\,  \frac{(\, (p_{1}\cdot p_{2})^{2}\, -\, \frac{1}{2}\, m_{1}^{2}\, m_{2}^{2}\, )}{l_{2}^{2}}
\end{array}
\end{flalign}
where $\kappa\, =\, \sqrt{32\pi G}$.

Using the approximate identity,  it can be readily checked that to leading order in momentum mis-match $l^{\mu}$ (and given order in frequency), the radiation kernel obtained from sub-leading soft graviton theorem can be written as,
\begin{flalign}\label{lmmssf}
\begin{array}{lll}
{\cal R}_{(0)}^{\mu\nu}(k) =&&\\
\frac{i\kappa}{2}\, \sum_{i}\, \frac{1}{p_{i}\cdot k}\, p_{i}^{(\mu}\, \hat{J}_{i}^{\nu)\rho}\, k_{\rho}\, \int\, \frac{d^{4} l_{2}}{(2\pi)^{4}}\, \{ \, \delta(2 p_{1} \cdot l_{2})\, \delta(2 p_{2} \cdot l_{2})\, e^{-\, \frac{i}{\hbar} b \cdot l_{1}}\,  {\cal M}_{4}^{\textrm{cl}}( \tilde{p}_{1},\, \tilde{p}_{2},\, p_{1},\, p_{2})\, \}&&
\end{array}
\end{flalign}
The soft radiation contribution contained in eqn.(\ref{lmmssf}) can now be analysed using exactly the same analysis as in QED case. 
That is, we consider the region of integration $\omega\, <<\, \vert l_{2}\vert\, <<\, b^{-1}$ and evaluate the above integral.  In this region, the phase factor trivialises  $e^{\frac{i}{\hbar} b \cdot (k - l)}\, =\, 1$  and (to leading order in $l^{\mu}$) the integral is given by,
\begin{equation}
{\cal R}_{(0)}^{\mu\nu}(k)\, =\, -\, \frac{i \kappa^{3}}{8}\, \sum_{i}\, \frac{1}{p_{i}\cdot k}\, p_{i}^{(\mu}\, \hat{J}_{i}^{\nu)\rho}\, k_{\rho}\, \int_{\omega}^{b^{-1}}\, \frac{d^{4} l_{2}}{(2\pi)^{4}}\, \{ \, \delta( p_{1} \cdot l_{2})\, \delta(p_{2} \cdot l_{2})\,  
\frac{(\, (p_{1}\cdot p_{2})^{2}\, -\, \frac{1}{2}\, m_{1}^{2}\, m_{2}^{2}\, )}{l_{2}^{2} + i\epsilon}\, \}
\end{equation}
This integral was essentially analysed in \cite{sahoo}. As in the case of QED, the pole corresponding to Feynman graviton propagator does not contribute as the initial states are on-shell. As shown in appendix \ref{appA},  adding contribution from all the matter poles, we get
\begin{equation}\label{slrkfab}
{\cal R}_{(0)}^{\mu\nu}(k)\, =\, -\, \ln\omega\, \frac{i \kappa^{3}}{8}\, \sum_{i}\, \frac{1}{p_{i}\cdot k}\, p_{i}^{(\mu}\, \hat{J}_{i}^{\nu)\rho}\, k_{\rho}\, [\, \frac{1}{2\pi}\, \frac{\{(p_{1}\cdot p_{2})^{2}\, -\, \frac{1}{2}\, m_{1}^{2}\, m_{2}^{2}\}}{\sqrt{\, (p_{1}\cdot p_{2})^{2} - m_{1}^{2}m_{2}^{2}}}\, ]
\end{equation}
We can now compare the LO radiation kernel with the classical log soft factor in \cite{aab} at leading order.\footnote{We remind the reader that $\kappa\, =\, \sqrt{32\pi G}$ in this paper. In \cite{aab}, $\kappa\, =\, \sqrt{8\pi G}$. Moreover the radiation kernel ${\cal J}^{\mu\nu}(k)$ in \cite{aab} is at order $\kappa^{2}$ as ${\cal R}^{\mu\nu}_{(0)}(k)\, =\, \kappa\, {\cal J}^{\mu\nu}(k)$.}  We see that 
the two results match upto an overall sign. The sign difference is due to the fact that we use mostly minus metric signature as opposed to mostly plus signature used in \cite{aab}. 
\subsection{The vanishing phase at leading order}\label{ptd4v}
At next to leading order in the coupling, there is an additional term in the classical log soft factor which is a pure phase and arises due to the Coulombic effect of gravitational potential on the out-going radiation itself. For a generic $2\ \rightarrow\ 2$ scattering this term is given by \cite{aab}
\begin{flalign}
{\cal R}_{\textrm{phase}}^{\mu\nu}(k)\, =\, \ln(\omega + i\epsilon)\, \sum_{b=1}^{2}\, p_{b} \cdot k\, S^{(0)}
\end{flalign}
where $S^{(0)}$ is the leading soft factor. As the Weinberg soft graviton factor vanishes at leading order in the coupling, the phase term vanishes.  It is nonetheless an interesting question to ask as to why such a term never appeared in our computation. We will now show that structurally such a term is indeed present, but it arises when we considered sub-subleading soft amplitude. We will first present a schematic argument and then give the detailed computation. This section has no direct relevance for rest of the paper,  and the reader may skip it in the first reading. Our purpose here is to show the existence of such a phase term in KMOC approach and why it vanishes at leading order in the coupling.\\ 
Consider a schematic integral of the following form.
\begin{equation}
{\cal I}\, =\, \int\, d^{4}l\, \hdelta(p_{1} \cdot (k - l))\, \hdelta(p_{2}\cdot l)\, \frac{1}{l^{2}}\, F(p_{1},\, p_{2},\, k,\, l)
\end{equation} 
Now let us suppose we consider region of integration $\vert l\vert\, <<\, \omega$, then this integral is trivial as,  
\begin{equation}
{\cal I}\, =\, \int_{\vert l\vert\, <<\, \omega}\, d^{4}l\, \hdelta(p_{1} \cdot k)\, \hdelta(p_{2}\cdot l)\, \frac{1}{l^{2}}\, F(p_{1},\, p_{2},\, k,\, l)
\end{equation} 
However as $p_{1}^{\mu}$ is time-like and $k^{\mu}$ is null,  $\hdelta(p_{1}\cdot k)\, =\, 0$, this term vanishes.

But if for a moment we ignore the triviality of delta function in this region, then it can be seen that ${\cal I}$ will have a non-trivial contribution at order $\ln\omega$ only if, 
\begin{equation}
F\, \approx\, O(k,\, l^{-1})
\end{equation}
Clearly such a contribution can only arise by considering the soft expansion at sub-subleading order. By examining all the contribution to the tree-level 5 pt. amplitude, it can be seen that the`` inverse dependence" on $l^{\mu}$ implies that $F$ must scale as $\frac{1}{l\cdot k}$ in the integration region $\vert l\vert\, <<\, \omega$ and this contribution arises when the graviton is emitted from the propagator.\footnote{This is why such a contribution is absent in the case of QED, but will be present if we considered classical gluon radiation \cite{golrid}.} The easiest way to compute such a contribution is to look at gravitational amplitude obtained via double copy \cite{jo, ot, wi, jo1, gb1}.


As was shown in \cite{wi}, the tree-level scalar QCD amplitude with two distinct scalar fields  naturally satisfies color kinematics duality  and the 5 point amplitude involving two scalars of masses $m_{1},\, m_{2}$ and a graviton in the external states is given by(\cite{ot, wi}), 
\begin{equation}\label{bcjdcs}
\begin{array}{lll}
{\cal A}_{5}(\tipo,\, \tipt,\,  \po,\, \pt,\, k)\, =\, \delta^{4}(l_{1} + l_{2} - k)\ {\cal M}_{5}(\tipo,\, \tipt,\,  \po,\, \pt,\, k)
\end{array}
\end{equation}
where the reduced amplitude obtained via color kinematics duality has the form \begin{equation}
\begin{array}{lll}
{\cal M}_{5}^{\mu\nu}(\tipo,\, \tipt,\,  \po,\, \pt,\, k)\, =\, -\, \frac{\kappa^{3}}{16}\, \sum_{I=1}^{5}\, \frac{n_{I}^{\mu}\, \otimes\, n_{I}^{\nu}}{d_{I}}
\end{array}
\end{equation}
The numerator kinematic factors $n_{I}$ as well as the corresponding propagators $d_{I}$ were computed in \cite{ot, wi}.\footnote{We deviate slightly from the usual convention in the literature and show the coupling constant dependence explicitly.} The terms corresponding to $I\, \neq\, 3$  arise due to  graviton emission from the four external legs and the third channel  corresponds to  graviton emission from the propagator.  Before proceeding we make a few cautionary remarks on the use of double copy in obtaining gravitational amplitudes with minimally coupled scalars.
\begin{itemize}
\item The double copied 5 point amplitude, contains graviton as well as dilaton and a B-field as an external state. The graviton is isolated simply by contracting the tensor with the symmetric traceless polarisation.
\item Even after ensuring that the external states do not contain a dilaton or the $B$-field, the amplitude obtained via double copy is not pure gravitational amplitude as the dilaton can propagate and mediate interaction between the two scalars\cite{plef, wi, jo1, gb1}. There have been many techniques developed to decouple the dilaton and obtain pure gravity amplitudes. Fortunately  as we will see below, for our purpose these subtleties will not be relevant.  However we emphasise that to do the first principal computation of soft radiation by using color kinematics duality will require that the dilaton is consistently decoupled from the amplitude. 
\end{itemize}
We begin from the well known BCJ representation of the tree-level amplitudes in scalar QCD where the numerator factors are given as,
\begin{equation}\label{bcjsqcd}
\begin{array}{lll}
n_{1}\, =\, (\, 4\, \po \cdot \pt\, -\, 2 \po \cdot l_{2} + 2 \pt\cdot k\, -\, l_{2} \cdot k\, +\, 2 l_{1} \cdot l_{2}\, ) (2 p_{1} - l_{1})^{\mu}\\
\hspace*{2.4in} +\, (\, 2 p_{1}\cdot l_{2}\, +\, l_{2}\cdot k\, 2 l_{1}\cdot l_{2}\, )\, (\, 2p_{2} - l_{2}\, )^{\mu}\\[0.4em]
n_{2}\, =\, (\, 2 p_{1} + l_{2}) \cdot (\, 2 p_{2} - l_{2}\, )\, 2 p_{1}^{\mu}\, +\, 2 p_{1}\cdot k\, (\, 2 p_{2} - l_{2} )^{\mu}\\[0.4em]
n_{3}\, =\, (\, 2p_{1} - l_{1} )^{\alpha} (\, 2p_{2} -  l_{2} )^{\beta}\, [\, (k - l_{2})_{\alpha}\, \eta_{\mu\beta}\, +\, (l_{1} - l_{2})_{\mu} \eta_{\alpha\beta}\, -\, ( k + l_{1} )_{\beta}\, \eta_{\alpha\mu}\, )\\[0.4em]
n_{4}\, =\, n_{1}\vert_{1\leftrightarrow 2}\\[0.4em]
n_{5}\, =\, n_{2}\vert_{1\leftrightarrow 2}
\end{array}
\end{equation} 
And the denominator factors are given by,\footnote{All the propagators are Feynman propagators, but we will suppress the $i\epsilon$ untill we compute the  momentum space integrals} 
\begin{flalign}\label{bcjdcd}
\begin{array}{lll}
d_{1}\, =\, l_{2}^{2}\, (\, (p_{1}\, -\, l_{1}\, + k)^{2} - m_{1}^{2}\, )\\[0.4em]
d_{2}\, =\, -\, 2 p_{1}\cdot k\, l_{2}^{2}&&\\[0.4em]
d_{3}\, =\, l_{1}^{2}\, l_{2}^{2}&&\\[0.4em]
d_{4}\, =\, d_{1}\vert_{1\leftrightarrow 2}&&\\[0.4em]
d_{5}\, =\, d_{2}\vert_{1\leftrightarrow 2}&&
\end{array}
\end{flalign}
The diagrammatic representation satisfying BCJ duality is shown in the figure below.
\begin{figure}[h]
    \centering
    \includegraphics[scale=0.4]{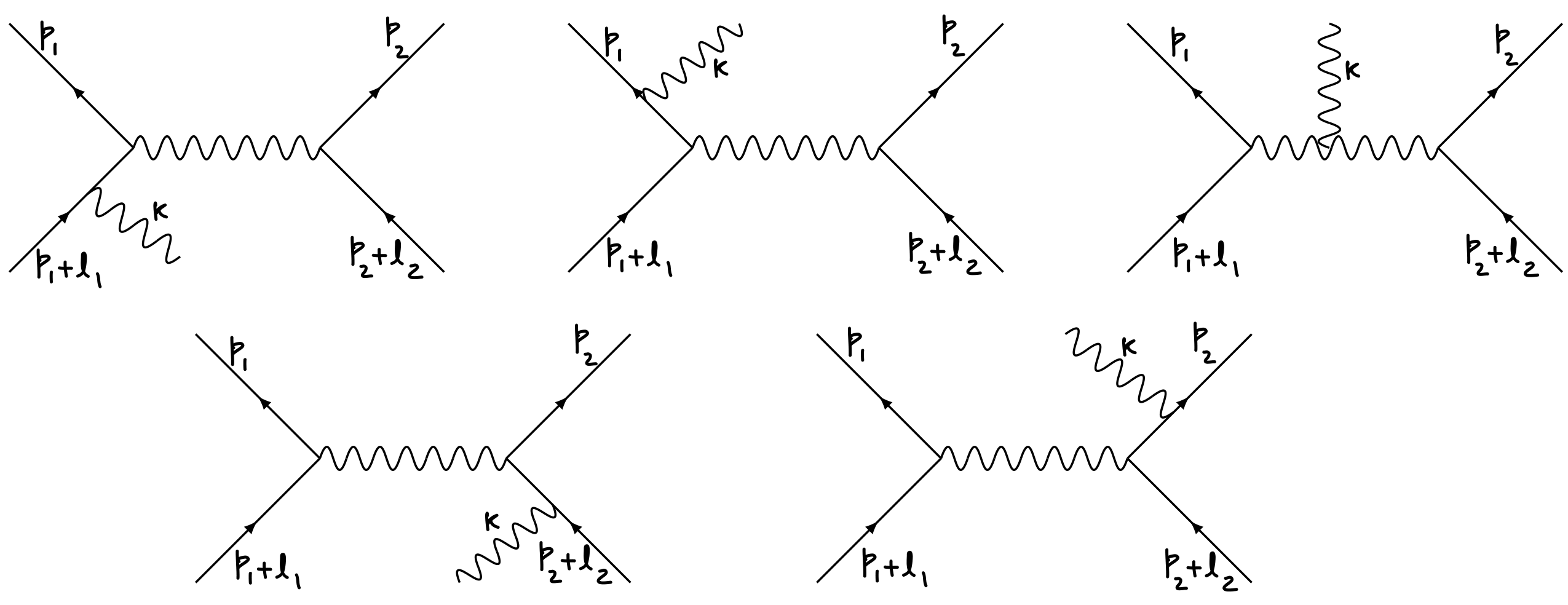}
    \caption{Diagrammatic representation satisfying colour-kinematics duality}
    \label{bcjdiag}
\end{figure}
In this case, the contribution to sub-subleading terms only arises from the third channel. 
To leading order in $l^{\mu}$ this  term can be computed as follows. 
\begin{equation}\label{tpof3}
( n_{3}\, \otimes\, n_{3} )^{\mu\nu}\, =\, \sum_{m=1}^{3}\, \alpha_{m}\, \alpha_{n} P_{m}^{\mu}\, \otimes\, P_{n}^{\nu}
\end{equation}
where $P_{1}\, =\, p_{1},\, P_{2}\, =\, p_{2}$ and $P_{3}\, =\, l_{2}$. The Co-efficients can be easily computed from eqn. (\ref{bcjsqcd}) and the fact that all the external states are on-shell (i.e. $p_{i}\ \cdot l_{i}\, =\, -\, l_{i}^{2}$) 
\begin{equation}\label{dcof3}
\begin{array}{lll}
\alpha_{1}\, =\,  -\, 4\, p_{2}\cdot (\, k + l_{1}\, ) + O(l_{2}^{2}) \\[0.4em]
\alpha_{2}\, =\, 4\, p_{1}\cdot (\, k + l_{2}\, )\, +\, O(l_{2}^{2}) \\[0.4em]
\alpha_{3}\ =\, -8\, p_{1}\cdot p_{2}\ +\ O(l_{2})
\end{array}
\end{equation}
Using eqn.(\ref{dcof3}) in eqn.(\ref{tpof3}), we get 
\begin{equation}\label{db3c}
\begin{array}{lll}
\frac{1}{d_{3}}\, ( n_{3}\, \otimes\, n_{3} )^{\mu\nu}\, =\\[0.4em]
\hspace*{0.4in}\frac{1}{l_{2}^{2}}\frac{1}{(l_{2}^{2}\, -\, 2 l_{2}\, \cdot k)}\,\left(\, [\, 16 (p_{2}\cdot (k + l_{1}))^{2}\, p_{1}^{\mu}\, p_{1}^{\nu}\, + 32\, (p_{1}\cdot p_{2})^{2}\, l_{1}^{\mu}\, l_{1}^{\nu}\right.\\[0.4em]
\hspace*{1.4in}-\, 8\, [\, p_{2}\cdot (k + l_{1})\, ]\, [p_{1}\cdot (k + l_{2})\, ]\, (\, p_{1}^{\mu}\, p_{2}^{\nu}\, +\, p_{2}^{\mu}\, p_{1}^{\nu}\, )\\[0.4em]
\hspace*{1.8in}\left.+\, 32\, (p_{1}\, \cdot\, p_{2})\, (p_{2}\cdot (k + l_{1}) )\, [\, p_{1}^{\mu}\, l_{2}^{\nu}\, +\, p_{1}^{\nu}\, l_{2}^{\mu}\, ]\, \right)\, +\,(\,1 \leftrightarrow 2
 \,)\end{array}
\end{equation}
The propagating dilaton ``infects" all the terms that involve $l_{i}^{\mu}$ in the numerator. However as we show below, the term relevant for us is precisely the term proportional to $\alpha_{1}^{2}$. This term is not affected by  the propogation of the dilaton and hence we do not have to worry about the more refined details of the double copy when obtaining gravity amplitudes.
We consider two separate contributions from the regions $\vert l_{1}\vert\, <<\, \vert k\vert\, <<\, b^{-1}$ and $\vert l_{2}\vert\, <<\, \vert k\vert\, <<\, b^{-1}$ respectively.
It can now be readily verified that with $l_{1}^{\mu}\, +\, l_{2}^{\mu}\, =\, k^{\mu}$, this leads to
\begin{equation}
\begin{array}{lll}
\frac{1}{d_{3}}\, ( n_{3}\, \otimes\, n_{3} )^{\mu\nu}\\[0.4em]
\hspace*{0.4in}=\, -\, 8\, \frac{1}{l_{2}^{2} + i\epsilon}\frac{1}{ l_{2}\, \cdot k - i\epsilon}\, (p_{2}\cdot k)^{2}\, p_{1}^{\mu}\, p_{1}^{\nu}\ \ \  \textrm{if}\ \vert k\vert\, >>\, \vert l_{1}\, \vert\\[0.4em]
\hspace*{0.4in}=\, -\, 8\, \frac{1}{l_{1}^{2} + i\epsilon}\frac{1}{l_{1}\, \cdot k - i\epsilon}\, (p_{1}\cdot k)^{2}\, p_{2}^{\mu}\, p_{2}^{\nu}\ \ \ \textrm{if}\ \vert k\vert\, >>\, \vert l_{2}\,\vert 
\end{array}
\end{equation}
Hence the corresponding contribution to the (un-stripped) 5 point amplitude is given by,
\begin{equation}
\begin{array}{lll}
{\cal I}_{5}\, &\approx\, (\frac{\kappa^{3}}{2})\, \delta^{4}(l_{2} - k)\, \frac{1}{l_{2}^{2} + i \epsilon}\frac{1}{l_{2}\, \cdot k - i\epsilon}\, (p_{2}\cdot k)^{2}\, p_{1}^{\mu}\, p_{1}^{\nu}\ \  \textrm{if}\ \vert k\vert\, >>\, \vert l_{1}\, \vert\\[0.4em]
&=\, (\frac{\kappa^{3}}{2})\, \delta^{4}(l_{1} - k)  \frac{1}{l_{1}^{2} + i\epsilon}\frac{1}{l_{1}\, \cdot k - i\epsilon}\, (p_{1}\cdot k)^{2}\, p_{2}^{\mu}\, p_{2}^{\nu}\ \ \textrm{if}\ \vert k\vert\, >>\, \vert l_{2}\, \vert 
\end{array}
\end{equation}
\begin{equation}
\begin{array}{lll}
{\cal R}^{\mu \nu}_{(1)}(k)\, &=\, 
\, \frac{\kappa^{3}}{8}\, \int_{l}\, \hdelta(p_{1}\cdot k)\, [\,\hdelta(p_{2} \cdot l)\, (p_{2}\cdot k)^{2}\, p_{1}^{\mu}\, p_{1}^{\nu}\, ]\, \frac{1}{l\, \cdot k - i\epsilon}\, \frac{1}{l^{2} + i\epsilon}\, 
+\, (\,1\, \leftrightarrow\, 2\,)\\[0.4em]
&=\, -i\, \frac{\kappa^{3}}{8}\, (p_{2}\cdot k)^{2}\, p_{1}^{\mu}\, p_{1}^{\nu}\, \hdelta(p_{1}\cdot k)\, \int_{l}\, ( \frac{1}{p_{2}\cdot l - i\epsilon}\, -\, \frac{1}{p_{2}\cdot l + i\epsilon} )\, \frac{1}{l\, \cdot k - i\epsilon}\, \frac{1}{l^{2} + \,i\epsilon}\, + (\,1\, \leftrightarrow\, 2\,)
\end{array}
\end{equation}
The contribution of poles from Feynman propagator $\frac{1}{l^{2} + i\epsilon}$ is zero as $\hdelta(\,E_{2}\,\vert\, l\vert\, -\, \vec{p}_{2}\cdot\vec{l}\,)\, =\, 0$.  Whence we focus on the other poles. We close the contour in the lower half plane.  
The resulting integral is,
\begin{flalign}\label{pfkqc}
{\cal R}^{\mu \nu}_{(1)}(k)\, =\, 
 \, \frac{\kappa^{3}}{8}\, (p_{2}\cdot k)^{2}\, p_{1}^{\mu}\, p_{1}^{\nu}\, \hdelta(p_{1}\cdot k)\, \int_{l}\, \frac{1}{p_{2}\cdot l + i\epsilon}\, \frac{1}{l\, \cdot k - i\epsilon}\, \frac{1}{l^{2}}\, +\, (\,1\, \leftrightarrow\, 2\,)
\end{flalign}
The integral in the above equation was evaluated in appendix A of \cite{aab} 
\begin{equation}\label{aaba10}
 \int_{l}\, \frac{1}{p_{2}\cdot l + i\epsilon}\, \frac{1}{l\, \cdot k - i\epsilon}\, \frac{1}{l^{2} - i\epsilon}\, =\, \frac{1}{4\pi}\ \ln(\omega + i\epsilon)\, \frac{1}{p_{2}\cdot k} 
\end{equation}
substituting eqn.(\ref{aaba10}) in eqn.(\ref{pfkqc}) we get,
\begin{flalign}\label{sslpcdc}
{\cal R}^{\mu \nu}_{(1)}(k)\, =\, 
 \frac{\kappa^{3}}{32\pi}\, \ln(\omega + i\epsilon)\, \{\, (p_{2}\cdot k)\, [\, \hdelta(p_{1}\cdot k)\, p_{1}^{\mu}\, p_{1}^{\nu} ]\, +\, (p_{1}\cdot k)\, [\, \hdelta(p_{2}\cdot k)\, p_{2}^{\mu}\, p_{2}^{\nu}\, ]\, \}
 \end{flalign}
As we emphasised before, this term is trivial but it's structure precisely matches with the phase term obtained in \cite{aab}. 
We can now substitute eqns. (\ref{sslpcdc}, \ref{slrkfab}) in eqn.(\ref{rkqfslssl}) and get 
\begin{equation}\label{frcrkglo}
\begin{array}{lll}
\hspace*{-0.5in}{\cal R}^{\mu\nu}(k)\, =\\
\ln\omega\, \frac{\kappa^{3}}{2}\, \bigg[ \frac{-i}{8\pi}\, \frac{1}{p_{1}\cdot k}\, p_{1}^{(\mu}\, \hat{J}_{1}^{\nu)\rho}\, k_{\rho}\, [\, \frac{\{(p_{1}\cdot p_{2})^{2}\, -\, \frac{1}{2}\, m_{1}^{2}\, m_{2}^{2}\}}{\sqrt{\, (p_{1}\cdot p_{2})^{2} - m_{1}^{2}m_{2}^{2}}}\, ]\,  \bigg]\, +\, (\,1\, \leftrightarrow\, 2\,)
\end{array}
\end{equation}
\subsection{Soft Gravitational radiation at NLO}\label{csgfde4nlo}
In this section, we repeat the analysis of section \ref{nloem4d}  and use loop corrected soft graviton theorem to obtain the radiative gravitational field at sub-leading order in soft expansion and next to leading order $\kappa^{5}$ in the coupling. Due to similarity with computations of section (\ref{nloem4d}),  we outline the main results and do explicit computation only for those terms which are qualitatively different than the ones in analysing loop corrected soft photon theorem. 

The loop corrected sub-leading soft graviton theorem for infra-red finite five point amplitude can be written as,
\begin{flalign}
{\cal A}^{\mu\nu}_{5}\, =\, \frac{\kappa^{3}}{8}\, (\, \frac{1}{\omega}\, S_{(0)}^{\mu\nu}\, +\, \ln\omega\, {\cal S}^{\mu\nu}_{\ln}\, )\, {\cal A}_{4}^{\textrm{tree}}\, +\, O(\omega^{0})
\end{flalign}
where we once again remind the reader that $\kappa\, =\, \sqrt{32\pi G}$.

The infrared sensitive loop effects generate a new universal factorisation at order $\ln\omega$ where $\cal S^{\mu\nu}_{\ln}$ only depends on the initial and final momenta of the scattering amplitude. Just as in the case of QED, the loop corrected soft factor can be decomposed into two terms which we denote as ${\cal S}^{\mu\nu}_{\ln \textrm{cl}},\, {\cal S}^{\mu\nu}_{\ln q}$. 
The classical log soft theorem derived in \cite{aab} shows how only ${\cal S}^{\mu\nu}_{\ln \textrm{cl}}$ contributes to classical radiation at $\ln\omega$ order, even though in the quantum soft theorem both the terms occur at the same order in $\hbar$. 
\begin{flalign}
{\cal S}^{\mu\nu}_{\ln}\, =\, {\cal S}^{\mu\nu}_{\ln \textrm{cl}}\, +\, {\cal S}^{\mu\nu}_{\ln q}
\end{flalign}
The expressions for ${\cal S}^{\mu\nu}_{\ln \textrm{cl}}$, ${\cal S}^{\mu\nu}_{\ln q}$ are not easy on the eye but their beauty lies in their universality.  
\begin{flalign}
\begin{array}{lll}
{\cal S}^{\mu\nu}_{\ln \textrm{cl}}\, &&=\\[0.5em]
&&\frac{1}{4\pi}\, \big[\, \frac{1}{2}\, \sum_{a =1}^{4}\, \frac{\tilde{p}_{a}^{(\mu} k_{\rho}}{\tilde{p}_{a} \cdot k}\, \sum_{b\vert \eta_{a} \cdot \eta_{b} = 1}\, \frac{\tilde{p}_{a} \cdot \tilde{p}_{b}}{{\cal D}(\tilde{p}_{a},\, \tilde{p}_{b})^{3}}\, (\, \tilde{p}_{b}^{\rho}\, \tilde{p}_{a}^{\nu)}\, -\, \tp_{a}^{\rho} \tp_{b}^{\nu)}\, )\, \{\, 2 (\tp_{a} \cdot \tp_{b})^{2}\, -\, 3\, \tp_{a}^{2} \tp_{b}^{2}\, \}\\[0.6em]
&&\hspace*{3.8in}+\, \sum_{a=3}^{4}\, (\tp_{a}\ \cdot k)\, S^{(0) \mu\nu}\, \big]
\end{array}
\end{flalign}
In the first line sum is over both the incoming as well as outgoing states with $\tilde{p}_{3}\, :=\, -\, p_{1}$ and $\tilde{p}_{4}\, =\, -\, p_{4}$. ${\cal D}$ is the (by now familiar) Jacobian  and $S^{(0)}$ is the Weinberg soft factor, 
\begin{flalign}
\begin{array}{lll}
{\cal D}(\tilde{p}_{a}, \tp_{b})\, =\, \sqrt{ (\tp_{a} \cdot \tp_{b})^{2} - \tp_{a}^{2} \tp_{b}^{2} }\\[0.4em]
S^{(0) \mu\nu}\, :=\, \sum_{a=1}^{4}\, \frac{\tilde{p}_{i}^{(\mu}\, \tilde{p}_{i}^{\nu )}}{\tp_{i} \cdot k}
\end{array}
\end{flalign}
Similarly, 
\begin{flalign}
\begin{array}{lll}
{\cal S}^{\mu\nu}_{\ln q}&&=\, 
\frac{i}{8 \pi^{2}}\, \big[\, \frac{1}{2} \, \sum_{a,b=1 \vert a \neq b}^{4}\, {S}^{(1) \mu\nu}(\tp_{a}, \hat{k})\, \frac{ \{\, 2 (\tp_{a} \cdot \tp_{b})^{2}\, -\, \tp_{a}^{2} \tp_{b}^{2}\, \}}{{\cal D}(\tp_{a},\, \tp_{b})}\, \ln[\frac{\tp_{a} \cdot \tp_{b} + {\cal D}}{\tp_{a} \cdot \tp_{b} - {\cal D}}\, ]\\[0.5em]
&&\hspace*{2.5in}+\, S^{(0) \mu\nu}\, \sum_{a=1}^{4}\,(\,\tp_{a} \cdot {k}\,)\, \ln\frac{\tp_{a}^{2}}{(\,\tp_{a} \cdot \hat{k}\,)^{2}}\, \big]
\end{array}
\end{flalign}
In order to simplify the analysis, we decompose the soft factors  further as
\begin{flalign}
\begin{array}{lll}
{\cal S}^{\mu\nu}_{\ln \textrm{cl}}\hspace*{-4.5in}&&=\, s^{\mu\nu}_{1,\, \textrm{cl}}(p_{1},\, p_{2})\, +\, s^{\mu\nu}_{2,\, \textrm{cl}}(p_{1},\, p_{2},\, l_{1},\, l_{2})\, +\, s^{\mu\nu}_{3,\, \textrm{cl}}(p_{1},\, p_{2},\, l_{1},\, l_{2})\\[0.4em]
{\cal S}^{\mu\nu}_{\ln q}&&=\, s^{\mu\nu}_{1,\, q}\, +\, s^{\mu\nu}_{2,\, q}
\end{array}
\end{flalign}
where
\begin{flalign}
\begin{array}{lll}
s^{\mu\nu}_{1\, \textrm{cl}}(p_{1},\, p_{2})\, =\, \frac{1}{4\pi}\, \sum_{a =1}^{2}\, \frac{p_{a}^{(\mu} k_{\rho}}{p_{a} \cdot k}\, \sum_{b\vert \eta_{a} \cdot \eta_{b} = 1}\, \frac{p_{a} \cdot p_{b}}{{\cal D}(p_{a},\, p_{b})^{3}}\, (\, p_{b}^{\rho}\, p_{a}^{\nu)}\, -\, p_{a}^{\rho} p_{b}^{\nu)}\, )\\[0.4em]
\hspace*{3.6in} \{\, 2 ( p_{a} \cdot p_{b})^{2}\, -\, 3\, p_{a}^{2} p_{b}^{2}\, \}\\[0.5em]
s^{\mu\nu}_{3, \textrm{cl}}(p_{1},\, p_{2},\, l_{1},\, l_{2})\, =\, \frac{1}{4\pi}\, \sum_{a=3}^{4} (\tp_{a}\ \cdot k)\, S^{(0) \mu\nu}\\[0.5em]
s^{\mu\nu}_{2\, \textrm{cl}}(p_{1},\, p_{2},\, l_{1},\, l_{2})\, =\, {\cal S}^{\mu\nu}_{\ln \textrm{cl}}\, -\, s^{\mu\nu}_{1, \textrm{cl}}\, -\, s^{\mu\nu}_{3, \textrm{cl}}
\end{array}
\end{flalign}

\begin{flalign}
\begin{array}{lll}
\hspace*{-1.5in} s^{\mu\nu}_{1,\, q}\, =\, \frac{i}{16 \pi^{2}}\, \sum_{a,b=1 \vert a \neq b}^{4}\,{S}^{(1) \mu\nu}(\tp_{a}, \hat{k})\, \big[\, \frac{ \{\, 2 (\tp_{a} \cdot \tp_{b})^{2}\, -\, \tp_{a}^{2} \tp_{b}^{2}\, \}}{{\cal D}(\tp_{a},\, \tp_{b})}\, \ln[\frac{\tp_{a} \cdot \tp_{b} + {\cal D}}{\tp_{a} \cdot \tp_{b} - {\cal D}}\, ]\, \big]\\[0.4em]
\hspace*{-1.5in} s^{\mu\nu}_{2,\, q}\, =\,   \frac{i}{8\pi^{2}}\, S^{(0) \mu\nu}\, \sum_{a=1}^{4}\, (\tp_{a} \cdot k) \ln\frac{\tp_{a}^{2}}{(\,\tp_{a} \cdot \hat{k}\,)^{2}}
\end{array}
\end{flalign}
We will analyse ${\cal S}^{\mu\nu}_{\ln \textrm{cl}/q}$ separately.  But we first do a dimensional analysis to analyze which terms contribute in the classical limit. We once again remind the reader that the classical limit of quantum radiation kernel can be written as,
\begin{flalign}
{\cal R}^{\mu\nu}\, \sim\, \lim_{\hbar \rightarrow\, 0}\, \hbar^{\frac{3}{2}}\, \int_{l_{1}, l_{2}}^{\textrm{on-shell}}\, \delta^{4}(l_{1} + l_{2})\, {\cal I}^{\mu\nu}(p_{1}, p_{2}, l_{1}, l_{2})
\end{flalign}
As,
\begin{itemize}
 \item $\kappa\, \sim\, \frac{1}{\sqrt{\hbar}}$, $\kappa^{5}\, \sim\, \frac{1}{\hbar^{\frac{5}{2}}}$ and
 \item $\int_{l_{1}, l_{2}}^{\textrm{on-shell}}\, \delta^{4}(l_{1} + l_{2})\, \frac{1}{l_{2}^{2} + i\epsilon}\, \sim\, \hbar^{0}$ 
 \end{itemize}
 ${\cal I}^{\mu\nu}$ must scale as $O(\hbar)$. If it scales at order $\hbar^{0}$, we will get a super-classical term and an ill-defined classical limit and all the terms which scale as $O(\hbar^{2})$ are purely quantum and will vanish in the classical limit. 
\subsubsection{Contribution of ${\cal S}^{\mu\nu}_{\ln q}$}
We first analyse the contribution of ${\cal S}^{\mu\nu}_{\ln q}$ to the classical radiation kernel at order $\ln\omega$.  Just as in the case of QED, $s^{\mu\nu}_{1,\, q}$ has a vanishing contribution at this order. As the computation is analogous to the analysis in section (\ref{nloem4d}), we do not repeat here. A direct computation reveals that,
\begin{flalign}
s^{\mu\nu}_{1,\, q}\, =\, O(l^{2})
\end{flalign}
It can also verified by a direct computation that $s^{\mu\nu}_{2,\, q}$ does not contribute at next to leading order in the coupling. $S^{(0) \mu\nu}$ depends linearly on $l^{\mu}$ and the sum $\sum_{a=1}^{4}\, (\tp_{a} \cdot k) \ln\frac{\tp_{a}^{2}}{(\,\tp_{a} \cdot \hat{k})^{2}}$ is also linear in $l \cdot k$, thus this term will not contribute to ${\cal R}^{\mu\nu}_{\ln}(k)$ and contributes at $\omega\ln\omega$ order in the soft expansion. 

\subsubsection{Contribution of ${\cal S}^{\mu\nu}_{\ln \textrm{cl}}$} 
The computation of $s^{\mu\nu}_{1, \textrm{cl}}(p_{a},\,p_{b})$ and $s^{\mu\nu}_{2, \textrm{cl}}$ proceeds exactly analogous to the QED computation given in section \ref{cofrlncl}. In the classical limit $s^{\mu\nu}_{1,\textrm{cl}}(p_{a},\, p_{b})$ contributes at order $\omega\ln\omega$.\\

Contribution of $s^{\mu\nu}_{2, \textrm{cl}}$ to the radiation kernel is, 
\begin{flalign}\label{s2conting}
\begin{array}{lll}
{\cal R}^{\mu\nu}_{\ln \textrm{cl}}(k)\, =\, 
\frac{\kappa^{3}}{64\pi}\, \ln\omega\, {\cal M}_{4}^{\textrm{cl}}\, \int \frac{d^{4} l}{(2\pi)^{4}}\, \hdelta(2 p_{1} \cdot l_{1})\, \hdelta (2 p_{2} \cdot l_{2})\, e^{-i \frac{b \cdot l_{1}}{\hbar}}\, \delta^{4}(l_{1} + l_{2})\\[0.4em]
\hspace*{1.5in} \sum_{a, b =1 \vert a \neq b}^{2}\,  \frac{p_{a} \cdot p_{b}}{{\cal D}(p_{a},\, p_{b})^{3}}\, \{\, 2 ( p_{a} \cdot p_{b})^{2}\, -\, 3\, p_{a}^{2} p_{b}^{2}\, \} {\cal I}(p_{1},\, p_{2}, l)\, +\, \dots
\end{array}
\end{flalign}
where $\dots$ denote remaining contribution due to $s^{\mu\nu}_{3, \textrm{cl}}$.  ${\cal I}(p_{1},\, p_{2},\, l)$ is defined as,
\begin{flalign}\label{s2contit}
\begin{array}{lll}
{\cal I}(p_{1},\, p_{2},\, l)\, =\\[0.4em]
\frac{1}{p_{a} \cdot k}\,  [\, l_{a}^{(\mu}\,  k_{\rho} (\, p_{b}^{\rho}\, p_{a}^{\nu)}\, -\, p_{a}^{\rho} p_{b}^{\nu)}\, )\, -\, \frac{l_{a} \cdot k}{p_{a} \cdot k}\, p_{a}^{(\mu}\,  k_{\rho} (\, p_{b}^{\rho}\, p_{a}^{\nu)}\, -\, p_{a}^{\rho} p_{b}^{\nu)}\, )\\[0.4em]
\hspace*{1.8in}+\, p_{a}^{(\mu}\,  k_{\rho}\, \{\, (l_{b}^{\rho} p_{a} ^{\nu} - p_{a}^{\rho} l_{b}^{\nu})\, +\, (p_{b}^{\rho} l_{a} ^{\nu} - l_{a}^{\rho} p_{b}^{\nu})\, \}\, ]
\end{array}
\end{flalign}
Each term in ${\cal I}$ is linear in $l_{a}^{\mu}$ and as 
\begin{flalign}
\begin{array}{lll}
 i {\cal M}_{4}^{\textrm{cl}}  \int \frac{d^{4} l}{(2\pi)^{4}}\, \frac{1}{l^{2} + i\epsilon}\, \hdelta(2 p_{1} \cdot l_{1})\, \hdelta (2 p_{2} \cdot l_{2})\, \delta^{4}(l_{1} + l_{2})\, e^{-i \frac{b \cdot l_{1}}{\hbar}}\, l_{a}^{\mu}\, =\, \triangle p_{a}^{\mu}
\end{array}
\end{flalign}
Hence contribution of $s^{\mu\nu}_{2, \textrm{cl}}$ is,
\begin{flalign}\label{s1s2contrt}
\begin{array}{lll}
{\cal R}^{\mu\nu}_{\ln \textrm{cl}}(k)\, =\, 
-\, \frac{ i\, \kappa^{3}}{64\pi}\, \ln\omega\, \sum_{a, b =1 \vert a \neq b}^{2}\,  \frac{p_{a} \cdot p_{b}}{{\cal D}(p_{a},\, p_{b})^{3}}\, \{\, 2 ( p_{a} \cdot p_{b})^{2}\, -\, 3\, p_{a}^{2} p_{b}^{2}\, \}\, {\cal I}(p_{1},\, p_{2}, \triangle p_{a})\, +\, \dots
\end{array}
\end{flalign}
We now analyse the contribution of $s^{\mu\nu}_{3, \textrm{cl}}$ to the radiation kernel.  As shown in section \ref{ptd4v}, at leading order ($\kappa^{3}$) in the coupling, there is no contribution of such a phase term. We note that this is consistent with the structural form of $s^{\mu\nu}_{3, \textrm{cl}}$ which has trivial contribution at $l^{0}$ order. The leading non-trivial contribution is in fact given by,
\begin{flalign}\label{s3lins}
s^{\mu\nu}_{3, \textrm{cl}}\, =\, \frac{1}{4\pi}\, \sum_{a=1}^{2} (p_{a} \cdot k)\, \sum_{b=1}^{2}\, [\, \frac{2 l_{b}^{(\mu} p_{b}^{\nu)}}{p_{b} \cdot k}\, -\, \frac{p_{b}^{\mu} p_{b}^{\nu}}{(p_{b} \cdot k)^{2}}\, l_{b} \cdot k\, ]
\end{flalign}
We can now substitute eqn. (\ref{s3lins}) in the integrand for ${\cal R}^{\mu\nu}_{\ln \textrm{cl}}(k)$ and just as it was seen in eqns. (\ref{s2conting}, \ref{s2contit}), the result is  simply a replacement of $l_{a}^{\mu}$ in eqn.(\ref{s3lins}) with $\triangle p_{a}^{\mu}$. 
Substituting this result in eqn.(\ref{s1s2contrt}), we determine the classical radiation kernel at next to leading order and at sub-leading order in frequency expansion.
\begin{flalign}\label{clsfgnlofr}
\begin{array}{lll}
{\cal R}^{\mu\nu}_{\ln \textrm{cl}}(k)\, =\\[0.4em]
-\, \frac{i\, \kappa^{3}}{32\pi}\, \ln\omega\, \big[\, \frac{1}{2}\, \sum_{a, b =1 \vert a \neq b}^{2}\,  \frac{p_{a} \cdot p_{b}}{{\cal D}(p_{a},\, p_{b})^{3}}\, \{\, 2 ( p_{a} \cdot p_{b})^{2}\, -\, 3\, p_{a}^{2} p_{b}^{2}\, \} {\cal I}(p_{1},\, p_{2}, \triangle p_{a})\\[0.5em]
\hspace*{1.3in}+\,  \sum_{a=1}^{2} (p_{a} \cdot k)\, \sum_{b=1}^{2}\, [\, \frac{2 \triangle p_{b}^{(\mu} p_{b}^{\nu)}}{p_{b} \cdot k}\, +\, \frac{p_{b}^{\mu} p_{b}^{\nu}}{(p_{b} \cdot k)^{2}}\, \triangle p_{b} \cdot k\, ]\, \big]
\end{array}
\end{flalign}
It can now be readily verified that ${\cal R}^{\mu\nu}_{\ln \textrm{cl}}(k)$ equals the classical log soft factor for gravity at NLO upto an overall sign. The equality (modulo sign) is for the same reason as in QED. Namely,  $\triangle p_{a}$ is transversal to both the final momenta. The relative sign is due to change in the metric signature. Combining eqn.(\ref{clsfgnlofr}) with 
eqn.(\ref{frcrkglo}) for the leading order result, we see that the NLO gravitational radiation kernel at sub-leading order in frequency is consistent with classical soft graviton theorem.\\
We end this section with a speculative remark. One of the most striking developments in the relationship between classical General Relativity and scattering amplitudes is the study of scattering of Kerr blackholes which are treated as point particles with universal coupling to (linearised) gravity as dictated by no hair theorem. The coupling of Kerr blackhole with linearised metric perturbation equals the minimal  3 point coupling of a finite mass particle with infinite spin with graviton. It was shown in \cite{cfq3.6} that this dictionary can be used in the KMOC formalism to compute classical observables such as momentum impulse involving scattering of Kerr blackholes. This essentially amounted to an imaginary shift in the impact parameter by the ring radius $\vec{b}\, \rightarrow\, \vec{b} - i\vec{a}$. This rather strikingly simple map (from Schwarzchild black hole to Kerr blackhole) leads us to speculate that even from the perspective of scattering amplitudes the classical log soft factor is insensitive to the spin of the black holes.  This is because the contribution to the soft radiation comes from $\omega\, <<\, \vert l\vert\, <<\, b^{-1}\, <<\, a^{-1}$ or $\vert l\vert\, <<\, \omega\, <<\, b^{-1}\, <<\, a^{-}$ regions, the complex shift which results in $e^{\frac{-i b \cdot l}{\hbar}}\, \rightarrow\, e^{-l \cdot a}e^{i b\cdot l}$ has no effect on the soft regions as the exponents become unity.  

Note that this result (if established by concrete computation) is in fact rather obvious from the analysis of (\cite{ashoke1906}, \cite{aab}), as in that derivation the higher multipoles do not effect the classical soft factors upto sub-leading order in the frequency. But it is pleasing that this fact may be verified in KMOC formlism as well.

\subsection{Generalisation to NNLO?}\label{nnloc}
Our derivation of classical log soft radiative field from infra-red finite amplitude does not admit a direct generalisation  to higher orders. At one loop ${\cal A}_{\textrm{IR-fin}}(p_{1},\, \dots,\, p_{n})\, =\, 0$, but this is not so at higher loops. If we consider the soft expansion of $L$-loop five point amplitude, then the quantum log soft theorem can be written as,
\begin{flalign}
\amp_{L}^{\textrm{IR-fin}}\, \sim\,  \ln\omega\, {\cal S}_{\ln}\, \usafour^{\textrm{IR-fin}}_{L-1}
\end{flalign}
where the infra-red finite four point amplitude has a rather intricate structure which has been investigated in \cite{knut}. As $\cal S_{\ln}$ is one loop exact, it's form remains the same but the higher loop four point amplitudes need to be treated with care in KMOC formalism. ${\cal S}_{\ln q}$ scales with momentum mismatch at $O(l^{2})$ and hence will also start contributing at this order in the coupling\footnote{$l^{\mu}$ scales linearly with $\hbar$ and increasing orders of $\hbar$ can be compensated by higher orders in the coupling as coupling scales as $\frac{1}{\sqrt{\hbar}}$.} and delicate cancellations will have to take place so that at any order in the coupling ${\cal S}_{\ln q}$ does not contribute at sub-leading order in soft expansion.\\
We note that it is at NNLO order that a new subtlety in the proof of classical soft theorem from loop corrected quantum soft theorem enters the picture. Till NLO, ${\cal S}_{\ln q}$  vanishes for a $2\, \rightarrow\, 2$ scattering in large impact parameter regime and the classical limit obtained from KMOC formalism is consistent with this result. At NNLO, ${\cal S}_{\ln q}$ is non-vanishing when final momenta are expanded in terms of initial momenta and impulse and hence it's cancellation in the classical limit would provide a highly non-trivial test on classical limit of quantum soft theorem.

 We expect that the final answer should agree with the classical log soft factor, when final momenta are expanded in terms of initial momenta and impulse at next to leading order \cite{kosower}.

\section{Open Issues}
There is now a large body of work which utilises the remarkable simplicity and power of on-shell techniques to compute classical observables  such as scattering angle or Impulse. However the main focuse so far has been on conservative dynamics and analysis of radiation and inelastic scattering in general remains in it's formative stages. Few  notable exceptions in this regard are (\cite{wi}, \cite{gb}, \cite{gb1})  and the  papers by Veneziano and his collaborators(\cite{ven1}, \cite{ven2}). These works have opened doors to analyse radiative sector of classical scattering processes using on-shell techniques.  On the classical side, Saha, Sahoo and Sen proved in complete generality that  the tail to the memory terms in any scattering process in four  dimensions have a universality and are completely determined by the asymptotic momenta of the scattering objects. These classical soft theorems were in turn motivated by loop corrected quantum soft theorems derived in \cite{ashoke1, ashoke2, sahoo}. Inspired by these results we attempted to prove the classical log soft theorem in \cite{aab} using formulation developed in \cite{kosower}.  Although our work merely verifies the established results upto next to leading order, we believe that it constitutes the first step in providing a  perturbative proof of the classical log soft theorem from scattering amplitudes in four dimensions.

Thus a rather obvious open issue is to extend this analysis to higher orders (NNLO) in the coupling. As we argued in section \ref{nnloc}, this could either involve applying KMOC formalism to 2-loop amplitudes or to use loop corrected soft theorems for bare amplitudes where the intra-red divergent factor has not been removed. It will also be extremely interesting to see if the sub-subleading soft factors in $D\, =\, 4$ dimensions which are conjectured to be universal \cite{aab} and occur at $O(\omega\, (\ln\omega)^{2})$ in the soft expansion can be related to soft expansion of scattering amplitudes. 

Throughout the paper, we analysed radiation emitted from spinless particles. From the perspective of  scattering of Kerr blackholes, inclusion of spin in the analysis will be interesting. In $D\, >\, 4$ dimensions, the sub-leading soft graviton factor is universal and has a term which is linear in spin of the particle. KMOC formalism can be used  to derive the soft radiation for spinning particles using the spin-part of sub-leading soft graviton theorem. \cite{deb-prep}. 

The relationship between log soft theorems and the double copy structure in scattering amplitudes remains to be explored.  Naive analysis indicates that soft gluon theorem is not loop corrected in any controllable way as loop correction induces a soft factor which diverges as $\frac{\ln\omega}{\omega}$. It will be extremely interesting to use the techniques developed in \cite{gb1} and check if the classical log soft factor for gravity can be derived using double copy relations. This may be more then just an academic exercise as the ``classical double copy" which relates radiative solutions in classical yang-Mills theory and a gravitational theory have aquired a central stage in recent developments.\footnote{We are grateful to Biswajit Sahoo for discussions on this issue.} 

The formalism developed by Kosower, Maybee and O'Connell is for $2\, \rightarrow\, 2$ scattering, but if the separation between any pair of particles remains large then we believe that this analysis can be generalised to $n\, \rightarrow\, m$ particle scattering. This is because the crucial requirement for the KMOC formalism is the existence of so-called ``Goldilock's zone" defined by $l_{c}\, <<\, l_{w}\, <<\, b_{ij}$. Such zones will exist as long as the inter-particle separation $b_{ij}$ between any pair of particles remains large. 

However as was shown in \cite{ashoke1801}, the classical soft theorem remains valid even when the system is not in large impact parameter regime. It continues to hold when, (1) there is plunge (two states colliding and merging into a single object), or fragmentation where a given body fragments under influence of internal forces, (2) In a generic classical scattering process, the outgoing states are not only described as point particles (with multipole moments) but also flux of finite energy  massless fields.\\
The KMOC formalism is not directly applicable to any of these scenarios as scattering process in such cases (such as plunge) is not described by perturbative amplitudes of asymptotic multi particle states. However the fact that emitted radiation satisfies classical soft theorem perhaps hints at a possibility  that there must be  generalisation of the KMOC framework to the scenario where the outgoing states are described not only by single particle states but by coherent states of say finite energy gravitons and where bound states can form during scattering.  We leave these and myriad of other questions with a hope of future investigations.
\section*{Acknowledgement}
We are indebted to Alfredo Guevara and Ashoke Sen for key discussions which led to the formulation of this project. We thank Ritabrata Bhattacharya for discussions and collaboration in the initial stages of the project. We thank Siddharth Prabhu, Suvrat Raju, Pushkal Srivastava and Arnab Priya Saha for many discussions on issues pertaining to soft theorems. We would especially like to thank Sayali Atul Bhatkar, Miguel Campiglia, Biswajit Sahoo and Ashoke sen for patiently clarifying our numerous doubts regarding classical and quantum soft theorems in four  dimensions and Alfredo Guevara for his comments on the earlier version of the manuscript.
 \appendix
\section{Classical soft log factor in $D\, =\, 4$}\label{d=4ssrqed}
In this appendix we review the derivation of the classical soft radiation in four  dimensions. Our analysis essentially follows that  in \cite{aab} with a minor technical difference being that  (1) we do not consider final and initial momenta to be independent, and (2)  as our set up is that of \cite{golrid, kosower, chen}, we impose  boundary condition that particles are free in the far past.\footnote{In \cite{aab}, the initial and final state particles were considered independent precisely as the soft theorems are phrased. Due to this, they had an additional boundary condition on incoming as well as outgoing particles at some finite time. As our final states are determined by equations of motion of the initial states, there are some small technical differences in the computation.}   The trajectories of the particles are hence parametrized as,
\begin{equation}
x^{\mu}_{i}(\sigma_{i})\, =\, b^{\mu}_{i}\,+\, v_{i}\, \sigma^{\mu}_{i}\, +\, z_{i}^{\mu}(\sigma_{i})\, \textrm{with}\, \lim_{\sigma_{i}\, \rightarrow\, -\infty}\, z^{\mu}_{i}(\sigma_{i})\, =\, 0
\end{equation}
The key difference in $D\, =\, 4$ and $D\, >\, 4$ dimensions is that generically particles are not  asymptotically free  and hence the specific boundary conditions imposed  in the far past play an important role in that the soft radiation is only emitted in the far future.\footnote{As we will argue below, these conditions  essentially mean that  $\hdelta(p_{1}\cdot (k-l))$ is replaced with $\frac{1}{(p_{1}\cdot (k-l)\, -\, i\epsilon)}$.}.  
We consider the radiative gauge field ${\cal R}^{\mu}_{1}(k)$ emitted by particle $1$ with mass and charge being $m_{1},\, q_{1}$.  The complete answer is obtained by interchanging particles $1$ and $2$ in the answer for ${\cal R}_{\mu}^{1}(k)$ to obtain ${\cal R}_{\mu}^{2}(k)$ and adding the two contributions. 
\begin{equation}
\begin{array}{lll}
{\cal R}^{\mu}_{1}(k)\, =\, q_{1}\, \int\, d\sigma_{1}\, e^{ik\, \cdot\, x_{1}(\sigma_{1})}\, [\, v_{1}^{\mu}\, +\, \dot{z}_{1}^{\mu}(\sigma_{1})\, ]\, +\, \textrm{Bnd-term}
\end{array}
\end{equation}
Where the boundary term is required to make the integral well defined. As was shown in \cite{ashoke1804}, addition of such a boundary term is tantamount to defining ${\cal R}^{\mu}_{1}(k)$ as, 
\begin{equation}
{\cal R}^{\mu}_{1}(k)\, =\, i\, q_{1}\, \int d\sigma_{1}\, e^{ik\cdot x_{1}(\sigma_{1})}\, \frac{d}{d\sigma_{1}}\, [\, \frac{ \, p_{1}^{\mu}\, +\, m_{1}\dot{z}_{1}^{\mu}(\sigma_{1})\, }{\, (p_{1}\, +\, m_{1}\dot{z}_{1})\cdot k}\, ]
\end{equation}
At the leading order in the coupling, we can re-write this equation in terms of,\\ $a_{1}^{\mu}(\sigma)\, :=\, \frac{d^{2}\, z_{i}^{\mu}}{d\sigma^{2}}$ as
\begin{equation}
{\cal R}^{\mu}_{1}(k)\, =\, i\, q_{1}\, \int\, d\sigma_{1}\, e^{ik\cdot x_{1}(\sigma_{1})}\, [\, \frac{1}{p_{1}\, \cdot k}\, m_{1}\, a_{1}^{\mu}(\sigma_{1})\, -\, \frac{1}{(p_{1}\cdot k)^{2}}\, m_{1}\, k\, \cdot\, a(\sigma_{1}) \, p_{1}^{\mu}\,  ]
\end{equation}
\begin{equation}
\begin{array}{lll}
{\cal R}^{\mu}_{1}(k)\, =\\
i\, q_{1}\, \int\, d\sigma_{1}\, \left(\, e^{ik\cdot x_{1}(\sigma_{1})}\, -\, 1\right)\, [\, \frac{1}{p_{1}\, \cdot k}\, m_{1}\, a^{\mu}(\sigma_{1})\, -\, p_{1}^{\mu}\, \frac{1}{(p_{1}\cdot k)^{2}}\, m_{1}\, k\, \cdot\, a(\sigma_{1})\, ]  \\[0.4em]
\hspace*{1.9in}\, +\; {i}\, \int\, d\sigma_{1}\,  [\, \frac{1}{p_{1}\; \cdot\; k}\, m_{1}\, a^{\mu}(\sigma_{1})\, -\, p_{1}^{\mu}\; \frac{1}{(p_{1}\cdot k)^{2}}\, m_{1}\, k\, \cdot\, a(\sigma_{1})\, ]
\end{array}
\end{equation}
It is easy to check that the second term produces leading order soft radiation and has no sub-leading terms. We thus focus on the first term and denote it as $\tilde{\cal R}^{\mu}_{1}(k)$.
\begin{equation}
\begin{array}{lll}
\tilde{{\cal R}}^{\mu}_{1}(k)\, =\, i\, q_{1}\, \int\, d\sigma_{1}\, \left(\, e^{ik\cdot x_{1}(\sigma_{1})}\, -\, 1\right)\, [\, \frac{1}{p_{1}\, \cdot k}\, m_{1}\, a^{\mu}(\sigma_{1})\, -\, p_{1}^{\mu}\, \frac{1}{(p_{1}\cdot k)^{2}}\, m_{1}\, k\, \cdot\, a(\sigma_{1})\, ]  \\
\end{array}
\end{equation}
To impose the boundary condition that the particles are free in the fast past, we use the $i\epsilon$ prescription in the exponent as \cite{chen}
\begin{flalign}
e^{i l^{\prime} \cdot x_{1}(\sigma_{1})}\, \rightarrow\, e^{i l^{\prime} \cdot x_{1}(\sigma_{1}  - i\epsilon)}
\end{flalign}
Using eqn.(\ref{zacc}) we can now write the classical radiation current as,
\begin{equation}
\begin{array}{lll}
\tilde{{\cal R}}^{\mu}_{1}(k)\, =\\[0.4em]
- \, q_{1}^{2}q_{2}\, \int\, \frac{d^{4}l}{(2\pi)^{4}}\, G_{r}(l)\, e^{-i l\cdot b}\, \hdelta(p_{2}\cdot l)\\[0.4em]
\left\{\,  e^{ik\cdot b}\, \hdelta(p_{1}\cdot (k-l)\, -\, i\, \epsilon)\, -\, \hdelta(p_{1}\cdot l + i\epsilon)\, \right\}\\[0.4em]
\hspace*{2.4in} [\, \frac{1}{p_{1}\, \cdot k}\, \tilde{f}^{\mu\nu}(p_{1},l)\, p_{1 \nu}\, -\, p_{1}^{\mu}\, \frac{1}{(p_{1}\cdot k)^{2}}\, k_{\alpha} \cdot \tilde{f}^{\alpha\beta}(p_{1},l)\, p_{1 \beta}\, ]
\end{array}
\end{equation}
where, $ \tilde{f}^{\alpha\beta}(p_{1},l)= [\, l\wedge\, p_{1}\, ]^{\alpha\beta}$.
We consider the contribution to the region determined by $\omega\, <<\, \vert l\vert\, <<\, b^{-1}$ due to which the exponentials can be set to one 
\begin{equation}
\begin{array}{lll}
\tilde{{\cal R}}^{\mu}_{1}(k)&=\\[0.4em]
& -\, q_{1}^{2}q_{2}\, \int\, \frac{d^{4}l}{(2\pi)^{4}}\, G_{r}(l)\, \hdelta(p_{2}\cdot l)
\left\{\, \hdelta(p_{1}\cdot (k-l)\, -\, i\, \epsilon)\, -\, \hdelta(p_{1}\cdot l + i\epsilon)\, \right\}\\[0.4em]
&\hspace*{2.0in} [\, \frac{1}{p_{1}\, \cdot k}\, \tilde{f}^{\mu\nu}(p_{1},l)\, p_{1 \nu}\, -\, p_{1}^{\mu}\, \frac{1}{(p_{1}\cdot k)^{2}}\, k_{\alpha} \cdot \tilde{f}^{\alpha\beta}(p_{1},l)\, p_{1 \beta}\, ]
\end{array}
\end{equation}
In this integration region we also have, 
\begin{equation}
\begin{array}{lll}
\{\, \hdelta(p_{1}\cdot (k-l)\, -\, i\, \epsilon)\, -\, \hdelta(p_{1}\cdot l + i\epsilon)\, \}\, =\\[0.4em]
\hspace*{1.3in}  \,-i\, \{\, \frac{1}{p_{1}\cdot (k-l) - i\epsilon}\, -\, \frac{1}{p_{1}\cdot l + i\epsilon}\, \}\, +\,  \, i\, \{\, P(\frac{1}{p_{1}\cdot (k-l)}) \, -\, P(\frac{1}{p_{1}\cdot l})\, \}
\end{array}
\end{equation}
In the integration region of interest, the second term vanishes. 
We now use the identities,
\begin{equation}
\frac{1}{(p_{1}\cdot (k-l))_{-}}\, -\, \frac{1}{(p_{1}\cdot l)_{+}}\, =\, \frac{2}{(p_{1}\cdot (k-l))_{-}}\, -\, \frac{ p_{1}\cdot k}{(p_{1}\cdot (k-l))_{-}\, (p_{1}\cdot l)_{+}}\\
\end{equation}
It can also be checked that in $\omega\, <<\, \vert l\vert\, <<\, b^{-1}$,  the first term will produce $O(\omega^{0})$ terms and hence we drop it in this computation as such term will contribute to the radiation at higher order in $\omega$.  
Hence we focus on the second term. 
\begin{equation}\label{class4d1}
\begin{array}{lll}
\tilde{{\cal R}}^{\mu}_{1}(k)\, =\\[0.4em]
 \, -\, i\, q_{1}^{2}q_{2}\, \int\, \frac{d^{4}l}{(2\pi)^{4}}\, G_{r}(l)\, \hdelta(p_{2}\cdot l)\,  \{\, \frac{1}{(p_{1}\cdot (k-l))_{-}\, (p_{1}\cdot l)_{+}}\, \}\\[0.4em]
\hspace*{2.4in} [\, \tilde{f}^{\mu\nu}(p_{1},l)\, p_{1 \nu}\, -\, p_{1}^{\mu}\, \frac{1}{(p_{1}\cdot k)}\, k_{\alpha} \cdot \tilde{f}^{\alpha\beta}(p_{1},l)\, p_{1 \beta}\, ]
\end{array}
\end{equation}
Using the fact that $G_{r}(l)\, =\, \frac{1}{(l_{0} + i\epsilon)^{2}\, -\, \vec{l}^{2}}$, it can be readily seen that if we write $\hdelta(p_{2}\cdot l)\, =\, -i\, [\, \frac{1}{p_{2}\cdot l - i\epsilon}\, -\, \frac{1}{p_{2}\cdot l + i\epsilon}\, ]$ then the second term will not contribute to eqn.(\ref{class4d1}) by closing the contour in upper half plane. So we finally get,
\begin{equation}
\begin{array}{lll}
\tilde{{\cal R}}^{\mu}_{1}(k)\, =\\[0.4em]
-\, q_{1}^{2}q_{2}\, \int\, \frac{d^{4}l}{(2\pi)^{4}}\, G_{r}(l)\,\frac{1}{p_{2}\cdot l - i\epsilon}\, \{\, \frac{1}{(p_{1}\cdot (k-l))_{-}\, (p_{1}\cdot l)_{+}}\, \}\\[0.4em]
\hspace*{2.4in} [\, \tilde{f}^{\mu\nu}(p_{1},l)\, p_{1 \nu}\, -\, p_{1}^{\mu}\, \frac{1}{(p_{1}\cdot k)}\, k_{\alpha} \cdot \tilde{f}^{\alpha\beta}(p_{1},l)\, p_{1 \beta}\, ]
\end{array}
\end{equation}
This formula matches the integral formula derived in section 4 in \cite{aab}, from where it was shown that soft radiation equals the classical log soft factor.  
\section{Proof of eqn. (\ref{eq1})}\label{appA}
In this section we prove identity used in eqns.(\ref{eq1}).
We first split the radiation current (denoted as ${\cal R}^{\mu}(k)$ in eq .(\ref{eq1}) ) in two parts. 
\begin{equation}
{\cal R}^{\mu}(k)\, =\, {\cal R}^{\mu}_{1}(k)\, +\, {\cal R}^{\mu}_{2}(k)
\end{equation}
where from sub-leading soft photon theorem in eqn.(\ref{sspt2t}) we have,
\begin{equation}\label{clossfd4i}
\begin{array}{lll}
{\cal R}^{\mu}_{1}(k)\, :=\, 
q_{1}^{2}\, q_{2}\, \int_{l\, \in\, {\cal S}}\, G_{F}(l)\, \hdelta(2p_{1}\cdot l)\, \hdelta(2 p_{2}\cdot l)\, \left[\, \{\, 4\frac{p_{2}\cdot k}{p_{1}\cdot k}\, p_{1}^{\mu} - 4\, p_{2}^{\mu}\; \}\, \right] +\, ( 
1\, \leftrightarrow\, 2\,) \\
\end{array}
\end{equation}
The above equation can also be written as 
\begin{equation}
\begin{array}{lll}
{\cal {R}}^{\mu}_{1}(k) =\\ q_{1}^{2}\, q_{2}\, \int_{l \in \mathcal{S}}\, G_{F}(l)\, \hat{\delta}(p_{1}\cdot l)\, \hat{\delta}(p_{2}\cdot l)\, \frac{k_{\nu}}{p_{1}\cdot k} \,\left[\, \{ \, p_{1}^{\mu} \, \frac{\partial}{\partial p_{1\nu}} \, -\,  p_{1}^{\nu} \, \frac{\partial}{\partial p_{1\mu}} \} \, \right] \, (p_{1}\cdot p_{2}) \, +\, (1 \leftrightarrow 2\, ).
\end{array}
\end{equation}
Similarly,
\begin{equation}
\begin{array}{lll}
{\cal R}^{\mu}_{2}(k)\, :=\\
q_{1}^{2}q_{2} \int_{l \in \mathcal{S}}\, G_{F}(l)\, \hat{\delta}'(2p_{1}.l)\, \hat{\delta}(2p_{2}.l)\, \left[ \, \{ \, \frac{l \cdot k}{p_{1}\cdot k} \,  p_{1}^{\mu}\,  -\,  \, l^{\mu} \, \} \, (4p_{1}\cdot p_{2}) \, \right]\, + \, ( \,1 \leftrightarrow 2 \, ).
\end{array}
\end{equation}
The prime on the delta function denotes derivative with respect to the argument.\\
This integral can also be written in terms of the sub-leading operator for the two particles by noting that 
\begin{equation}
\hat{\delta}'(p_{i}\cdot l) \, l^{\mu}\, =\, \frac{\partial}{\partial p_{i\mu}}\,  \hat{\delta}(p_{i}\cdot l).
\end{equation}
Using the above trick the second integral can be written as 
\begin{equation}
\begin{array}{lll}
\mathcal{R}^{\mu}_{2}(k) = \\ q_{1}^{2}\, q_{2}\, (p_{1}\cdot p_{2})\,  \frac{k_{\nu}}{p_{1}\cdot k} \,  \left[ \, \big( \,  p_{1}^{\mu} \, \frac{\partial}{\partial p_{1\nu}} \, - \, p_{1}^{\nu} \, \frac{\partial}{\partial p_{1\mu}} \, \big) \, \right]  \, \int_{l \in \mathcal{S}}\, G_{F}(l)\, \hat{\delta}(p_{1}\cdot l)\, \hat{\delta}(p_{2}\cdot l) + \, ( \, 1 \leftrightarrow 2\, ).
\end{array}
\end{equation}
We can add the two integrals and get
\begin{equation} \label{eqn1}
\begin{array}{lll}
{\cal R}^{\mu}(k)\, =\, \mathcal{R}^{\mu}_{1}(k) + \mathcal{R}^{\mu}_{2}(k)  =\\ q_{1}^{2}\, q_{2}\, \frac{k_{\nu}}{p_{1}\cdot k}\,  \left[ \, \big( \, p_{1}^{\mu} \, \frac{\partial}{\partial p_{1\nu}} \, - \,   p_{1}^{\nu} \, \frac{\partial}{\partial p_{1\mu}} \, \big) \, \right] \, 
 \bigg\{ \, (p_{1}\cdot p_{2})\, \int_{l \in \mathcal{S}}\, G_{F}(l)\, \hat{\delta}(p_{1}\cdot l)\, \hat{\delta}(p_{2}\cdot l) \, \bigg\}  \,+  \,(1 \leftrightarrow 2\, ) 
\end{array}
\end{equation} 
\section{Proof of identity in eqn (\ref{appiddiff})}\label{appiddiffp}
In this section we verify eqn. (\ref{appiddiff}) written below for convenience. 
\begin{flalign}\label{appc1}
\begin{array}{lll}
\hspace*{-0.9in}\int\, \frac{d^{4} l_{1}}{(2\pi)^{4}}\, \frac{d^{4} l_{2}}{(2\pi)^{4}}\, \prod_{i}\, \theta(l_{i}^{0})\, \hdelta(\tilde{p}_{i}^{2} - m_{i}^{2})\, e^{\frac{i}{\hbar} b \cdot l_{1}}\, {S}^{(1) \mu\nu}\, {\cal A}_{4}(\tilde{p}_{1},\, \tilde{p}_{2},\, p_{1},\, p_{2})\, \approx\\[0.4em]
\frac{i\kappa}{2}\, \sum_{i=1}^{2}\, \frac{p_{i}^{(\mu} \hat{J}_{i}^{\nu)\lambda} k_{\lambda}}{p_{i} \cdot k}\, \int\,  \frac{d^{4} l}{(2\pi)^{4}}\, \prod_{i}\,  \hdelta(2 p_{i} \cdot l)\, e^{\frac{i}{\hbar} b \cdot l}\,  {\cal M}^{\textrm{cl}}_{4}(p_{1},\, p_{2},\, l_{2})\,
\end{array}
\end{flalign}
The computation of L.H.S involves evaluation of the sub-leading soft operator on ${\cal M}_{4}$ and $\delta^{4}(l_{1} + l_{2})$.  To evaluate the action on ${\cal M}_{4}$ we note that, 
A direct verification shows that,
\begin{flalign}\label{firterappc}
{S}^{(1) \mu\nu}\, {\cal M}_{4}(\tilde{p}_{1},\, \tilde{p}_{2},\, p_{1},\, p_{2})\, =\, \frac{i\kappa}{2}\, \sum_{i=1}^{2}\, \frac{p_{i}^{(\mu} \hat{J}_{i}^{\nu)\lambda} k_{\lambda}}{p_{i} \cdot k}\,  {\cal M}^{\textrm{cl}}_{4}(p_{1},\, p_{2},\, l_{2}) + O(l^{\mu}) 
\end{flalign} 
L.H.S of eqn(\ref{appc1}) also involves action of the sub-leading soft operator on $\delta^{4}(l_{1} + l_{2})$ and this can be easily computed.
\begin{flalign}\label{s1ondel}
\begin{array}{lll}
{S}^{(1) \mu\nu}\, \delta^{4}(l_{1} + l_{2})&&=\\[0.4em]
&&\hspace*{-1.3in}\frac{i\kappa}{2}\, \sum_{i}\, [\, \frac{2 p_{i}^{(\mu} l_{i}^{\nu)}}{p_{i} \cdot k}\, k \cdot \frac{\partial}{\partial l_{1}}\, \delta^{4}(l_{1} + l_{2})\, -\,  \frac{p_{i}^{\mu} p_{i}^{\nu}}{(p_{i} \cdot k)^{2}}\, (l_{i} \cdot k)\, k \cdot \frac{\partial}{\partial l_{1}}\, \delta^{4}(l_{1} + l_{2})\, - l_{i}^{(\mu}\, \frac{\partial}{\partial l_{i}^{\nu)}}\delta^{4}(l_{1} + l_{2})\, ]
\end{array}
\end{flalign}
On substituting eqn. (\ref{s1ondel}) in L.H.S of the eqn.(\ref{appc1}), integrating by parts and keeping terms which are leading order in $l^{\mu}$, we get,
\begin{flalign}\label{secterappc}
\begin{array}{lll}
\hspace*{-0.2in}\int\, \frac{d^{4} l_{1}}{(2\pi)^{4}}\, \frac{d^{4} l_{2}}{(2\pi)^{4}}\, \prod_{i}\, \theta(l_{i}^{0})\, \hdelta(\tilde{p}_{i}^{2} - m_{i}^{2})\, e^{\frac{i}{\hbar} b \cdot l_{1}}\, {\cal M}_{4}\, \hat{S}^{(1) \mu\nu}\, \delta^{4}(l_{1} + l_{2}) \approx\\[0.4em]
\frac{i\kappa}{2}\, {\cal M}_{4}^{\textrm{cl}}\, \sum_{i=1}^{2}\, \frac{p_{i}^{(\mu} J_{i}^{\nu)\lambda} k_{\lambda}}{p_{i} \cdot k}\, \int\,  \frac{d^{4} l}{(2\pi)^{4}}\, \prod_{i}\,  \hdelta(2 p_{i} \cdot l)\, e^{\frac{i}{\hbar} b \cdot l} + \textrm{terms linear in $\frac{b^{\mu}}{\hbar}$}
\end{array}
\end{flalign}
where the remainder term (that is, terms which are linear in $\frac{b^{\mu}}{\hbar}$) appear to be super-classical and we need to be careful while taking classical limit. As a result, we obtain two types of contribution to the remainder term.\\
(1) Either replacing $\frac{1}{l_{2}^{2}}$ with $\frac{1}{-2 p_{2} \cdot l_{2}}$  before taking the classical limit or by keeping terms in ${\cal M}_{4}$ which are linear in $l$. Both of these terms are sub-leading in $\omega$. Hence using eqns. (\ref{firterappc}, \ref{secterappc}) and the argument presented above, proof of the approximate identity follows.

 \end{document}